\documentclass[useAMS,usenatbib]{mn2e}
\usepackage{times}
\usepackage{graphicx, latexsym, amssymb, amscd, psfrag, cancel}
\usepackage{epsfig}
\usepackage{float}
\usepackage{color}
\title[The Complete Local Volume Groups Sample II]{The Complete Local Volume Groups Sample - II. A study of the Central Radio Galaxies in the High--Richness Sub-sample}

\author[Konstantinos Kolokythas et al.]{Konstantinos Kolokythas,$^{1}$\thanks{e-mail: kkolok@iucaa.in}  Ewan O'Sullivan,$^{2}$ Somak Raychaudhury,$^{1,3}$ 
\newauthor{Simona Giacintucci,$^{4}$ Myriam Gitti,$^{5,6}$ Arif Babul$^{7,8}$} \\ \\
$^{1}$Inter-University Centre for Astronomy and Astrophysics, Pune University Campus, Ganeshkhind, Pune, Maharashtra 411007, India\\
$^{2}$Harvard-Smithsonian Center for Astrophysics, 60 Garden Street,
Cambridge, MA 02138, USA\\
$^{3}$School of Physics and Astronomy, University of Birmingham, Birmingham B15~2TT, UK\\
$^{4}$Naval Research Laboratory, 4555 Overlook Avenue SW, Code 7213, Washington, DC 20375, USA\\
$^{5}$Dipartimento di Fisica e Astronomia, Universit\`a di Bologna, via Gobetti 93/2, 40129 Bologna, Italy\\
$^{6}$INAF, Istituto di Radioastronomia di Bologna, via Gobetti 101, 40129 Bologna, Italy\\
$^{7}$Department of Physics and Astronomy, University of Victoria, Victoria, BC V8P 1A1, Canada\\
$^{8}$Center for Theoretical Astrophysics and Cosmology, Institute for Computational Science, University of Zurich, Winterthurerstrasse 190, 8057 Zurich, Switzerland }

\newcommand\kmsmpc{km~s$^{-1}$~Mpc$^{-1}$}

\begin{document}

\date{Accepted 2018 July 26; Received 2018 July 17; in original form 2018 April 11}

\pagerange{\pageref{firstpage}--\pageref{lastpage}} \pubyear{2018}
\maketitle

\label{firstpage}
 \begin{abstract}
We present a study of the radio properties of the dominant early--type galaxies in 26 galaxy groups, the high-richness sub-sample of the Complete Local-volume Groups Sample (CLoGS). Combining new 610~MHz and 235~MHz observations of 21 groups from the Giant Metrewave Radio Telescope (GMRT) with archival GMRT and Very Large Array (VLA) survey data, we find a high detection rate, with 92\% of the dominant galaxies hosting radio sources. The sources have a wide range of luminosities, 10$^{20}$$-$10$^{24}$~W~Hz$^{-1}$ in the 235 and 610~MHz bands. The majority (54\%) are point-like, but 23\% have radio jets, and another 15\% are diffuse radio sources with no clear jet/lobe morphology. Star formation may dominate the radio emission in 2 of the point-like systems and may make a significant contribution to a further 1$-$3, but is unlikely to be important in the remaining 21 galaxies. The spectral index of the detected radio sources ranges from very flat values of $\sim$0.2 to typical radio synchrotron spectra of $\sim$0.9 with only two presenting steep radio spectra with $\alpha_{235}^{610}>$1. We find that jet sources are more common in X-ray bright groups, with radio non-detections found only in X-ray faint systems. Radio point sources appear in all group environments irrespective of their X-ray properties or spiral fraction. We estimate the mechanical power (P$_{\rm{cav}}$) of the jet sources in the X-ray bright groups to be 10$^{41}$ $-$ 10$^{43}$ erg s$^{-1}$, with the two large-scale jet systems (NGC~193 and NGC~4261) showing jet powers two orders of magnitude greater than the radiative losses from the cool cores of their groups. This suggests that central AGN are not always in balance with cooling, but may instead produce powerful periodical bursts of feedback heating.

 \end{abstract}
 

 
 \section{Introduction} 

Much of the evolution of galaxies takes place in the group environment. Despite the small number of galaxies a typical group contains compared to a cluster, the role of galaxy groups in the construction of large-scale structures is fundamental. With approximately 70\% of galaxies located in groups \citep{Tully87}, they are the most common environment in the local Universe \citep{GellerHuchra83,Eke05}. In hierarchical structure formation, galaxy clusters are built up through the merger of groups \citep[e.g.][]{vandenBosch14,Haines17}, and the early evolution of rich cluster galaxies \citep[e.g.][]{Bekki99,MossWhittle00} is thus closely connected with the evolution of galaxies in groups. The study of groups is therefore essential in order to acquire a complete understanding of the evolution of galaxies along with the environmental processes involved \citep[e.g.][]{Forbes06}. 

Galaxy groups are an ideal laboratory to study the most efficient process in the morphological transformation of galaxies: merging \citep[e.g.][]{ToomreToomre72}. \citet{Mamon00} showed that galaxies in groups merge roughly two orders of magnitude more often than in rich clusters. Although groups represent shallower potential wells, gravity plays an important role due to the low velocity dispersion of these systems and the close proximity of members. The high rate of mergers and tidal interactions makes the group environment a locus of galaxy evolution and enhanced star formation \citep[e.g.,][]{MulchaeyZabludoff98,HashimotoOemler00}.

Many galaxy groups also possess hot gaseous haloes which, given the ubiquity of groups, make up a significant fraction of the baryonic component of the Universe \citep{Fukugita98}. However, whereas in virialised clusters the hot gas is always the dominant baryonic component, in groups the gas content varies, with numerous examples of systems in which the gas and stellar components are equal \citep{Lagana11} or in which the stellar component dominates  \citep[e.g.][]{Giodini09}. Hence, groups are the natural environment to study the origin and nature of this important mass component and its close relationship with galaxies and their evolution \citep{Forbes06,Liang16}. 

One aspect of this relationship which is of particular interest is the thermal regulation of the intra-group medium (IGM) via star formation or active galactic nuclei (AGN) fuelled by cooling gas (generally referred to as `feedback'). The most common observational evidence for the AGN mechanism are the radio bubbles and the X-ray cavities that the radio galaxies create in the hot IGM or the gaseous haloes of their host galaxies \citep{McNamara00,Fabian06}. While many studies have focused  on the most powerful AGN in massive clusters of galaxies \citep[e.g.,][]{McNamara05}, groups of galaxies are an important environment, where feedback may have the greatest impact on galaxy evolution and formation. 

Like the brightest cluster galaxies (BCGs), brightest group-dominant early-type galaxies (BGEs) are ideal targets for the study of the evolution of groups and massive galaxies. They are typically highly luminous, old elliptical galaxies, located near the centres of the IGM and dark matter halo \cite[e.g.,][]{Linden07,Stott10}. Their stellar kinematic properties cover the full range from field elliptical-like to BCG-like \citep{Loubser18} and their nuclei host super-massive black holes (e.g. \citealt{Rafferty06}), with many of the BGEs revealing the activity of their nuclei by radio emission and even with radio jets that deposit their energy back to the IGM (see \citealt{McNu}).

Earlier studies have shown that $\sim30$\% of the most massive galaxies exhibit radio continuum emission \citep[e.g.,][]{Best05,Shabala08}, with relevant group/cluster studies of central BGEs/BCGS in the local Universe suggesting a high detection rate (80-90\%) in radio \citep[e.g.,][]{Magliocchetti07,Dunn10}. It is also known that radio AGN in early-type galaxies are more common in higher density environments (groups or clusters; \citealt{Lilly09,Bardelli10,Malavasi15}). In such environments, the radio activity in cluster and group-central AGN is most effectively enhanced by cluster/group merging, or in the case of lower density environments by `inter-group' galaxy-galaxy interactions and mergers  \citep{Miles04,TaylorBabul05}. It appears that mergers and interactions direct gas to the AGN in early type galaxies, resulting in radio emission and the launching of jets, whereas in late-type galaxies, the corresponding processes would trigger or increase radio emission from star formation (SF) \citep[e.g.,][]{Vollmer01}.

Galaxy clusters may also host diffuse structures associated with the ICM \citep[see][for a general classification of cluster radio sources]{Kempner04}. These include  \textit{mini-halos}, typically found at the center of cool-core clusters around powerful radio galaxies, and thought to be powered by turbulence in the cooling region or perhaps by minor mergers \citep{Ferrari08,Feretti12,Brunetti14,Gitti15,Giacintucci17}, \textit{radio relics}, narrow, linear arcs of radio emission found far from the cluster core, associated with large-scale shocks and steep spectral indices, and \textit{radio halos}, which are thought to be produced through turbulent re-acceleration of electrons by a cluster merger event. In galaxy groups the diffuse radio sources seen are mainly associated with the central galaxy and not the IGM but their production mechanism is still uncertain.



In this paper we present a study of the radio properties of the dominant galaxies of the 26-group CLoGS high-richness sub-sample, including new Giant Metrewave Radio Telescope (GMRT) 235 and 610~MHz observations of 21 systems. We present the properties of the central radio sources, examine their environment and provide qualitative comparison between the GMRT radio data and the X-ray data for each group. The CLoGS sample and the X-ray properties of the groups are described in more detail in \citet[hereafter Paper~I]{OSullivan17}. The paper is organized as follows: In Section 2 we present the sample of galaxy groups, whereas, in Section 3 we describe the GMRT and X-ray observations along with the approach followed for the radio data reduction. In Section 4 we present the radio detection statistics of the BGEs, and in Section 5 their radio properties (morphology and power) along with information on their spectral index and the possible contribution from star formation on their radio emission. Section 6 contains the discussion of our results focusing on detection statistics, environmental properties of the radio sources and the energetics of jet systems with cavities. The summary and the conclusions are given in Section 7. The radio images and information on the central galaxies of this sample work are presented in Appendix~A, the values of star formation rates (SFR$_{FUV}$) and expected radio power due to star formation for the related BGEs are shown in Appendix~B and the information on the flux density distribution for each system is given in Appendix~C. Throughout the paper we adopt the $\Lambda$CDM cosmology with $H_o=71$ \kmsmpc, $\Omega_m$ = 0.27, and $\Omega_\Lambda$ = 0.73. The radio spectral index $\alpha$ is defined as $S_\nu \propto \nu^{-\alpha}$, where $S_\nu$ is the flux density at the frequency $\nu$.


\section{The Complete Local-Volume Groups Sample}
We present in this section only a short general description of the Complete Local-Volume Groups Sample (CLoGS). CLoGS is an optically-selected, statistically-complete sample of groups in the nearby universe, chosen to facilitate studies in the radio, X-ray and optical bands, with the detection of an X-ray luminous IGM providing confirmation that groups are gravitationally bound systems. The sample is intended to facilitate investigations of a number of scientific questions, including the role of AGNs in maintaining the thermal balance of the IGM.

A detailed description of the CLoGS sample selection criteria and the X-ray properties of the high--richness sub-sample can be found in Paper~I. CLoGS is an optically selected sample of 53 groups in the local Universe ($\leq$80 Mpc), drawn from the shallow, all-sky Lyon Galaxy Group catalog sample (LGG; \citealt{Garcia93}). Groups were selected to have a minimum of 4 members and at least one luminous early-type galaxy (L$_B$ $>$ 3$\times$10$^{10}$ L$_\odot$). Declination was required to be $>$30$^\circ$ to ensure visibility from the GMRT and Very Large Array (VLA). 

Group membership was expanded and refined using the HyperLEDA catalog \citep{Paturel03}, and group mean velocity dispersion and richness $R$ were estimated from the revised membership, where richness is defined as the number of member galaxies with log L$_B$ $\geq$ 10.2. Systems with $R>10$ correspond to known galaxy clusters and were excluded, as were six groups that have $R$ = 1, since they are not rich enough to give a trustworthy determination of their physical parameters. This selection process produced a 53--group statistically complete sample, which we then divided into two sub-samples: i) the 26 high-richness groups with $R$ = 4$-$8 and ii) the 27 low-richness groups $R$ = 2$-$3. The BGE was assumed to lie at or near the centre of the group, and used as the target for observations. In this paper we examine the radio properties of the high--richness sub-sample.

\begin{table*}
\begin{minipage}{\linewidth}
 \caption{Details of our GMRT observations analysed here, along with information on data used in the previous study of \citet{Simona11} and  \citet{David09}. For each source the first line displays the details for the 610~MHz and the second line for the 235~MHz. The columns give the LGG (Lyon Groups of Galaxies) number for each group, the BGE name, observation date, frequency, time on source, beam parameters and the rms noise in the resulting images.}
 \centering
 \label{GMRTtable}
\begin{tabular}{|c|c|c|c|c|c|c|}
 
 \hline \hline
  Group Name & BGE       & Observation & Frequency &    On source      &   Beam, P.A.                  & rms \\
 
    LGG      &           &    Date    &  (MHz)    &  Time (minutes)    & (Full array, $''\times'',{}^{\circ}$) & mJy beam$^{-1}$ \\

 \hline
  18  &NGC 410  &  2011 Jul       &    610    &     218      &  $6.97\times4.09$, 71.48   & 0.05 \\ 
      &      &   2011 Jul      &    235    &     218      &  $17.87\times10.95$, 50.40   & 0.40 \\ 
  27   &NGC 584  &   2011 Jul      &    610    &     200      &  $6.82\times3.22$, 59.80   & 0.20 \\ 
      &    &   2011 Jul      &    235    &     200      &  $17.87\times10.95$, 50.40   & 1.20 \\ 
  31   &NGC 677  &   2013 Oct      &    610    &     200      &  $5.26\times4.34$,  68.59   & 0.04 \\ 
      &     &   2013 Oct      &    235    &     200      &  $12.94\times10.76$, 66.90   & 1.20 \\ 
  42   &NGC 777   &   2010 Dec      &    610    &     338      &  $5.71\times4.76$, -70.53   & 0.15 \\ 
      &     &   2010 Dec      &    235    &     338      &  $12.96\times19.94$, 65.78  & 0.40 \\ 
  58   &NGC 940    &   2010 Dec      &    610    &     360      &  $5.81\times4.76$, -75.11   & 0.06 \\ 
      &     &   2010 Dec      &    235    &     360      &  $13.27\times11.16$, 66.88  & 0.30 \\  
  61   & NGC 924  &   2011 Jul      &    610    &     125      &  $5.52\times3.74$,  56.25   & 0.05 \\ 
      &     &   2011 Jul      &    235    &     125      &  $12.40\times10.43$, 59.19   & 0.30 \\ 
  66   &NGC 978    &   2010 Dec      &    610    &     341      &  $5.81\times4.65$, -81.63   & 0.06 \\ 
      &     &   2010 Dec      &    235    &     341      &  $14.13\times11.69$, 76.74  & 0.40 \\ 
  72   &NGC 1060   &   2010 Dec      &    610    &     286      &  $3.97\times3.54$, -86.98   & 0.09 \\ 
      &     &   2010 Dec      &    235    &     286      &  $13.97\times11.87$, 81.36  & 0.50 \\ 
  80   &NGC 1167   &   2011 Jul      &    610    &     139      &  $7.41\times4.04$, 68.27  & 0.06 \\ 
      &     &   2011 Jul      &    235    &     139      &  $16.08\times10.40$, 66.80  & 0.6 \\ 
  103   &NGC 1453   &   2011 Nov      &    610    &     187      &  $7.99\times4.60$, 48.98  & 0.06 \\ 
      &     &   2013 Oct      &    235    &     258      &  $16.06\times11.34$, 58.47  & 0.60 \\ 
  158   &NGC 2563   &   2010 Dec      &    610    &     344      &  $5.10\times4.60$, -57.85   & 0.07 \\ 
      &     &   2010 Dec      &    235    &     344      &  $11.88\times9.93$, 53.00  & 0.30 \\ 
  185   &NGC 3078   &   2010 Dec      &    610    &     369      &  $6.61\times4.71$,  2.38  & 0.20 \\ 
      &     &   2010 Dec      &    235    &     369      &  $16.02\times10.99$, 20.72  & 0.50 \\
  262   &NGC 4008   &   2011 Apr      &    610    &     168      &  $5.76\times4.52$, 78.93   & 0.05 \\ 
      &     &   2011 Apr      &    235    &     168      &  $14.38\times11.78$, 89.31  & 1.30 \\  
  276   &NGC 4169   &   2011 Apr      &    610    &     168      &  $5.47\times4.43$, 61.96   & 0.08 \\ 
      &     &   2011 Apr      &    235    &     168      &  $13.94\times10.78$, 72.62   & 1.20 \\ 
  278   &NGC 4261   &   2009 Feb      &    610    &     270      &  $7.32\times4.77$, 76.62  & 1.00 \\ 
      &     &   2009 Feb      &    235    &     270      &  $15.30\times11.00$, 67.17  & 1.60 \\ 
  310   &ESO 507-25 &   2011 Apr      &    610    &     167      &  $9.40\times8.43$,  2.59   & 0.10 \\ 
      &     &   2011 Apr      &    235    &     167      &  $15.59\times12.22$, 7.69   & 0.50 \\ 
  345   &NGC 5084   &  2011 Jul       &    610    &     150      &  $5.91\times5.06$, 29.68   & 0.09 \\ 
      &     &   2011 Jul      &    235    &     150      &  $15.71\times12.62$, -4.35   & 0.65 \\ 
  351   &NGC 5153   &   2011 Apr      &    610    &     167      &  $7.63\times4.81$,  26.32   & 0.06 \\ 
      &     &   2011 Apr      &    235    &     167      &  $16.01\times11.25$, 13.68   & 0.30 \\ 
  363   &NGC 5353   &   2011 Apr      &    610    &     207      &  $5.92\times4.44$,  -76.02   & 0.06 \\ 
      &     &   2011 Apr      &    235    &     207      &  $16.91\times12.00$, -79.87   & 0.60 \\ 
  393   &NGC 5846   &   2012 Mar      &    235    &     145      &  $14.02\times11.27$, 85.70  & 0.50 \\   
  402   &NGC 5982  &   2011 Apr      &    610    &     237      &  $7.99\times6.47$,  -81.04   & 0.09 \\ 
      &     &   2011 Apr      &    235    &     237      &  $19.42\times14.43$, -63.66   & 0.40 \\ 
  421   &NGC 6658   &   2011 Apr      &    610    &     236      &  $5.03\times4.35$, 55.03   & 0.05 \\ 
      &     &   2011 Apr      &    235    &     236      &  $12.82\times10.30$, 69.71   & 0.60 \\ 
  \hline

  Previous work &   &           &          &      &          &      \\

 \hline
 9  &NGC 193\footnote{\cite{Simona11}}  & 2007 Aug  &     610        &    110     &  $7.0\times6.0$, 0                  & 0.08    \\
    &                                   & 2008 Aug  &     235        &    120       &  $13.5\times12.9$, -55        & 0.80   \\
 117&  NGC 1587$^a$                     & 2006 Aug &     610        &       200      &  $5.7\times4.7$, 67              & 0.05     \\
     &                                  & 2008 Aug  &     235        &      120      &  $17.2\times11.0$, 46                    & 1.00    \\
 338& NGC 5044\footnote{\cite{David09}} & 2008 Feb &     610       &      130       &  $6.2\times4.4$, 41.9                   & 0.10   \\
     &                                  & 2008 Feb  &     235       &      140      &  $15.9\times11.3$, -4.6                    & 0.80   \\
 393 &    NGC 5846$^a$                  &   2006 Aug &      610      &      140      &    $6.0\times5.5$, -84             & 0.06   \\
 473 &    NGC 7619$^{a}$                &   2007 Aug   &     610       &      100        &  $6.1\times4.6$, 39                   & 0.10    \\
     &                                  &   2008 Aug   &      235       &     120       &  $34.6\times11.3$, -41.4                    & 1.20   \\
 \hline
 \end{tabular}
\end{minipage}
 \end{table*}

\section{OBSERVATIONS AND DATA ANALYSIS}

\subsection{GMRT radio observations and data analysis} 

With the exception of systems for which suitable archival data was already available, the CLoGS galaxy groups were observed using the GMRT in dual-frequency 235/610 MHz mode during 2010 December, 2011 April, July and November, 2012 March and 2013 October. The average total time spent on each source was approximately 4 hours. The data for both frequencies were recorded using the upper side band correlator (USB) providing a total observing bandwidth of 32 MHz at both 610 and 235~MHz. 
The data for both frequencies were collected with 512/256 channels at a spectral resolution of 65.1/130.2 kHz per channel and were analysed and reduced using the NRAO Astronomical Image Processing System (\textsc{aips}) package. In Table~\ref{GMRTtable} we summarize the details of the observations, where we report the observing date, frequency, total time on source, the half-power beamwidth (HPBW) and the rms noise level (1 $\sigma$) at full resolution. 

The procedure of the GMRT data analysis followed is the same as described in \citet{Kolokythas14}. The data were edited to remove radio-frequency interference (RFI) by initially removing data with extreme phase difference from the phase and flux calibrators, and then from the source. The task \textsc{setjy} was used and the flux density scale was set to ``best VLA values (2010)\footnote{http://www.vla.nrao.edu/astro/calib/manual/baars.html}''. The data were then calibrated using the task \textsc{calib} using uniform weighting, and the flux density scale was defined using the amplitude calibrators observed at the beginning and end at each of our observations. Then a calibrated, RFI-free channel was used for bandpass calibration, in order to relate the amplitude difference between all channels. Using task \textsc{splat}, the channels were averaged in order to increase signal-to-noise while minimizing the effects of band smearing. 

For imaging purposes, the field of view of the GMRT was split into multiple facets approximating a plane surface in the imaging field. The normal number of facets created at 610 MHz was 56 (with cellsize 1.5$''$) and 85 at 235 MHz (with cellsize 3$''$). The final images give a field of view of $\sim 1.2^{\circ}\times1.2^{\circ}$ at 610 MHz and $\sim 3^{\circ}\times3^{\circ}$ at 235 MHz. Task \textsc{imagr} was then used repeatedly, applying phase-only self-calibration before every iteration. The majority of the noise remaining in our images arises from calibration uncertainties with phase errors at the lowest frequencies originating from rapidly varying ionospheric delays. The presence of bright sources in the field can also affect the noise level in our final images, reducing the dynamic range.

The final images were corrected for the primary beam pattern of the GMRT using the task \textsc{pbcor}. The mean sensitivity (1$\sigma$ noise level) of the analysis achieved in the final full resolution images is $\sim$0.08 mJy at 610 MHz and $\sim$0.6 mJy at 235~MHz (see Table~\ref{GMRTtable}). The theoretical values of noise calculated for our observations are 29 $\mu$Jy for 610 MHz and 80 $\mu$Jy for 235 MHz\footnote{Calculated using the rms noise sensitivity equation in \S 2.1.6 from the GMRT Observer's Manual: http://gmrt.ncra.tifr.res.in/gmrt$\_$hpage/Users /doc/manual/Manual$\_$2013/manual$\_$20Sep2013.pdf}. The mean sensitivity achieved for 610 MHz is comparable to the theoretical noise level while at 235~MHz the mean sensitivity is considerably higher than the theoretical level. We note that the full resolution of the GMRT is $\sim13''$ at 235~MHz and $\sim6''$ at 610 MHz with the \textit{u-v} range at 235~MHz being $\sim0.05 - 21$ k$\lambda$ and $\sim0.1 - 50$ k$\lambda$ at 610~MHz. The error associated with the flux density measurements is the typical uncertainty in the residual amplitude calibration errors. That is $\sim$5\% at 610~MHz and $\sim$8\% at 235~MHz \citep[e.g.,][]{Chandra04}.


Out of the 26 BGEs in the CLoGS high-richness sub-sample, 20 were analyzed in this study at both GMRT 610/235~MHz and one (NGC~5846) at 235~MHz. The GMRT observations and images for NGC~193, NGC~7619, NGC~1587 at both 235/610 MHz and NGC~5846 at 610~MHz, are drawn from the earlier study of \citet{Simona11}, those for NGC~5044 from \citet{David09, David11} and for NGC~4261 from \citet{Kolokythas14}. Analysis of those observations is described in detail in those works. For sources with resolved structure at both GMRT frequencies (see Table~\ref{corespix}), we created spectral index maps using the task \textsc{comb} in \textsc{aips} with images having the same \textit{uv} range, resolution and cellsize. 1400 MHz data were drawn primarily from the NRAO VLA Sky Survey (NVSS, \citealt{Condon98}) and the Faint Images of the Radio Sky at Twenty centimeters Survey (FIRST, \citealt{Becker95}). For some sources, flux densities were drawn from \citet{Brown11}, who used measurements from the NVSS, Green Bank Telescope, and Parkes Radio Telescope, as well as from \citet{Kuhr81} and \citet{Condon02} studies for NGC~4261 and NGC~193 respectively.

\subsection{X-ray observations}
\label{sec:Xray}
Paper~I describes the X-ray properties of the high-richness group sample, based on observations from the \textit{Chandra} and \textit{XMM-Newton} observatories. We summarize here only the main results of the X-ray analysis, as our focus is the comparison between the radio and X-ray structures.

Of the 26 groups of the high-richness sub-sample, 19 were observed by \textit{XMM-Newton}, 13 by \textit{Chandra}, and 8 have data from both X-ray observatories. An IGM is detected in 14/26 systems ($\sim$54\%) with a further 3/26 systems ($\sim$12\%) containing galaxy-scale haloes of X-ray emitting gas (extent $<65$~kpc, luminosity $<10^{41}$~erg~s$^{-1}$) associated with the dominant early-type galaxy. The remaining nine groups show only point-like X-ray emission ($\sim$35\%) from AGN and stellar sources in member galaxies. In the groups with detected gas haloes, the typical halo temperatures are found to be in the range $\sim0.4-1.4$ keV, corresponding to masses in the range $M_{500}\sim0.5-5\times 10^{13} M_{\odot}$. The X-ray luminosities of the systems were estimated in the $0.5-7$ keV band to be between  $L_{X,R500}\sim2-200\times10^{41}$ erg s$^{-1}$. Paper~I classes systems that have a central decline in their X-ray temperature profile (at greater than 3$\sigma$ significance) as cool core. By this definition, roughly one third (5/14) of the X-ray confirmed groups are non-cool-core systems, while the remaining two thirds (9/14) have a cool core. 

Cavities in the IGM, associated with the activity of central radio jet sources, have been identified in four systems (see e.g., \citealt{David09,David11,David17} for NGC~5044, \citealt{Allen06}, \citealt{Dong10} and \citealt{Machacek11} for NGC~5846, \citealt{Bogdan14} for NGC~193, \citealt{ewan4261} for NGC~4261). In this paper, we use cavity properties to examine the power output of AGN jets. Where necessary, cavity power estimates are made using methods similar to those described in \citet{OSullivan11}. The cooling luminosity (L$_{cool}$), defined as the luminosity of the gas in the volume within which the cooling time is $\leq$3 Gyr, was calculated in the 0.5-7 keV band, based on the luminosity and cooling time profiles measured in Paper~I, using a PROJCT*WABS*APEC model in \textsc{Xspec}.

\section{Radio detections in brightest group galaxies} 

Using the 235 and 610~MHz GMRT images, and the available NVSS and FIRST catalog data, we identified sources in the BGE of each high-richness group, determined their morphology and measured their flux densities and extent. Measured flux densities are shown in Table~\ref{Sourcetable}. GMRT images of the groups and more details of these sources are presented in Appendix~\ref{AppA}.

\begin{figure}
\centering
\includegraphics[width=0.49\textwidth]{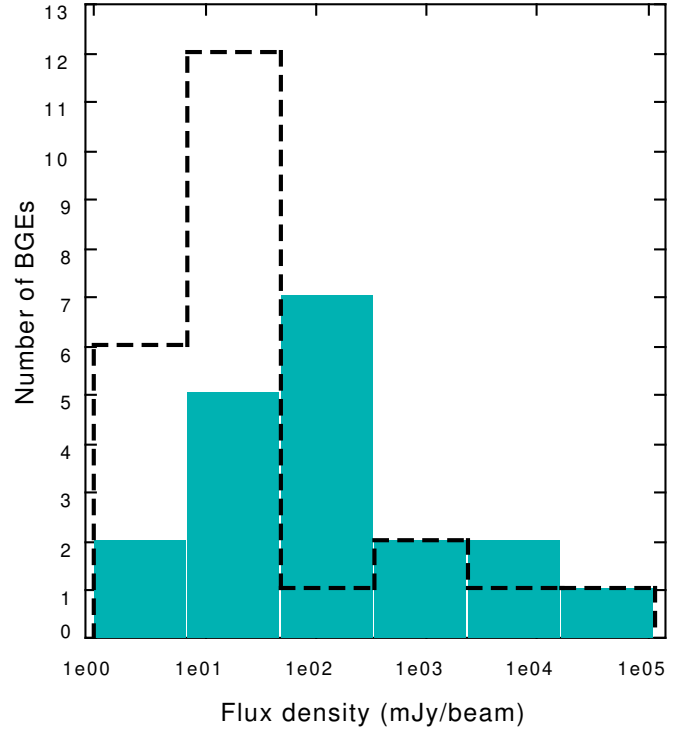}
\caption{Flux density distribution of the central brightest group early-type galaxies of the high-richness sample at 610~MHz (black dashed line) and 235~MHz (cyan columns) in CLoGS high-richness sample.}
\label{235Flux}
\end{figure}

We find a high radio detection rate of 92\% (24 of 26 BGEs), with only two galaxies (NGC~5153 and NGC~6658) being undetected at 235, 610 and 1400 MHz. Considering only the GMRT data,
23 of the 26 BGEs are detected at 235 and/or 610~MHz (89\%), with NGC~584 being the only galaxy detected at 1.4~GHz alone. In Figure~\ref{235Flux}, we present the flux density distribution at 610~MHz and 235~MHz. The majority of the BGEs have flux densities in the range $10-100$~mJy with the flux density distribution at 610~MHz exhibiting a narrow sharp peak compared to the 235~MHz where the spread in flux densities is bigger.


\begin{figure}
\centering
\includegraphics[width=0.49\textwidth]{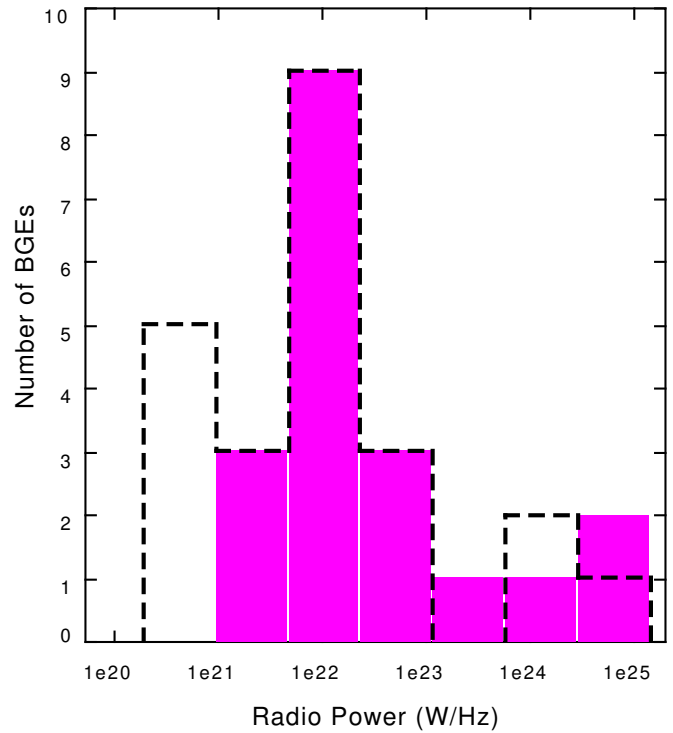}
\caption{Radio power of the detected BGEs at 610~MHz (black dashed line) and 235~MHz (pink columns) in CLoGS high-richness sample. }
\label{Lradio}
\end{figure}

We estimate the limiting sensitivity of our sample based on the typical noise level of our observations and the maximum distance for the observed groups, 78~Mpc. We define radio sources as detected in our GMRT data if they reach a 5$\sigma$ level of significance above the noise. Using the mean r.m.s. from our GMRT observations for each frequency ($\sim$80 $\mu$Jy beam$^{-1}$ at 610~MHz and $\sim$600 $\mu$Jy beam$^{-1}$ at 235~MHz), we find that we should be sensitive to any source with power $>2.9\times10^{20}$ W Hz$^{-1}$ at 610~MHz or $>2.2\times10^{21}$ W Hz$^{-1}$ at 235~MHz. For comparison, the equivalent limit for NVSS 1400~MHz power sensitivity at this distance and level of significance is $>1.7\times10^{21}$ W Hz$^{-1}$. We note that sources in nearby groups may be detected at $>$5$\sigma$ significance with lower powers; these limits reflect the sensitivity of the sample as a whole.

\begin{table*} 
\caption{Radio flux densities and spectral indices for our targets. The columns list the BGE name, redshift, flux density of each source at 235 and 610~MHz, the 235$-$610~MHz spectral index, the flux density at 1.4~GHz (drawn from the literature), the 235$-$1400~MHz spectral index, and the radio power at 235 and 610~MHz. All upper limits shown here from our analysis are 5 $\times$ r.m.s. Three galaxies show no radio emission detected from our GMRT observations. The sensitivity of the GMRT observations is on average $\sim$80 $\mu$Jy beam$^{-1}$ at 610~MHz and $\sim$600 $\mu$Jy beam$^{-1}$ at 235~MHz. The references for the 1.4~GHz flux densities and the GMRT measurements from previous works are listed at the bottom of the table. \label{Sourcetable}}
\begin{center}
\begin{tabular}{lcccccccc}
\hline 
 Source& Redshift&S$_{235 MHz}$ & S$_{610 MHz}$ & $\alpha_{235 MHz}^{610 MHz}$ & S$_{1.4GHz}$ & $\alpha_{235 MHz}^{1400 MHz}$ & P$_{235 MHz}$ & P$_{610 MHz}$ \\ 
      & ($z$) & $\pm8\%$ (mJy) & $\pm5\%$ (mJy) &   ($\pm$0.04)      &   (mJy)     &      & (10$^{23}$ W Hz$^{-1}$) & (10$^{23}$ W Hz$^{-1}$)  \\
\hline
 NGC 410   &  0.017659 & 28.5      &      13.6 &  0.78 & 6.3$\pm0.6^a$   & 0.85$\pm$0.05 &  0.201 &  0.096 \\
 NGC 584   &  0.006011 & $\leq$6.0 & $\leq$1.0 &   -    & 0.6$\pm0.5^b$   &       -        &  -      &     -   \\
 NGC 677   &  0.017012 & 99.2      &      45.6 &  0.81 & 20.6$\pm1.6^a$  & 0.88$\pm$0.05 &  0.720 &  0.331 \\
 NGC 777   &  0.016728 & 20.9      &      10.2 &  0.75 & 7.0$\pm0.5^b$   & 0.61$\pm$0.05 &  0.133 &  0.065 \\
 NGC 940   &  0.017075 &  3.3      &       4.3 & -0.28 &  -               &         -      &  0.021 &  0.028 \\
  NGC 924   &  0.014880 & $\leq$1.5 &       1.7 &    -   &   -              &         -      &    -    &  0.008 \\
 NGC 978   &  0.015794 & $\leq$2.0 &       1.3 &   -    &   -              &          -     &    -    &  0.007 \\
 NGC 1060  &  0.017312 & 28.5      &      12.4 &  0.87 & 9.2$\pm0.5^b$   & 0.63$\pm$0.04 &  0.197 &  0.086 \\
 NGC 1167  &  0.016495 & 4018.4    &    2295.3 &  0.59 & 1700$\pm100^b$  & 0.48$\pm$0.04 & 24.751 & 14.138 \\
 NGC 1453  &  0.012962 & 47.1      &      40.4 &  0.16 & 28.0$\pm1^b$    & 0.29$\pm$0.04 &  0.221 &  0.190 \\
 NGC 2563  &  0.014944 & $\leq$1.5 &       1.3 &  -     & 0.3$\pm0.5^b$   &     -          &    -    &  0.007 \\
 NGC 3078  &  0.008606 & 582.8     &     384.5 &  0.44 & 310$\pm10^b$    & 0.35$\pm$0.04 &  0.802 &  0.529 \\
 NGC 4008  &  0.012075 & 24.8      &      16.4 &  0.43 & 10.9$\pm0.5^b$  & 0.46$\pm$0.04 &  0.086 &  0.057 \\
 NGC 4169  &  0.012622 & $\leq$6.0 &       3.0 &   -    & 1.07$\pm0.15^c$ &       -        &    -    &  0.007 \\
 NGC 4261$^d$  &  0.007465 & 48500     &     29600 &  0.52 & 19510$\pm410^e$ & 0.51$\pm$0.04 & 59.193 & 36.126 \\
 ESO 507-25 & 0.010788 & 55.2      &      46.6 &  0.18 & 24.0$\pm2^b$    & 0.47$\pm$0.05 &  0.133 &  0.112 \\
 NGC 5084  &  0.005741 & 53.9      &      36.2 &  0.42 & 46.6$\pm1.8^a$  & 0.08$\pm$0.04 &  0.034 &  0.023 \\
 NGC 5153  &  0.014413 & $\leq$1.5 & $\leq$0.3 &   -    &     -              &       -      &     -     &   -     \\
 NGC 5353  &  0.007755 & 65.8      &      45.8 &  0.38 & 41.0$\pm1.3^a$  & 0.27$\pm$0.04 &  0.096 &  0.067 \\
 NGC 5846  &  0.005717 & 58.5      &    -       &  0.52 & 21.0$\pm1^b$    & 0.57$\pm$0.04 &  0.047 &    -    \\
 NGC 5982  &  0.010064 & 4.5       &       1.9 &  0.90 &  0.5$\pm0.5^b$  & 1.23$\pm$0.04 &  0.010 &  0.004 \\
 NGC 6658  &  0.014243 & $\leq$3.0 & $\leq$0.3 &     -  &      -           &      -         &    -    &     -   \\
\hline
 \multicolumn{9}{l}{Previous work}\\
\hline
 NGC 193$^f$  & 0.014723 & 5260   &      3184 & 0.53    & 1710$\pm102^g$ & 0.62$\pm$0.04 & 34.217 & 20.710 \\
 NGC 1587$^f$ & 0.01230  &  655    &       222 & $>$1.13 &  131$\pm5^b$   & 0.90$\pm$0.04 &  2.042 &  0.692 \\
 NGC 5044$^h$ & 0.009280 &  229    &      38.0 & 1.88    &   35$\pm1^b$   & 1.05$\pm$0.04 &  0.399 &  0.066 \\
 NGC 5846$^f$ & 0.005717 &   -      &      36.0 &   -      &      -          &       -        &    -    &  0.029 \\
 NGC 7619$^f$ & 0.012549 & 56.3    &      32.3 & 0.58    &   20$\pm1^b$   & 0.58$\pm$0.04 &  0.195 &  0.112 \\
\hline
\end{tabular}
\end{center}
$^a$ \citet{Condon98}, $^b$ \citet{Brown11}, $^c$ \citet{Becker95}, $^d$ \citet{Kolokythas14} $^e$ \citet{Kuhr81}, $^f$ \citet{Simona11}, $^g$ \citet{Condon02}, $^h$ \citet{David09}
\end{table*}

\section{Radio properties of the brightest group galaxies}
\subsection{Radio morphology} 

The radio-detected BGEs of the CLoGS high-richness sample exhibit a rich variety of structures that differ dramatically in size, radio power and morphology. The physical scale of their radio emission ranges from a few kpc (point sources; galactic scale) to several tens of kpc (large jets; group scale). The radio structures observed in central galaxies consist of the normal double-lobed radio galaxies, small-scale jets, point sources and irregular-diffuse structures with no clear jet/lobe structure.  

Our morphology classification is based on our dual-frequency GMRT analysis, the NVSS and FIRST data surveys and where needed, on the literature. Including the BGEs that are non-radio emitting at any frequency, we classify the radio emission seen from the BGEs in the following categories: i) \textit{point-like} - unresolved radio point source, ii) \textit{diffuse} emission - extended but amorphous with no preference in orientation,  iii) \textit{small-scale jets} - confined within the stellar body of the host galaxy, with extent $<20$~kpc, and iv) \textit{large-scale jets} - extending $>20$~kpc, beyond the host galaxy and into the intra-group medium. In this class of radio sources we also include the subclass of \textit{remnant jet} systems. These are systems which previous studies have shown to be the products of past periods of jet activity, and which are now passively ageing, without significant input from the AGN. Lastly, the systems that have no radio source detected are classed as category v) \textit{absence of radio emission}. Table~\ref{t:listtable2} lists the morphology of each source.


We note that some of our systems can be placed in more than one of the radio morphology classes described above. For example, the radio source in NGC~1167 appears as a \textit{small-scale jet} in our 610~MHz image, at 235~MHz the emission is \textit{point-like}, but deeper 1.4~GHz observations have revealed the presence of a pair of old radio lobes extending $>100$~kpc \citep{Shulevski12,Brienza16}. Since our interest is in the interaction between radio sources and their environment, in cases where the source can be placed in more than one category, we list the most extended class.

We find that 4 BGEs host large-scale jets (of which 2 are remnant jets), 2 host small-scale jets, 4 possess diffuse emission, while the remaining 14 radio-detected BGEs host point-like radio sources.  

The two currently-active large-scale jet systems are NGC~193 and NGC~4261. Both are  Fanaroff-Riley type I (FR I; Fanaroff \& Riley 1974) radio galaxies (4C+03.01 and 3C~270 respectively) with their jet/lobe components extending several tens of kpc away from the host galaxy. The two sources present similar morphology with roughly symmetrical double-lobed structure and bright, straight jets. Of the remnant sources, NGC~1167 (4C+34.09) is also roughly symmetrical \citep{Shulevski12}, while NGC~5044 has a single-sided bent jet and detached lobe. 

Two BGEs, NGC~5153 and NGC~6658, lack radio emission in any of the frequencies we examined. 
A third galaxy, NGC~584, is undetected in our GMRT data, but \citet{Brown11} find a marginal detection of a faint point source at 1.4~GHz ($0.6\pm0.5$~mJy). The noise levels of the GMRT observations ($\leq1$~mJy at 610 and $\leq6$ mJy at 235~MHz; 5$\sigma$ significance) are consistent with the non-detection of such a faint source.


\subsection{Radio power in CLoGS BGEs}

Using the measured radio flux densities, we calculated the radio power in each BGE for each frequency by: 

\begin{equation}
\label{Power610}
P_{\nu}=4 \pi D^2 (1+\rm z)^{(\alpha-1)}S_{\nu},
\end{equation}

where D is the distance to the source, $\alpha$ is the spectral index, $z$ the redshift and S$_{\nu}$ is the flux density of the source at frequency $\nu$. In the case where no spectral index is known for the source (i.e., it is only detected at one frequency), $\alpha$ is by default set to $0.8$ \citep{Condon92}, which is the typical value for extragalactic radio sources. 

In figure~\ref{Lradio}, we show the radio power distribution of the BGEs in our sample at 235 and 610~MHz.
The majority of the galaxies are having low powers in the range 10$^{21}$ $-$ 10$^{23}$ W Hz$^{-1}$. 
Only one of our BGEs exhibits power in the range $10^{23}-10^{24}$ W Hz$^{-1}$ (NGC 1587), but the sample contains three high-power sources with P$_{235MHz}>10^{24}$ W Hz$^{-1}$. All three are bright jet hosting radio galaxies (NGC~4261, NGC~193 and NGC~5044). We note that typical relevant values of radio power at 235 and 610 MHz in BCGs ranges between $\sim10^{23}- 5\times10^{26}$ W Hz$^{-1}$ \citep[see e.g. Table 2,][]{Yuan16}. 

\subsubsection{Radio-loudness in CLoGS BGEs}  

Only three of our BGEs can be considered radio-loud. We follow \citet{Best05} in defining radio-loud systems as having P$_{1.4GHz}>10^{23}$~W~Hz$^{-1}$, and note that the NVSS and FIRST sensitivity limits mean that those of our systems which are undetected at 1.4~GHz must have radio powers below this limit. The three radio-loud systems are shown in Table~\ref{radioloud23}.


\citet{LinMohr07}, using the same threshold of radio-loudness,
found that while in high-mass clusters (M$_{200}>10^{14.2}$~M$_\odot$) $\sim$36\% of the BCGs have P$_{1.4GHz}>10^{23}$ W Hz$^{-1}$, this fraction drops to only $\sim$13\% in low-mass clusters and groups (10$^{13}<M_{200}<10^{14.2}$). The latter percentage is very similar to that observed in our sample, $\sim$12\% (3/26).

\subsection{Contributions from star formation}

While AGN are clearly the origin of the radio jet sources in our sample, we must consider whether star formation might contribute to the radio emission of the other sources, particularly those with low radio luminosities. We discuss the diffuse sources in detail in \S6.5, but note here that their luminosity and morphology makes star formation unlikely to be their dominant source of radio emission. For the point-like sources, the beam sizes limit the emission to regions a few kiloparsecs across, but except in the most luminous systems, it is possible that we could mistake a compact central star forming region for AGN emission. Our CO survey of CLoGS BGEs confirms that the sample includes galaxies whose radio emission would be consistent with the star formation expected from their molecular gas reservoirs, if they were forming stars at rates similar to those in spiral galaxies (O'Sullivan et al., in prep.)

We will investigate star formation in these galaxies more thoroughly in a later paper (Kolokythas et al., in prep.), but for now we need only  determine whether it is likely to be a significant source of radio emission. To do this, we use Far-Ultraviolet (FUV) fluxes from the Galaxy Evolution Survey \citep{Martin05} GR6 catalog\footnote{$http://galex.stsci.edu/GR6/?page=mastform$} to estimate the star formation rate (SFR) using the calibration from \citet{Salim07}. We then calculate the expected 610~MHz radio emission from star formation at that rate using the relation of \citet{Garn09}. A table of SFR and expected radio power for the BGEs with point-like and diffuse radio emission is provided in Appendix~B. We find that in most cases, the detected radio emission is 1$-$2 orders of magnitude more luminous than would be expected from the FUV SFR. For three galaxies (NGC~924, NGC~2563 and NGC~5982) SF may contribute 20$-$40\% of the radio emission, and in one (NGC~584) it may be the dominant source of radio emission. There is also one galaxy (NGC~4169) for which no FUV flux is available. We therefore consider that SF dominates at most 2 of the 14 BGEs with point-like emission, though it could have some impact on our measurements of radio properties in 1$-$3 more.

\begin{table*}
 \caption{Morphological properties of CLoGS high-richness groups and their central radio sources. For each group we note the LGG number, the BGE name, the angular scale, the largest linear size (LLS) of the radio source, measured from the 235 MHz radio images unless stated otherwise, the radio and X-ray morphology class (X-ray class drawn from paper~I), the energy output of any radio jets, estimated from 235~MHz power or from the X-ray cavities (P$_{\mathrm{cav}}$; see also paper~I), and lastly the relevant cooling X-ray luminosity L$_{\mathrm{cool}}$.
}
 \label{t:listtable2}
\begin{center}
\begin{tabular}{llccccccc}
\hline 
Group &  BGE & Scale & LLS & Radio morphology & X-ray morphology & Energy output (radio) & P$_{\mathrm{cav}}$ & L$_{\mathrm{cool}}$ \\ 
   &   & (kpc/$''$) & (kpc) & & & (10$^{42}$ erg s$^{-1}$) & (10$^{42}$ erg s$^{-1}$) & (10$^{40}$ erg s$^{-1}$) \\
\hline 
 LGG 9 & NGC 193   & 0.359 & 80$^a$ & large-scale jet & Group  & 43.57$^{+22.18}_{-14.70}$ &  4.66$^{+3.55}_{-3.63}$ &   2.35$^{+0.13}_{-0.08}$ \\
 LGG 18 & NGC 410   & 0.373 & $\leq$11     &        point    & Group  & - & -& - \\ 
LGG 27 & NGC 584   & 0.121 & $\leq$3$^b$  &  point          & Galaxy & - & - & -\\
LGG 31 & NGC 677   & 0.378 & 30     &     diffuse     & Group  & - & - & -\\ 
LGG 42 & NGC 777   & 0.354 & $\leq$8      &  point          & Group  & - & - & -\\ 
LGG 58 & NGC 940   & 0.359 & $\leq$6      &    point        & Point  & - & - & -\\ 
LGG 61 & NGC 924   & 0.310 & $\leq$4$^c$  &    point        & Point  & - & - & -\\
LGG 66 & NGC 978   & 0.334 & $\leq$3$^c$  &  point          & Galaxy & - & - & -\\
LGG 72 & NGC 1060  & 0.368 & 14     & small-scale jet & Group  & 1.13$^{+0.16}_{-0.19}$ &  0.18$^{+0.31}_{-0.03}$ &   45.70$^{+1.40}_{-1.30}$\\
LGG 80 & NGC 1167  & 0.349 & 240$^d$   &  remnant jet    & Point  & 34.68$^{+15.84}_{-10.87}$ & -& - \\
LGG 103 & NGC 1453  & 0.305 & $\leq$11     &  point          & Group  & - & - & -\\ 
LGG 117 & NGC 1587  & 0.252 & 22$^a$ & diffuse         & Group  & - & -& - \\
LGG 158 & NGC 2563  & 0.315 & $\leq$3      &   point         & Group  & - & -& - \\
LGG 185 & NGC 3078  & 0.165 & 26     &  diffuse        & Galaxy & - & -& - \\ 
LGG 262 & NGC 4008  & 0.262 & $\leq$7      &    point        & Galaxy & - & - & -\\ 
LGG 276 & NGC 4169  & 0.218 & $\leq$3$^c$  &  point          & Point  & - & - & -\\
LGG 278 & NGC 4261  & 0.155 & 80     & large-scale jet & Group  & 64.32$^{+38.79}_{-24.20}$ &   21.45$^{+4.37}_{-3.04}$ &   8.00$^{+0.40}_{-0.01}$\\
LGG 310 & ESO 507-25 & 0.218 & 11    &    diffuse      & Galaxy &  - & -& - \\ 
LGG 338 & NGC 5044  & 0.184 & 63$^a$ &  remnant jet    & Group  & 1.85$^{+0.14}_{-0.15}$ &   2.00$^{+1.74}_{-0.33}$ &   310.00$^{+4.00}_{-9.00}$ \\
LGG 345 & NGC 5084  & 0.112 & $\leq$4      &      point      & Galaxy & - & -& - \\  
LGG 351 & NGC 5153  & 0.291 & -      &  -              & Galaxy & - & -& - \\
LGG 363 & NGC 5353  & 0.170 & $\leq$10     &  point          & Group  & - & - & -\\ 
LGG 393 & NGC 5846  & 0.126 & 12$^a$ & small-scale jet & Group  & 0.41$^{+0.11}_{-0.15}$ &  1.45$^{+0.42}_{-0.38}$ &   34.50$^{+0.70}_{-0.80}$ \\
LGG 402 & NGC 5982  & 0.213 & $\leq$3      &  point          & Group  & - & - & -\\ 
LGG 412 & NGC 6658  & 0.305 & -      &     -      & Point  & - & -& - \\
LGG 473 & NGC 7619  & 0.262 & $\leq$6$^a$  &  point          & Group  & - & -& - \\ 
\hline 
\end{tabular}
\end{center}
$^a$ \citet{Simona11}, $^b$ Measured from the 1.4~GHz image, $^c$ Measured from the 610~MHz image, $^b$ \citet{Shulevski12} (1.4~GHz)\\

\end{table*}

\subsection{235~MHz radio power and largest linear size}
\begin{figure}
\centering
\includegraphics[width=0.48\textwidth]{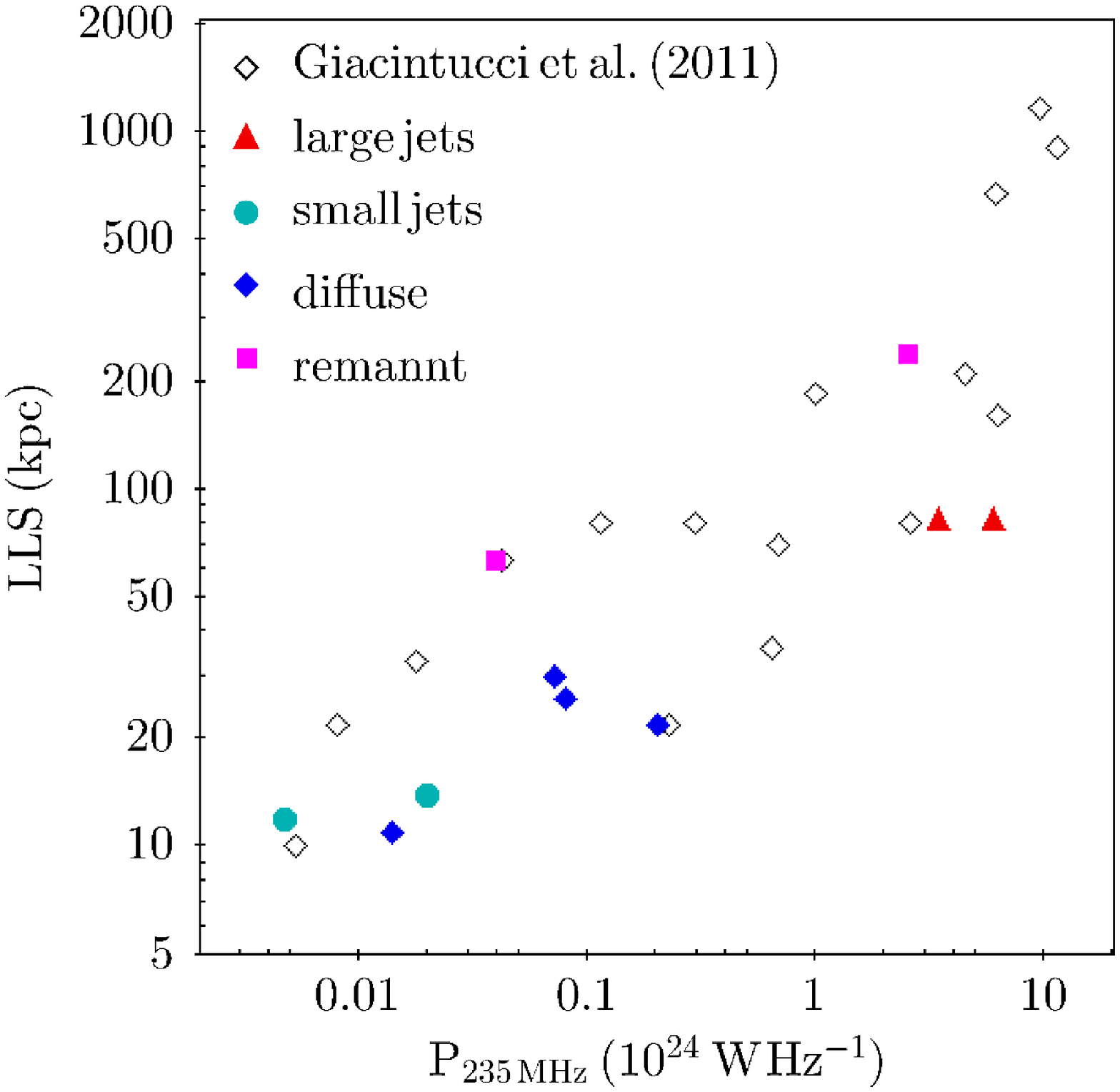}
\caption{Radio power at 235 MHz plotted against the largest linear size of radio sources associated with the central brightest group ellipticals of the CLoGS high-richness sample, with data points from \citet{Simona11} for comparison. Different symbols indicate the radio morphology of the CLoGS sources.}
\label{LLS235Power}
\end{figure}

Figure~\ref{LLS235Power} shows the 235~MHz power of our sources plotted against their largest linear size (LLS) at any radio frequency, with data points from the study of \citet{Simona11} for comparison. For all resolved radio sources (i.e., excluding point-like sources) the largest linear size was measured across the maximum extent of the detected radio emission, at the frequency at which the emission is most extended. The resolved radio sources of the CLoGS central galaxies cover a large spatial scale from $\sim$12~kpc (small scale jets; NGC~5846) to $>200$~kpc (large scale jets; NGC~1167) with 235~MHz radio powers in the range $\sim5\times10^{21}$ W Hz$^{-1}$ to $\sim5\times10^{24}$ W Hz$^{-1}$.

Our CLoGS radio sources seem to follow the same linear correlation between size and power noted by \citet{Ledlow02} and \citet{Simona11}. Although the sample in \citet{Simona11} included only cavity systems in groups, and the work of \citet{Ledlow02} included radio-sources that resided in rich clusters, we see that this linear correlation holds over 3 orders of magnitude at 235~MHz and extends to low power radio sources.


\begin{table}
 \centering
 \caption{Radio-loud AGN (P$_{1.4GHz}>$10$^{23}$ W Hz$^{-1}$) in CLoGS group dominant early-type galaxies. Columns show the LGG number, the BGE name and the 1.4 GHz power.}
\begin{tabular}{lcc}
 \hline
 Group   & BGE &     P$_{1.4GHz}$        \\
         &     & (10$^{23}$ W Hz$^{-1}$) \\   
 \hline
 LGG 9    & NGC 193  &    11.1$\pm$0.6    \\             
 LGG 80   & NGC 1167 &    10.5$\pm$0.6    \\
 LGG 278  & NGC 4261 &    23.2$\pm$0.5    \\             
 \hline
 \end{tabular}
 \label{radioloud23}
 \end{table}

\subsection{Spectral index} 


The spectral index of synchrotron emission corresponds to the index of the energy distribution of the relativistic electrons in the source which, in the absence of energy injection, will steepen over time owing to synchrotron and inverse Compton losses \citep[e.g.,][]{deZotti10}.
In our sample, 19/26 galaxies are detected at both 610 and 235~MHz. Considering only these two frequencies, we find that 2/20 sources have a steep radio spectra with $\alpha_{235}^{610}>$1 (NGC~1587 and NGC~5044) while the majority of the sources range from very flat values of $\alpha_{235}^{610}\sim$0.2 to typical radio synchrotron spectra of $\alpha_{235}^{610}\sim$0.9. Only NGC~940 presents a flux density greater at 610~MHz than in 235~MHz giving an inverted spectral index of $\alpha_{235}^{610}=-0.28\pm0.04$. Leaving the two steep spectrum outliers out, we find that the mean spectral index value for the 17/19 radio sources that reside in the central galaxies of our groups sample is $\alpha_{235}^{610}=0.53\pm0.16$. 

We can also measure the spectral index between 235 and 1400~MHz for 18/26 BGEs. The plots of the spectral distribution of the BGEs detected at both GMRT 235/610~MHz and 1.4~GHz from NVSS are shown in Appendix~C, from which we see that in most cases a simple powerlaw provides a reasonable representation of the spectrum as a whole for each central galaxy. Only small deviations from a powerlaw are observed in some systems that can be attributed mainly to the small flux density offsets at specific frequencies either due to a difference in morphology between the two GMRT frequencies (e.g., NGC~5044, ESO~507-25, NGC~1060) or the lower sensitivity at 235~MHz. The mean value of $\alpha_{235}^{1400}$ calculated for the 18/26 BGEs with both 235 and 1400 MHz data, is $0.60\pm0.16$.

In the case of the two steep spectra systems (NGC~1587 and NGC~5044), we observe a deviation from a simple powerlaw. This is caused by differences in morphology between frequencies, with both the remnant jet in NGC~5044 and the diffuse emission around NGC~1587 being only visible at 235~MHz. The steep spectra of these sources is an indicator that we are observing radio emission from different outbursts \citep[see also][]{Simona11}. 



Table~\ref{radiospix} shows the mean values of the spectral indices for each radio morphology class. Naturally, the steepest mean indices are found for the remnant jets. The mean indices for the other classes are all comparable within the uncertainties, although there is a hint that as expected, steeper spectra are found in diffuse sources and small jet systems, and flatter spectra in the point-like sources.

For the four sources where extended emission was observed at both 235 and 610~MHz we also calculated separate spectral indices for the core and the extended components by creating images with matched resolution. Table~\ref{corespix} lists these spectral index values. In three systems (NGC~193, NGC~677, NGC~3078) we find that the spectral index does not differ between the two components. Only NGC~4261 exhibits a flatter core index, probably as a result of the `cosmic conspiracy' (see \citealt{Cotton80}), or the presence of free-free emission \citep[see][for a more in-depth discussion]{Kolokythas14}.


\begin{table}
 \centering
 \caption{Mean spectral indices $\alpha_{235}^{610}$  and $\alpha_{235}^{1400}$ for the different radio morphologies of our sources.}
\begin{tabular}{lcc}
 \hline
 Radio Morphology      &  Mean $\alpha_{235}^{610}$    &  Mean $\alpha_{235}^{1400}$  \\
            
 \hline
 Point-like            &  0.46$\pm$0.11    &    0.54$\pm$0.11       \\
 Small-scale jet       &   0.70$\pm$0.06   &    0.60$\pm$0.06       \\      
 Remnant jet           &   1.24$\pm$0.06   &    0.77$\pm$0.06       \\            
 Large-scale jet       &   0.53$\pm$0.06   &    0.57$\pm$0.06      \\                 
 Diffuse emission      &   0.64$\pm$0.07   &    0.65$\pm$0.09      \\ 
 \hline
 \end{tabular}
 \label{radiospix}
 \end{table}

\begin{table}
 \centering
 \caption{235-610~MHz spectral indices for the cores and extended emission of those radio sources in our sample with their morphology resolved at both GMRT frequencies. Columns show the LGG number, the BGE name, the radio morphology and the spectral index $\alpha_{235}^{610}$ of the core and the surrounding emission of each source.}
\begin{tabular}{llcc}
 \hline
 Group   & BGE &  Core $\alpha_{235}^{610}$    &  Surrounding $\alpha_{235}^{610}$  \\
            
 \hline
 LGG 9 & NGC 193 &   0.54$\pm$0.04  & 0.53$\pm$0.04   \\
 LGG 31 &  NGC 677 &   0.79$\pm$0.04    &    0.82$\pm$0.04     \\
 LGG 185 &  NGC 3078 & 0.43$\pm$0.04  &    0.45$\pm$0.04   \\
 LGG 278 & NGC 4261 &  0.20$\pm$0.04 & 0.52$\pm$0.04   \\

 \hline
 \end{tabular}
 \label{corespix}
 \end{table}

\section{DISCUSSION}


\subsection{Detection statistics and comparisons} 



The comparison between group and cluster samples in radio requires caution, since the most relevant studies have not used exactly the same selection criteria (eg. distance, radio power etc). Several studies have investigated the radio detection fraction of brightest cluster galaxies, group-dominant galaxies and large ellipticals in the local Universe. \citet{Dunn10}, using a sample of nearby bright ellipticals, most of which reside in galaxy groups or clusters, found that $\sim$81\% (34/42) are detected using the NVSS and Sydney University Molonglo Sky Survey (SUMSS, 843~MHz), rising to $\sim$94\% (17/18) for those systems with \textit{ROSAT} All-Sky Survey $0.1-2.4$~keV X-ray fluxes $>3\times10^{-12}$~erg~s$^{-1}$~cm$^{-2}$ (equivalent to L$_{0.1-2.4keV}\sim2.3\times10^{42}$~erg~s$^{-1}$ at the distance limit of CLoGS). For comparison with CLoGS we detect 21/26 BGEs from NVSS ($\sim$81\%) whereas all of our groups above the equivalent X-ray flux limit from \citet{Dunn10} sample are detected (5/5; 100\%). Using the X-ray selected groups sample of \citet{Eckmiller11}, \citet{Bharadwaj14} found that 20/26 BGEs (77\%), are detected in radio using the NVSS, SUMSS and VLA Low frequency Sky Survey (VLSS, 74~MHz) radio catalogs. In the X-ray selected local, low mass cluster sample of \citet{Magliocchetti07}, the radio detection rate in central BCGs is 20\% when using a radio luminosity limit of L$_{1.4GHz}>10^{22}$ W Hz$^{-1}$ (our detection rate with CLoGS above this limit is 7/26, $\sim$27\%), rising to 92\% (11/12 of BCGs) when this limit is lowered to L$_{1.4GHz}>10^{20}$ W Hz$^{-1}$, comparable to the sensitivity of our sample. These results agree with our own in suggesting that, with sufficiently deep observations, almost all group or cluster dominant galaxies will be found to host a central radio source. 

Comparison with more massive clusters is difficult, since they tend to be much more distant than our groups, with radio surveys usually identifying only the brightest central radio sources. We note that numerous earlier X-ray selected cluster samples suggest a radio detection rate for BCGs $\sim$50\% (e.g., B55 sample \citealt{Edge90,Peres98}, Highest X-ray Flux Galaxy Cluster Sample, HIFLUGCS \citealt{Reiprich02,Mittal09}), for sources of sufficient power to be detected in volumes extending out to $z=0.3$. A similar radio detection rate for BCGs was also found by \citet{Ma13} in a combined sample of X-ray selected galaxy clusters matched with NVSS ($\sim$52\% for radio sources $>3$~mJy) and \citet{Kale15} ($\sim$48\%; matched with NVSS and FIRST catalogs) in a combined sample of 59 X-ray selected BCGs extracted from the Extended Giant Metrewave Radio Telescope (GMRT) Radio Halo Survey (EGRHS; \citealt{Venturi07, Venturi08}). However, \citet{LinMohr07} using a cluster sample from two large X-ray cluster catalogs drawn from the ROSAT All-Sky Survey (RASS), NORAS \citep{Bohringer00} and REFLEX \citep{Bohringer04}, detected 122 BCGs in 342 clusters (36\%) that have a radio component with flux density greater than 10~mJy at 1.4~GHz. The selection criterion of a cut-off at 10~mJy for the NVSS catalog matching is most probably the reason for this lower radio detection rate.

Another recent study by \citet{Hogan15} using a combination of X-ray selected BCG samples matching with NVSS, SUMSS and FIRST radio catalogs finds a detection rate of $61.1\pm5.5\%$ for the extended Brightest Cluster Sample (eBCS; \citealt{Ebeling00}), $62.6\pm5.5\%$ for the REFLEX-NVSS sample (ROSATESO Flux Limited X-ray; \citealt{Bohringer04}) and $60.3\pm7.7\%$ for the REFLEX-SUMSS sample. These detection percentages are slightly higher compared to previous radio-BCG detection rates found in X-ray selected clusters, but are in good agreement within errors.

Comparing by visual inspection the maxBCG sample with FIRST, \citet{MeiLin10} identify  552 double-lobed central radio galaxies in the $\sim$13000 cataloged clusters ($\sim$4\%). In our sample, only NGC~193 and NGC~4261 are large and bright enough to be comparable, though NGC~1167 probably was, and NGC~5044 may have been, in the recent past. Within the large uncertainties, this suggests a comparable fraction of large, double-lobed sources between groups and clusters, though clearly larger samples of groups are needed for a more accurate comparison.

\subsection{The X-ray environment of CLoGS central radio galaxies} 

While in cluster environments an X-ray emitting intra-cluster medium is almost always present (e.g., 72\%; \citealt{Balogh11}, 89\%; \citealt{Andreon11}) in our optically-selected groups sample we find that only $\sim$55\% of systems exhibit extended X-ray emission with luminosities and temperatures typical of group-scale haloes (see Paper~I). In these X-ray bright groups, a variety of radio morphologies is observed, including group-group mergers (NGC~1060 and NGC~7619) and two recently tidally-disturbed `sloshing' systems (NGC~5846 and NGC~5044, \citealt{Gastaldello13}). In other cases, disturbances in the X-ray-emitting gas are caused by a central radio source (e.g., NGC~193). The radio jets seen in these systems can potentially heat the IGM \citep[through a variety of mechanisms, see e.g.,][]{Fabian12}, balancing radiative energy losses and helping to maintain the long-term thermal equilibrium of the IGM.

In examining interactions between the central radio galaxy and the IGM, we divided our groups into X-ray bright and X-ray faint subsets; X-ray bright groups are those which in Paper~I we find to have group-scale haloes (extending $>65$~kpc with L$_{X,R500}>10^{41}$~erg~s$^{-1}$), while X-ray faint groups are those that either have galaxy-scale diffuse or point-like X-ray emission. Using this classification, 14/26 of our groups are X-ray bright. Table~\ref{t:listtable2} lists the radio and X-ray morphology of the central BGEs in the high-richness sub-sample.  

\begin{table}
 \centering
 \caption{Fraction of radio sources with a given radio morphology detected in X-ray bright and X-ray faint CLoGS high-richness groups.}
\begin{tabular}{lcc}
 \hline
 Radio Morphology  & X-ray Bright & X-ray Faint \\
 \hline
 Point-like            & 7/14 & 7/12 \\
 Small/Large scale jet & 5/14 & 1/12 \\             
 Diffuse emission      & 2/14 & 2/12 \\ 
 No detection          &  0   & 2/12 \\ 
 \hline
 \end{tabular}
 \label{radiomorphxrays}
 \end{table}

Table~\ref{radiomorphxrays} shows the fraction of X-ray bright and faint groups which host each class of central radio source. 
There is a clear environmental difference for the jet radio sources between X-ray bright and faint groups. All but one of the jet sources are found in X-ray bright systems, confirming that the X-ray bright groups or clusters are the preferred environment for radio jets \citep[as found by, e.g.,][]{Magliocchetti07,McNu}, even down to the relatively low mass range covered by our sample. This suggests that the presence of a group scale IGM provides a richer gas supply which is more likely to fuel an outburst by the central AGN.
On the other hand, all galaxies that lack radio emission are found in X-ray faint systems. With only two non-detections, this is not a statistically significant result, but it hints at an environmental trend worth further investigation with larger samples. These X-ray faint groups may have shallow gravitational potentials that are unable to heat the available gas to temperatures where X-ray emission becomes detectable, (though this is perhaps unlikely given the sensitivity of our X-ray data and the velocity dispersions of the groups), or they may be deficient in diffuse intra-group gas and hence not bright enough, as has been suggested for other X-ray faint optically-selected groups \citep[e.g.,][]{Rasmussen06}. 


All radio jet systems in X-ray bright groups (5/5) reside in cool cores \citep[in agreement with the study of][]{Sun09}, suggesting that a cool core is a necessary requirement for jet activity in such systems. NGC~1167, the only jet source identified in an X-ray faint group, has an extensive cold gas reservoir, which has likely fueled the AGN outburst. Of the five X-ray bright jet-hosting systems, four show clear correlations between the GMRT low-frequency radio and X-ray structures, in the form of cavities and/or rims of enhanced X-ray emission surrounding the radio lobes. This provides proof of interactions between the central radio galaxy and the IGM. Of these four systems, the two large-scale, currently active radio jet systems (NGC~193 and NGC~4261) have large cavities or cocoon-like structures excavated and filled by the radio lobes. NGC~5846 and NGC~5044 possess smaller-scale cavities commensurate with their current radio activity, but there is evidence of a larger cavity correlated with the old, detached radio lobe of NGC~5044, from abundance mapping \citep{OSullivan14}. In the fifth system, NGC~1060, the jets are too small for cavities to be resolved in the \textit{XMM-Newton} observation described in Paper~I.

However, the predominant radio morphology across all our groups is the point-like, found in $\sim$50\% of X-ray bright and $\sim$58\% of X-ray faint groups. In X-ray bright systems, point-like radio sources are common in both cool-core and non-cool core groups. Considering an AGN as the dominant source of radio activity (as most likely is the case in almost all of our point-like systems; see \S 5.3), this reveals that not only the environment, but also the efficiency of the processes involved plays a significant role in the morphology of a radio source. While in X-ray faint groups the lack of a significant quantity of gas to fuel the central engine may explain the presence of only low-power activity, in X-ray bright cool-core systems we must either invoke inefficiencies in fueling, or duty cycle effects (e.g., the need to build up a reservoir of cold, dense gas via cooling from the IGM after a previous round of feedback heating) to explain the lack of ongoing jet activity. Another possible explanation for low power point-like radio sources in both environments is a contribution from the stellar population.

We also note that diffuse radio sources appear equally common in both X-ray bright and X-ray faint groups. We discuss these sources further in \S~6.5.

\subsection{Environment and spectral index} 

In dense environments, such as the centres of galaxy clusters, it is known that radio galaxies usually present a steep synchrotron spectral slope compared to galaxies that reside in the field \citep[e.g.,][]{Govoni01,Bornancini10}. As the radio morphology of a galaxy depends on the environment in which it is embedded, we examine the radio spectral index distribution of the central galaxies in our sample in order to unveil any relation that may exist between spectral index and environment, comparing also with what is known for BCGs. 

The typical range of spectral index values $\alpha$, of radio galaxies at the centers of cool-core clusters is $1-2$ \citep{Simona11}. The simplest interpretation for such steep radio spectra is the restriction of the relativistic radio plasma in the central regions of the cluster by the high pressure and dense external medium, which may slow the expansion and consequent fading of the radio source \citep[e.g.,][]{Fanti95,Murgia99}. \citet{Simona11} observed 15 X-ray bright galaxy groups hosting extended radio sources with the GMRT at 235 and 610~MHz, and found that 9/15 had a steep spectral index of $\alpha_{235}^{610}>1$. This is the opposite of what we find from our CLoGS sample, where only 2/26 BGEs host a steep-spectrum radio source. Clearly our inclusion of X-ray faint groups and point-like central radio sources, some of which also have contribution from star-formation in their radio emission, has a significant impact on the spectral index distribution. 

Our sources are more comparable to those found in the general population of galaxy clusters, both cool-core and non-cool-core. \citet{Bornancini10}, using a sample of maxBCG clusters, calculated the mean value of the spectral index for BCGs between 325~MHz and 1.4~GHz to be $\alpha_{325}^{1400}=0.65$. This is similar to our mean spectral index ($\alpha_{235}^{1400}=0.60\pm0.16$) using almost the same frequency range. The mean spectral index in cool-core groups alone is calculated to be $\alpha_{235}^{1400}=0.72\pm0.13$, with the equivalent mean value in the non-cool-core ones being $\alpha_{235}^{1400}=0.60\pm0.16$. We find that cool-core groups present steeper mean spectral index but taking into account the uncertainties in the values, the difference does not seem to be so strong. 

\begin{figure}
\centering
\includegraphics[width=0.48\textwidth]{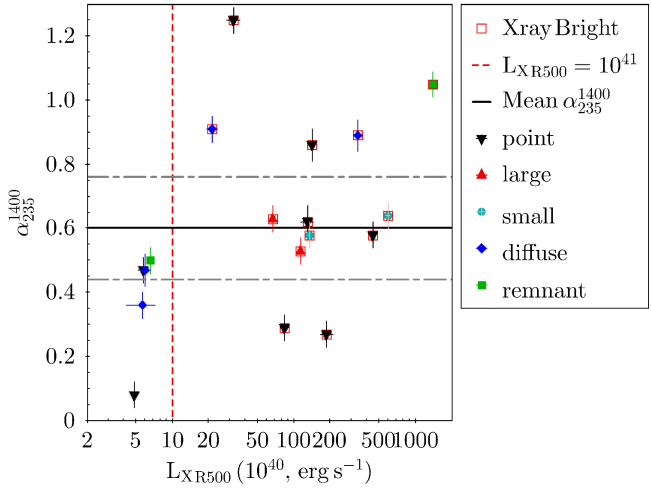}
\caption{Spectral index distribution $\alpha_{235}^{1400}$ of BGEs compared to the group X-ray luminosity within R$_{500}$. The solid horizontal line indicates the average spectral index value of 0.60 at this frequency range, with the associated error bars shown as dot-dashed lines. The red dashed vertical line shows the threshold in X-ray luminosity (L$_{X R_{500}}$=10$^{41}$ erg s$^{-1}$) above which we consider groups as X-ray bright.}
\label{LXspix2351400}
\end{figure}

Figure~\ref{LXspix2351400} shows spectral index $\alpha_{235}^{1400}$ plotted against L$_{X R_{500}}$, the X-ray luminosity of the group within the fiducial radius R$_{500}$. We find that radio sources with spectral indices steeper than the average value of $\alpha_{235}^{1400}=0.60$ reside in X-ray bright groups, regardless of their radio morphology. These steeper-than-average sources include not only jets and known remnants of past outbursts, but also systems with point-like or diffuse radio morphology. This supports the interpretation suggested above, that radio sources remain visible for longer periods in environments with a dense IGM, while those in X-ray faint systems expand and fade relatively rapidly.

\subsection{Radio emission and group spiral fraction}


The environment in which a galaxy resides is known to play an important role on its properties. In galaxy groups, the fraction of early-type galaxies ranges from $\sim$25\% (as in the field) to $\sim$55\% \citep[as in rich clusters,][]{MulchaeyZabludoff98}. A high fraction of spiral galaxies can be an indicator of a group's dynamical youth. Systems whose collapse and virialisation occurred further in the past are expected to have higher fractions of early-type galaxies, since there has been a longer history of interactions and mergers. Studies in a sample of compact groups have suggested a correlation between extended X-ray emission and low spiral fraction \citep[e.g.,][]{Pildis95} suggesting that dynamical age is linked to the ability to build and retain a hot IGM. We therefore explore whether any correlation between the close environment of a BGE and its radio morphology or the group's X-ray emission exists, by looking at the spiral galaxy content of the member galaxies in our groups.

Table~\ref{t:listtable3} shows the spiral fractions in each group. We took galaxy morphologies from HyperLEDA\footnote{http://leda.univ-lyon1.fr/}, classifying them as either early-type (morphological T-type $<$0), late-type (T-type $\geq$ 0) or unknown. As the objects with unknown morphology are typically faint and small, we consider them as dwarf galaxies and include them with the late-type galaxies, defining spiral fraction F$_{sp}$ as the number of late-type or unknown morphology galaxies over the total number of group members.

We follow \citet{Bitsakis10} in considering groups with F$_{sp}>0.75$ as spiral-rich, and potentially dynamically young. 
We find that 11/26 (42\%) of our groups are classified as spiral-rich and 15/26 (58\%) as spiral-poor. \citet{Bitsakis14}, considering a sample of 28 Hickson compact groups \citep[HCGs,][]{Hickson92} which are likely to have a higher rate of galaxy interaction than most of our groups \citep{Hickson97}, finds a similar fraction to our sample with 46\% (13/28) being spiral-rich and 54\% (15/28) being spiral-poor.

\begin{figure}
\centering
\includegraphics[width=0.48\textwidth]{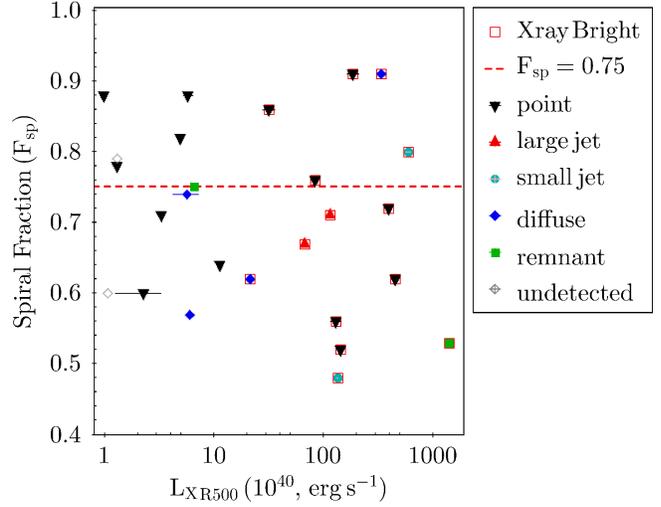}
\caption{Spiral fraction F$_{sp}$, i.e. the number of late type+unknown galaxies over the total number of galaxies in each group, of different radio morphologies in relation to the X-ray luminosity at R$_{500}$ (L$_{XR_{500}}$). The X-ray bright groups are marked in red square, different radio morphologies are shown in different symbols, whereas the red line indicates a spiral fraction of 0.75 following \citet{Bitsakis10}.}
\label{Fspunknown}
\end{figure}

Fig~\ref{Fspunknown} shows the relation between F$_{sp}$ and the X-ray luminosity calculated at R$_{500}$, L$_{X R_{500}}$. Point-like radio sources appear to be equally divided between dynamically young and old groups therefore showing no preference in the group environment they reside. We further find that in high spiral fraction groups the radio point-like systems are almost evenly distributed between X-ray bright (3/7) and X-ray faint groups (4/7). In \citet{Ponman96} a mild trend between spiral fraction systems and diffuse X-ray luminosity is suggested using ROSAT observations over a sample of HCGs. In our sample however, we see that spiral-rich systems have no particular preference in the X-ray environment, since 6/11 spiral-rich groups are X-ray faint and 5/11 X-ray bright.

\begin{table}
\begin{minipage}{\linewidth}
 \caption{Spiral fraction for the high-richness sub-sample CLoGS groups. We note the brightest group early-type galaxy and the spiral fraction F$_{sp}$, which is the ratio of the number of late-type plus unknown galaxies over the total number of galaxies.}
\centering
 \label{t:listtable3}
\begin{tabular}{|c|c|}
\hline 
 BGE   &    Spiral fraction   \\ 
       &    F$_{sp}$  \\
\hline 
  NGC 193   &     0.67   \\
  NGC 410   &     0.52  \\
  NGC 584   &     0.64   \\
  NGC 677   &     0.91    \\
  NGC 777   &     0.56 \\
  NGC 940   &     0.78  \\
  NGC 924   &     0.71  \\
  NGC 978   &     0.60  \\
  NGC 1060  &     0.80   \\
  NGC 1167  &     0.75  \\
  NGC 1453  &     0.76    \\
  NGC 1587  &     0.62  \\
  NGC 2563  &     0.72   \\
  NGC 3078  &     0.74   \\
  NGC 4008  &     0.88   \\
  NGC 4169  &     0.88 \\
  NGC 4261  &     0.71  \\
 ESO 507-25 &     0.57    \\
  NGC 5044  &     0.53 \\
  NGC 5084  &     0.82    \\  
  NGC 5153  &     0.79  \\
  NGC 5353  &     0.91    \\
  NGC 5846  &     0.48  \\
  NGC 5982  &     0.86  \\
  NGC 6658  &     0.60  \\
  NGC 7619  &     0.62  \\
  
\hline 
\end{tabular}
 \end{minipage}
 \end{table}

For fig~\ref{Fspunknown} a Spearman correlation coefficient $\rho=-0.17$ with $P=0.42$ is found, showing that the association between F$_{sp}$ and L$_{XR500}$ for our groups cannot be considered statistically significant. However, we point out that the majority of spiral-rich systems with L$_{XR500}<10^{42} \mbox{ } erg \mbox{ } s^{-1}$, 75\% (6/8), appear to exhibit point-like radio emission and that 67\% (4/6) of the jet systems resides in dynamically old groups.
 
\begin{figure}
\centering
\includegraphics[width=0.48\textwidth]{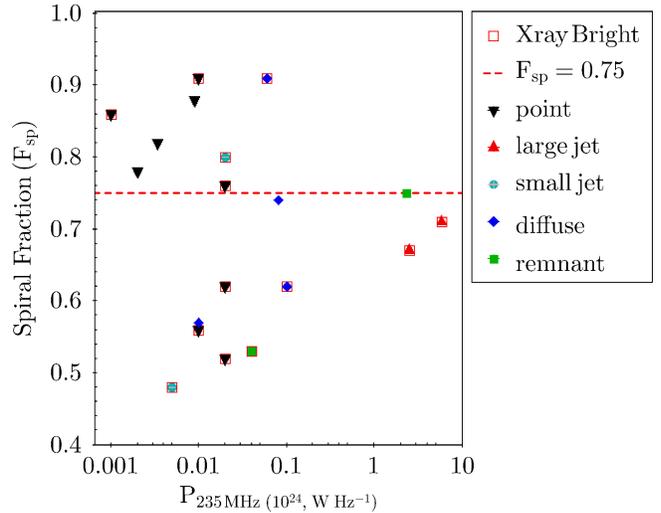}
\caption{Spiral fraction F$_{sp}$ as in fig.~\ref{Fspunknown}, in relation to radio power at 235~MHz (P$_{235 MHz}$). The radio morphology of each group can also be seen here in different symbols, with the X-ray bright groups being marked in red square.}
\label{Fspunknown2}
\end{figure}

Fig~\ref{Fspunknown2} shows the relation between F$_{sp}$ and the radio power at 235~MHz (P$_{235 MHz}$) for the systems with radio emission detected at 235~MHz. We find that the majority of radio point-like systems (6/9) detected at 235~MHz resides in groups with higher spiral fractions (dynamically young) with a mean spiral fraction of 0.73$\pm$0.12. Comparing with the mean value for jet sources (0.66$\pm$0.21) we find that point radio sources have a marginally higher mean F$_{sp}$ but, given the large errors this result cannot be regarded as statistically significant. 

While the suggestion that BGEs with point-like radio emission may be more common in X-ray faint, dynamically young groups is interesting, the lack of a clear trend is perhaps more telling. Point-like emission might be expected from the low accretion rates of AGN in hot-gas-poor systems, and in some cases star formation fueled by cold gas (perhaps more common in X-ray faint groups) may make a contribution, but low radio luminosities and point-like emission would also be expected in the interludes between outbursts of jet activity in BGEs at the centres of cooling flows (group relevant timescales $\sim$10$^7$ $-$ 10$^8$ yr; see \citealt{David11}, \citealt{Machacek11}, \citealt{Randall11}, \citealt{Rafferty13}). The presence of jets in dynamically young systems also indicates the diverse factors which can affect nuclear activity, with our sample including at least one example of powerful remnant jets likely fueled by interactions with a cold gas rich neighbor.

\subsection{Diffuse radio sources in group central galaxies} 
\label{sec:diffuse}


The origin of the diffuse radio sources that we find in our groups is somewhat mysterious. They have some similarities to the radio mini-halos seen in cool core clusters \citep[e.g.,][]{MazzottaGiacintucci08,Doria12,Kale13}. The scale of the diffuse emission in our groups is small, only a few tens of kpc, whereas mini-halos in clusters can extend up to a few hundreds of kpc in radius. Turbulence in groups is considerably weaker than in clusters, and is generally thought not to be capable of providing the energy necessary to accelerate the electron population and produce radio emission. It is also notable that the groups in which we see evidence of mergers and tidal disturbances do not host diffuse radio sources. It therefore seems unlikely that our sources are radio halos or mini-halos.

Of our four diffuse radio sources, in three cases the diffuse emission is only observed at 610~MHz. This is a result of the significantly superior sensitivity achieved at 610~MHz, $5-30$ times better than that at 235~MHz in these systems. We find no clear correlation between X-ray and radio structure for the diffuse sources; only two of the four reside in X-ray bright groups, and in those two (NGC~677 and NGC~1587) no cavities or shock fronts are observed. The spectral indices of these two sources are relatively steep, and the radio emission is more extended than the stellar population of the galaxies, ruling out star formation as the dominant source of emission. 

The spectral indices of the two diffuse sources in X-ray faint groups are relatively flat. They are therefore unlikely to be radio phoenixes or relic radio galaxies. Of the two X-ray faint systems, NGC~3078 presents the more interesting morphology. Located in a galaxy-scale X-ray halo, at both 235 and 610~MHz it shows a clear, relatively symmetric east-west extension, roughly matching the minor axis of its host galaxy and confined within the stellar body of the galaxy (at least in projection). This morphology could indicate a galactic wind or an outflow associated with a failed or decollimated jet source. A second possibility is star-formation. However, using the relation of \citet{Garn09} we find the expected star-formation rate only from the diffuse 610~MHz emission (excluding the central point source) to be $\sim$2.5~M$_{\odot}$ yr$^{-1}$, which is an order of magnitude higher than the rate expected from mid-infrared \textit{Wide-field Infrared Survey Explorer} (\textit{WISE}) W3 luminosity ($\sim$0.2 M$_{\odot}$ yr$^{-1}$). ESO~507-25, located in another galaxy-scale X-ray halo, is less structured, with diffuse emission extending in all directions, again on scales smaller than the stellar extent. As in NGC~3078, star formation can be ruled out as the main origin of the radio emission in this system, as the predicted rate ($\sim$1.20 M$_{\odot}$ yr$^{-1}$) is again almost an order of magnitude greater than what would be consistent with the WISE L$_{W3}$ rate ($\sim$0.17 M$_{\odot}$ yr$^{-1}$). In addition, for both of these systems the SFR$_{FUV}$ rates are also too low (see Appendix~B).  The flat spectral indices of these sources ($\alpha^{1400}_{235}=0.35\pm0.14$ and $0.47\pm0.05$) suggest ongoing activity. In general, it seems that while star formation may make a contribution to some of our diffuse radio sources, it cannot be the dominant source of emission in any of them.

We are left to conclude that there is no clear single origin for these sources, and that we may be observing sources with a similar appearance but different formation mechanisms, depending on their environments. A group-scale IGM can plausibly confine old radio lobes in the group core and, over time, IGM gas motions may alter their morphology until the lobe structure is lost. In NGC~1587, an ongoing encounter with a companion galaxy, NGC~1588, could raise the possibility of shock-driven re-acceleration, but the lack of a clear correlation between the radio and X-ray morphologies (i.e., no clear shock fronts or correlated surface brightness edges) is a problem for this hypothesis. In the X-ray faint groups, the flat spectral indices make relic radio lobes an unlikely explanation, but a disrupted jet may provide an explanation for the emission in NGC~3078.

\subsection{Power output of jet systems} 

For the BGEs that host jet sources, we estimate the mechanical power output of the jets using an approach similar to that described in \citet{OSullivan11}. For NGC~193, NGC~4261, and NGC~5044 and NGC~5846 cavity size was estimated from the X-ray images. For NGC~1060, the spatial resolution of the available \textit{XMM-Newton} data were insufficient to resolve cavities on the scale of the small-scale jet seen only at 610~MHz. We therefore estimated the cavity size from the 610~MHz extent. We defined the mechanical power output estimated from the cavities P$_\mathrm{cav}$ as:

\begin{equation}
\label{PcavX}
{\mathrm{P}}_{\mathrm{cav}}=\frac{4p_{th}V}{t_{cav}},
\end{equation}

where $p_{th}$ is the value of the azimuthally-averaged pressure profile at the radius of the cavity centre, $V$ is the cavity volume, and $t_{cav}$ is an estimate of the age of the cavity. We chose to use the sonic timescale (i.e., the time required for the cavity to rise to its current position at the local sound speed in the IGM) so as to facilitate comparison with estimates from the literature. The sonic timescale was calculated by

\begin{equation}
\label{tcav}
{\mathrm{t}}_{\mathrm{cav}}=\frac{r_{cav}}{\sqrt{\frac{\gamma kT}{m}}},
\end{equation}

where $r_{cav}$ is the mean radius of the cavity from the nucleus, $\gamma$ is the adiabatic index of the IGM (taken to be 5/3), $k$ is Boltzmann's constant, $T$ is the IGM temperature, and $m$ is the mean particle mass in the IGM. We also note that for NGC~5044, we exclude the detached radio lobe from consideration, since its filling factor and age cannot be accurately estimated, and only include the cavities identified in \citet{David17}. Our estimates of P$_\mathrm{cav}$ are listed in Table~\ref{t:listtable2}. We find mechanical power values in the range $\sim10^{41}-10^{43}$ erg s$^{-1}$, typical for galaxy groups.

\begin{figure}
\centering
\includegraphics[width=0.48\textwidth]{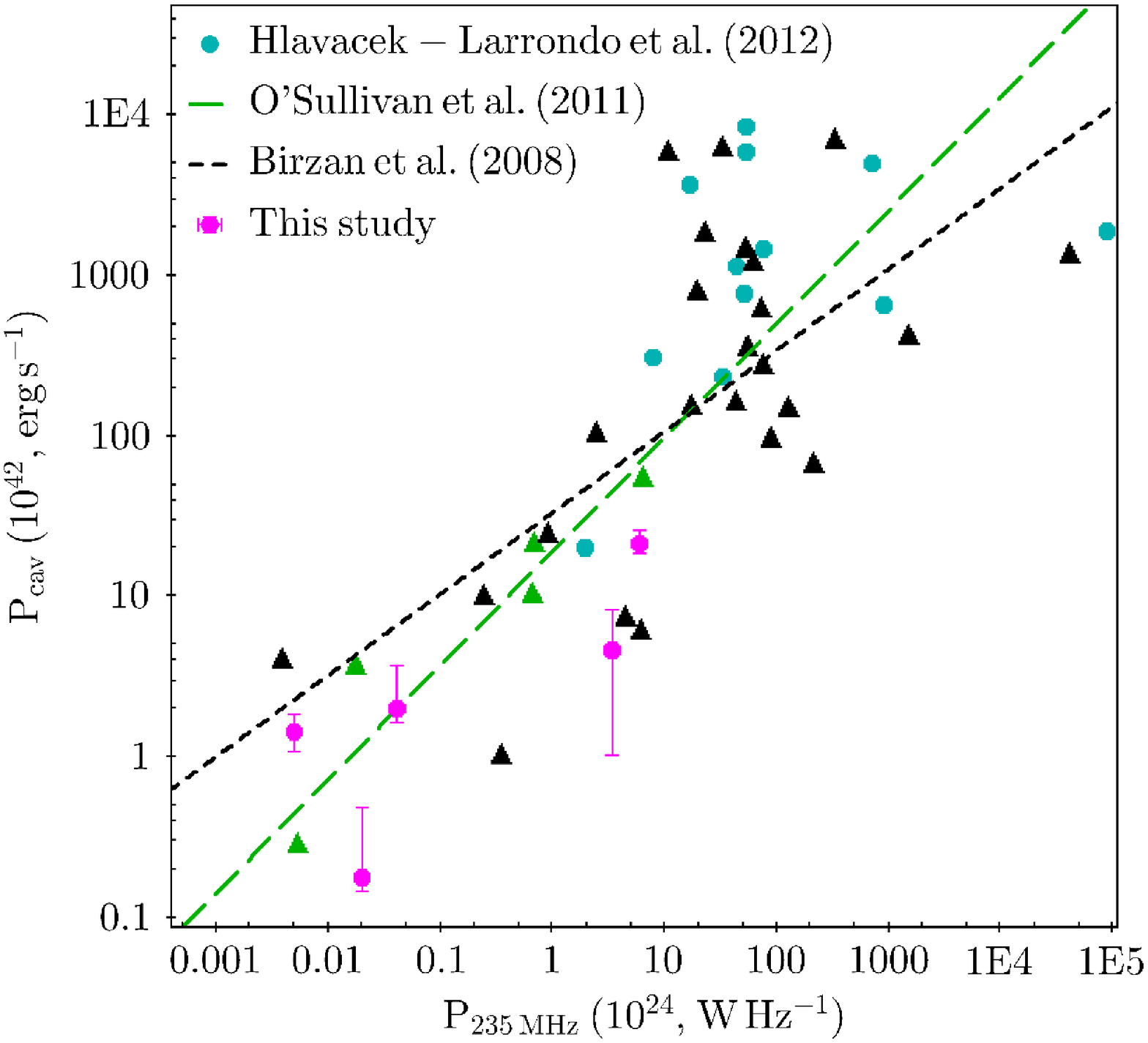}
\caption{Cavity power P$_\mathrm{cav}$ calculated from X-rays vs. radio power at 235~MHz. Systems of our groups sample are marked by pink circles with errors, members from \citet{Hlavacek2012} are shown by cyan circles, members from \citet{OSullivan11} are in green triangles and members from \citet{Birzan08} in black triangles. Note that for the three systems in common between CLoGS and \citet{OSullivan11} (NGC~193, NGC~5044 and NGC~5846) we use the CLoGS measurements. The black dotted line indicates the relation found by B\^{i}rzan et al., and the green dashed that of O'Sullivan et al.}
\label{PcavXP235}
\end{figure}

Fig.~\ref{PcavXP235} shows the relation between the mechanical power output (P$_\mathrm{cav}$) and the radio power at 235~MHz (P$_{235}$) for the 5 jet systems in the CLoGS high-richness sub-sample. Our galaxy groups fall in the lower end of the range covered by the 24 groups and clusters described by \citet{Birzan08} and are in good agreement with the 7 groups of \citet{OSullivan11}. This is unsurprising given the overlap between the samples (three systems in common; NGC~193, NGC~5044 and NGC~5846). NGC~1060 has one of the lowest observed cavity powers, suggesting that either the cavities are larger than the 610~MHz emission, or that its small lobes are still young and are as yet over-pressured with respect to the IGM.

One goal of measuring the relationship between radio power and cavity power is to allow estimates of feedback in systems where high-quality X-ray data are not available. The surveys performed by the eROSITA mission are expected to detect tens of thousands of groups and poor clusters \citep[]{Merloni12,Pillepich12}, but are unlikely to provide reliable estimates of cavity size. The use of radio data as a proxy estimator of AGN power output will therefore be important when studying this population.

We can consider the likely effectiveness of such an approach by applying it to our data, using the  P$_\mathrm{cav}$ $-$ P$_{235}$ relation measured by \citet{OSullivan11} to estimate a cavity power based on the radio power. The 235~MHz radio power is used since low frequencies are least likely to be affected by spectral aging. One immediate benefit is that we can estimate a mechanical power output for NGC~1167, where no X-ray luminous IGM is detected. As expected for such an extended source, the old radio lobes of NGC~1167 imply that during the period of activity, the jets in this systems had a large mechanical power output, $\sim3.5\times10^{43}$~erg~s$^{-1}$.

Table~\ref{t:listtable2} lists the radio-estimated jet powers for the six groups. We find that these estimates are in agreement with the X-ray estimate of P$_\mathrm{cav}$ for NGC~5044, but disagree for the small-scale jet sources and the large-scale jet systems with high P$_{235MHz}$. For the small jets, we find both under- and over-estimates, probably arising from the relatively large scatter in the relation. We note that this scatter appears to be a product on the intrinsic uncertainties in cavity measurements, and is not significantly improved at other frequencies \citep[see, e.g.,][]{Birzan08,Kokotanekov17}.

For the large scale jets, the radio estimate of cavity power is greater than the X-ray measurement, though the large uncertainties on the radio estimate make the differences only $\sim$2$\sigma$ significant. Both NGC~4261 and NGC~193 have shells of X-ray emission surrounding their radio lobes, indicating that the lobes are still over-pressured and expanding trans-sonically. It is therefore possible that for these systems the X-ray cavity power is actually an underestimate of the true jet power. We might expect the radio and X-ray estimates to come into agreement as these sources age, expand, and reach equilibrium with the surrounding IGM.

\subsubsection{Heating vs Cooling}

\begin{figure}
\centering{
\includegraphics[width=0.48\textwidth]{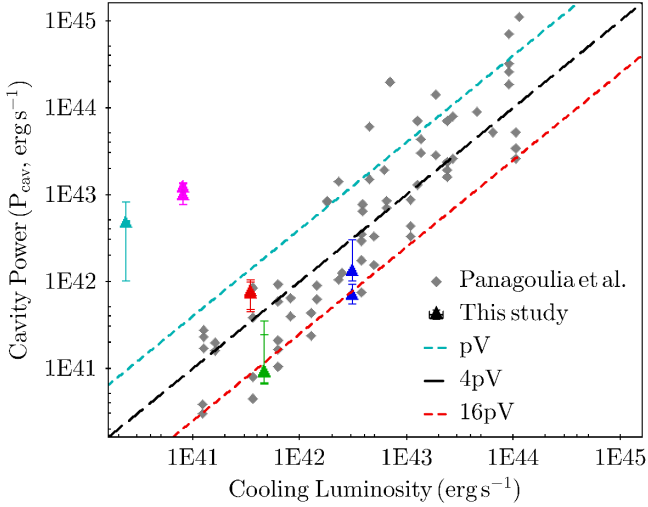}
}
\caption{Cavity power from the X-rays in relation to the cooling luminosity, L$_{cool}$, calculated in the 0.5$-$7 keV band for our CLoGS groups (colored triangles; see Table 4) in comparison to the groups and clusters used in the sample study of \citet{Panagoulia14}). In order from left to right for the data from this study, cyan and pink triangles represent the large-scale jet systems (NGC~193 and NGC~4261), whereas red, green and blue triangles represent the small-scale jet ones (NGC~5846, NGC~5044 and NGC~1060).} 
\label{PcavLxbolvsPanagoulia}
\end{figure}

We can also examine the balance between heating by the central AGN source and radiative cooling of the IGM core. Fig.~\ref{PcavLxbolvsPanagoulia}, shows the ratio P$_\mathrm{cav}$ compared with the cooling luminosity L$_{cool}$ (see \S~\ref{sec:Xray}), over-plotted on data from the sample of \citet{Panagoulia14}.

Comparing our results to \citet{Panagoulia14} we find that the small-scale jet systems NGC~1060 and NGC~5846, along with the remnant jet in NGC~5044, lie within the scatter about the relation, indicating approximate thermal balance. The large-scale jet systems NGC~193 and NGC~4261 fall about two orders of magnitude above the relation, suggesting that they are significantly over-powered. It is notable that in both systems the size of the jets and lobes greatly exceeds that of the cooling region. NGC~4261 has a small cool core, only $\sim$10~kpc in radius \citep{ewan4261} which corresponds to the cooling region as defined here. Its jets extend $\sim$40~kpc on each side (see also \citealt{Kolokythas14} for a detailed radio study), and seem to primarily impact the surrounding IGM, having excavated wedge-shaped channels to exit the core. The radio lobes of NGC~193 seem to have formed a cocoon or series of cavities with compressed gas rims, leaving little cool gas in the galaxy core. We note that the exceptional power of the outburst is not dependent on our assumption of a single cavity; using the jet powers estimated by \citet{Bogdan14} for a pair of cavities produces similar results. This may indicate that in both systems, cooling and heating have become detached, with the jets heating the IGM without significantly impacting the material which is fueling the AGN. In this case, absent some disturbance, we would expect these outbursts to continue until they exhaust the ability of the cool core to provide fuel.

\subsection{AGN feedback in CLoGS groups}

Previous observational studies have suggested that feedback in most galaxy groups and clusters operates in a near-continuous `bubbling' mode \citep[e.g.,][]{Birzan08,Birzan12,Panagoulia14}, in which feedback heating is relatively gentle. This model has been supported by theoretical studies \citep[e.g.,][]{Gaspari11}. Our small-scale and remnant jet systems seem to fit in with this picture. However, there are counter-examples. \citet{Nulsenetal07,Nulsenetal09} found that the cavity power of a number of giant ellipticals (some of which are the dominant galaxies of groups and clusters) exceeds their cooling luminosity by up to an order of magnitude. \citet{Rafferty06} give the example of the exceptionally powerful outburst in the cluster MS~0735+7421, which again greatly exceeds the cooling luminosity, even though using a longer cooling time threshold and consider bolometric X-ray luminosity. Investigating the effect of heating in the same cluster, \citet{Gitti07} found that powerful outbursts are likely occurring $\sim$ 10\% of the time in most cool core clusters. NGC~193 and NGC~4261 seem to fall in to this category, with cavity powers greatly exceeding the cooling luminosity. Such systems may imply an intermittent feedback mechanism, with strong outbursts heating their surroundings enough that long periods of cooling will be required before a new fuel supply can be built up. Modeling suggests that this type of feedback can be produced by more chaotic accretion processes \citep[e.g.,][and references therein]{Prasad15, Prasad17}.

It is interesting to note that both our large-scale, high-power jet systems are found in groups whose X-ray characteristics present difficulties to analysis. NGC~4261 has an X-ray bright core, but its IGM has a relatively low surface brightness, its cavities extend beyond the \textit{Chandra} ACIS-S3 field of view, and its cavity rims were only recognised in \textit{XMM-Newton} observations \citep{ewan4261,Croston05}. The outburst in NGC~193 has severely disturbed its inner X-ray halo, making the usual radial analysis difficult and somewhat uncertain. Neither system is included in the majority of X-ray selected samples of groups used to study group properties and AGN feedback. This raises the question of whether the biases implicit in X-ray selection \citep[e.g., toward centrally-concentrated, relaxed cool-core systems,][]{Eckert11} may tend to exclude the more extreme cases of AGN feedback. NGC~4261 and NGC~193 are by no means the most extreme radio galaxies hosted by groups and poor clusters. In the nearby universe, NGC~383 and NGC~315 host the giant radio sources 3C~31 and B2~0055+30, both of which have jets extending hundreds of kiloparsecs \citep[see, e.g.,][]{Simona11}. The impact of such sources on the cooling cycle is not well understood, but they suggest that caution should be exercised before concluding that feedback in groups and clusters is always a gentle process with only mild effects on IGM properties.

 \section{Conclusions}
 \label{conclusions}
 
In this paper we have presented, for the first time, the GMRT radio images of the brightest group early-type galaxies from the high-richness CLoGS sub-sample at 235 and 610~MHz. 

A high radio detection rate of 92\% (24 of 26 BGEs) is found at either 235, 610 and 1400~MHz, with the majority of the BGEs exhibiting low radio powers between 10$^{21}$ $-$ 10$^{23}$ W Hz$^{-1}$ with only three radio sources in the sample being characterized as radio loud (P$_{235~MHz}$ $>$10$^{24}$ W Hz$^{-1}$ - NGC~4261, NGC~193 and NGC~5044). In agreement with previous studies \citep[e.g.,][]{Magliocchetti07,Dunn10} we confirm the trend suggesting that nearly all dominant galaxies in groups or clusters are hosting a central radio source. 

The extended radio sources in our sample have spatial scales spanning the range $\sim10-240$~kpc with a variety of morphologies extending over 3 orders of magnitude in power, from ~10$^{21}$ W Hz$^{-1}$ (NGC~5982), to ~6 $\times$ 10$^{24}$ W Hz$^{-1}$ (NGC~4261). The majority of our systems (14/26, $\sim$53\%) exhibit a point-like radio morphology, while 6/26 groups ($\sim$23\%) host currently or recently active small/large scale jets, and 4/26 groups (15\%) host diffuse radio sources.

We find that the unresolved point-like radio sources are mainly AGN dominated with the stellar population most probably contributing in the radio emission between $20-40\%$ in 3 systems (NGC~940, NGC~2563, NGC~5982) and most likely dominating in another one (NGC~584).


Comparing the X-ray environment in which the BGEs reside with their radio emission we find a distinction between central radio sources in X-ray bright and faint groups. All but one of the jet sources are found in X-ray bright cool-core systems, confirming that the X-ray bright groups or clusters are the preferred environment for radio jets to appear even down to the lower mass range covered by our sample. On the other hand, all galaxies that lack radio emission are found in X-ray faint systems. Low power radio point-like sources are found to be common in both environments. 

We find that 19/26 galaxies are detected at both 610 and 235~MHz with 2/19 sources exhibiting steep radio spectra with $\alpha_{235}^{610}>1$ (NGC~1587 and NGC~5044). The spectral indices of the remaining 17/19 sources cover the range $\alpha_{235}^{610}\approx0.2-0.9$ with a mean of $\alpha_{235}^{610}=0.53\pm0.16$. The equivalent mean value 235-1400~MHz spectral index for the 18/26 BGEs that exhibit emission at both frequencies, is measured to be $\alpha_{235}^{1400}=0.60\pm0.16$. In X-ray bright groups, the radio sources are found to have steeper spectral indices than the mean $\alpha_{235}^{1400}$.

The mean spiral fraction (F$_{sp}$) per radio morphology shows that radio point-like emission appears preferably in group systems where the majority of galaxies is spirals with this mild trend being stronger towards lower X-ray luminosity systems.

Considering the origin of the radio emission in our diffuse sources, their range of spectral indices, morphologies and scales means that no single mechanism seems able to explain all the sources. We rule out star formation as the dominant factor, but it could, along with disrupted jets, aged and distorted radio lobes from past outbursts, and material shocked or compressed by galaxy interactions, potentially make a contribution.

Considering the energetics of the radio jet sources, we find mechanical power values (P$_\mathrm{cav}$) typical for galaxy groups, in the range $\sim10^{41}-10^{43}$ with our systems being in good agreement with other group-central sources \citep[e.g.,][]{Birzan08,OSullivan11}.

Lastly, we find that small scale jet systems are able to balance cooling in the central region of the group provided that the AGN continuously injects energy, in agreement with \citet{Panagoulia14}, whereas the mechanical power output of the two large scale systems in our sample (NGC~193 and NGC~4261) appears to exceed the cooling X-ray luminosity of their environment by a factor of $\sim$100. This suggests that while in some groups thermal regulation can be achieved by a relatively gentle, `bubbling', feedback mode, considerably more violent AGN outbursts can also take place, which may effectively shut down the central engine for long periods. 




 \section{Acknowledgments}
The GMRT staff are gratefully acknowledged for their assistance during data acquisition for this project. GMRT is run by the National Centre for Radio Astrophysics of the Tata Institute of Fundamental Research. E. O'Sullivan acknowledges support for this work from the National Aeronautics and Space Administration through Chandra Awards GO6-17121X and GO6-17122X, issued by the Chandra X-ray Observatory Center, which is operated by the Smithsonian Astrophysical Observatory for and on behalf of NASA under contract NAS8-03060, and through the Astrophysical Data Analysis programme, award NNX13AE71G. S. Giacintucci acknowledges support for basic research in radio astronomy at the Naval Research Laboratory by 6.1 Base funding. M. Gitti acknowledges partial support from PRIN-INAF 2014. A. Babul acknowledges support by NSERC (Canada) through the discovery grant program. Some of this research was supported by the EU/FP7 Marie Curie award of the IRSES grant CAFEGROUPS (247653). We acknowledge the usage of the HyperLeda database (http://leda.univ-lyon1.fr). This research has made use of the NASA/IPAC Extragalactic Database (NED) which is operated by the Jet Propulsion Laboratory, California Institute of Technology, under contract with the National Aeronautics and Space Administration. 

 
\newpage
\appendix

\section{GMRT RADIO CONTINUUM IMAGES}
\label{AppA}

In this appendix we provide \textit{Chandra} or \textit{XMM} X-ray and Digitized Sky Survey optical images of our BGEs, with 610 (cyan) and 235~MHz (green) contours overlaid, as well as notes on relevant features of each galaxy and radio source. In all figures, the panels on the left represent the X-ray images overlaid by the 235~MHz contours, and the panels on the right, the optical images overlaid by the 610~MHz contours.  







\subsection{NGC 410} 

\begin{figure*}
\centering
\includegraphics[width=1.00\textwidth]{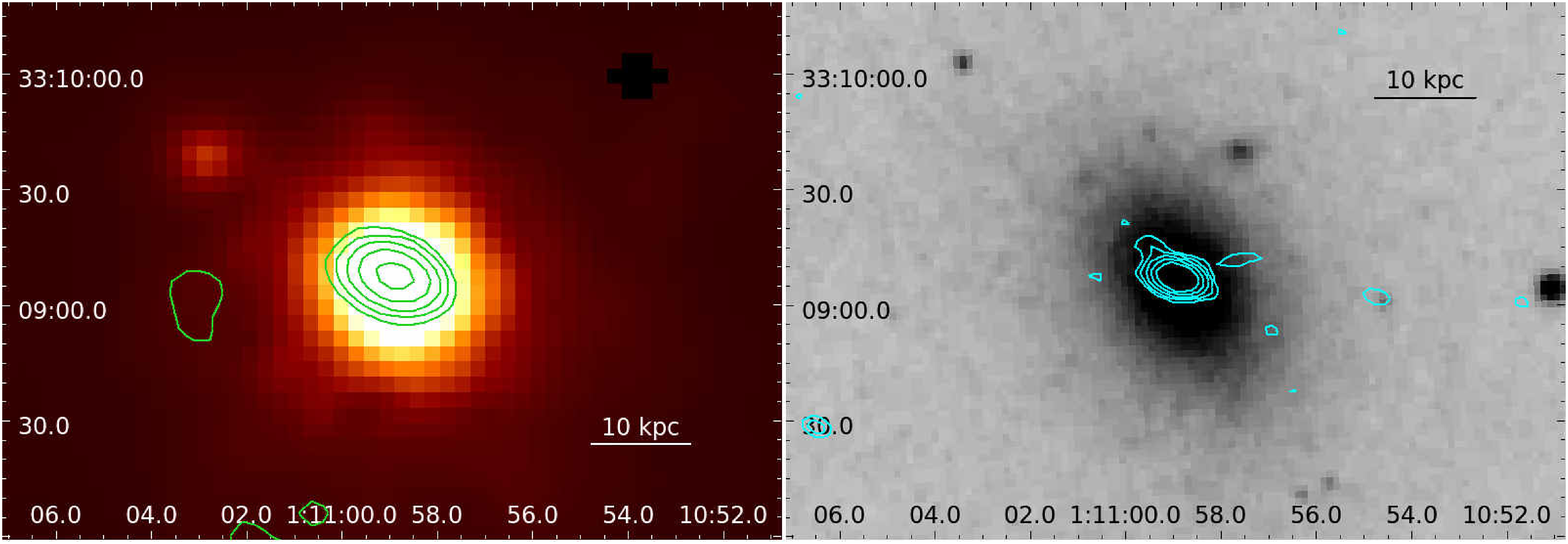}
\vspace{-3mm}
\caption{LGG 18 / NGC 410. \textit{Left:} GMRT 235 MHz contours in green (1$\sigma$ = 0.4 mJy beam$^{-1}$), overlaid on the adaptively smoothed 0.3$-$2.0 keV \textit{XMM-Newton} image. \textit{Right:} GMRT 610~MHz contours in cyan (1$\sigma$ = 50 $\mu$Jy beam$^{-1}$), overlaid on the \textit{Digitized Sky Survey (DSS)} optical image. In both panels the radio contours are spaced by a factor of two, starting from 3$\sigma$. For this source the scale is 0.373 kpc arcsec$^{-1}$.}
\label{fig:410}
\end{figure*}

NGC~410 is the BGE of the X-ray luminous LGG~18 group. A central point source was detected at 1.4~GHz \citep[in the NVSS catalog][]{Condon98} and at 8.4~GHz \citep{Filho02}, with flux densities of 6.3~mJy and 1.7~mJy respectively. Figure~\ref{fig:410} shows the point-like radio source detected in both GMRT frequencies, located coincident with the optical centroid, with some hint of a small extension to the northeast in the 610~MHz emission.

\citet{Gonzalez09} class the galaxy nucleus as a low-ionization emission region (LINER), but argue that it is not an AGN based on a combination of X-ray, UV, H$\alpha$ and radio data.

\subsection{NGC 584}

NGC~584 is the dominant member of a small group of galaxies, LGG~27, and is thought to have recently undergone a merger \citep{Trager00}, with evidence of ongoing star formation \cite{Schweizer92}. \citet{Madore04} argue that NGC~584 does not lie at the gravitational centre of its group. 

\begin{figure*}
\centering
\includegraphics[width=1.00\textwidth]{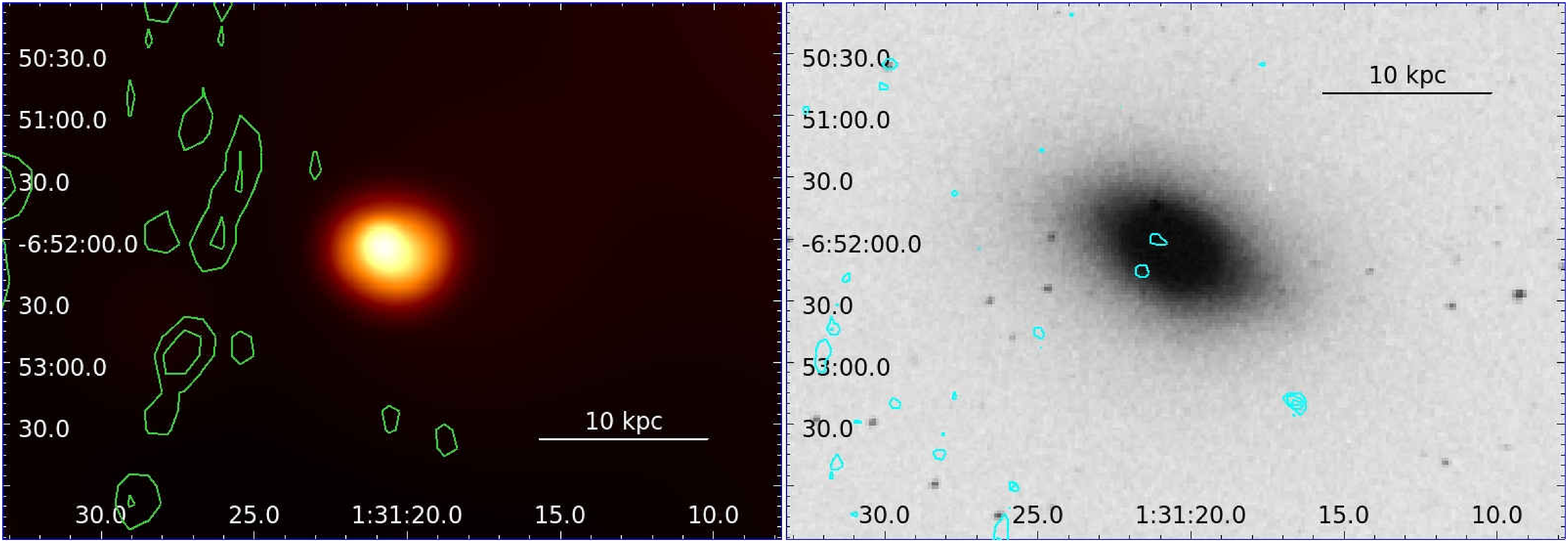}
\vspace{-3mm}
\caption{LGG 27 / NGC 584. \textit{Left:} GMRT 235 MHz contours in green (1$\sigma$ = 1.2 mJy beam$^{-1}$), overlaid on the adaptively smoothed 0.3$-$2.0 keV \textit{Chandra} image. \textit{Right:} GMRT 610~MHz contours in cyan (1$\sigma$ = 200 $\mu$Jy beam$^{-1}$), overlaid on the \textit{DSS} optical image. In both panels the radio contours are spaced by a factor of two, starting from 3$\sigma$. For this source the scale is 0.121 kpc arcsec$^{-1}$. }
\label{fig:584}
\end{figure*}

Although identified with a low-power source in NVSS, NGC~584 is undetected in our GMRT observations. Two nearby bright background sources ($\sim$5 and $\sim$7~Jy at 235~MHz) produce an enhanced noise level in the central part of our images, leading us to set the detection limit at 5$\sigma$. The radio contours overlapping the optical body of the galaxy in figure~\ref{fig:584} appear to be noise rather than detected sources. Our X-ray observations detected no hot gas associated with the group or galaxy.





\subsection{NGC 677} 

NGC~677 is located at the centre of the X-ray luminous group LGG~31. The galaxy has a LINER nucleus \citep{Zhao06} and was detected in the NVSS and both our GMRT observations. We observe a central point source surrounded by irregular diffuse emission extending $\sim$30~kpc at 610~MHz. The morphologies at 235 and 610~MHz are inconsistent, with the 235~MHz emission presenting an asymmetric bipolar morphology overlapping the more rounded 610~MHz emission. These morphological differences may arise from differences in relative sensitivity to low surface-brightness emission at the two frequencies particularly if spectral index varies across the source (i.e., if the regions detected at 235~MHz represent only the most luminous, steepest spectrum regions of the more extended source seen at 610~MHz).

\subsection{NGC 777} 

NGC 777, BGE of the LGG 42 group, is classed by \citet{Ho97} as a slow-rotating Seyfert 2 galaxy, having a population of very old stars \citep{Jarrett13}. Earlier observations by \cite{Ho01} reveal a marginally resolved radio source in the core, slightly extended to hint at jets. NGC~777 is at the centre of a group-scale X-ray halo and is detected at both GMRT frequencies, presenting an asymmetric structure of $\sim$6 kpc only at 235~MHz, south-west from the centre.

\begin{figure*}
\centering{
\includegraphics[width=1.00\textwidth]{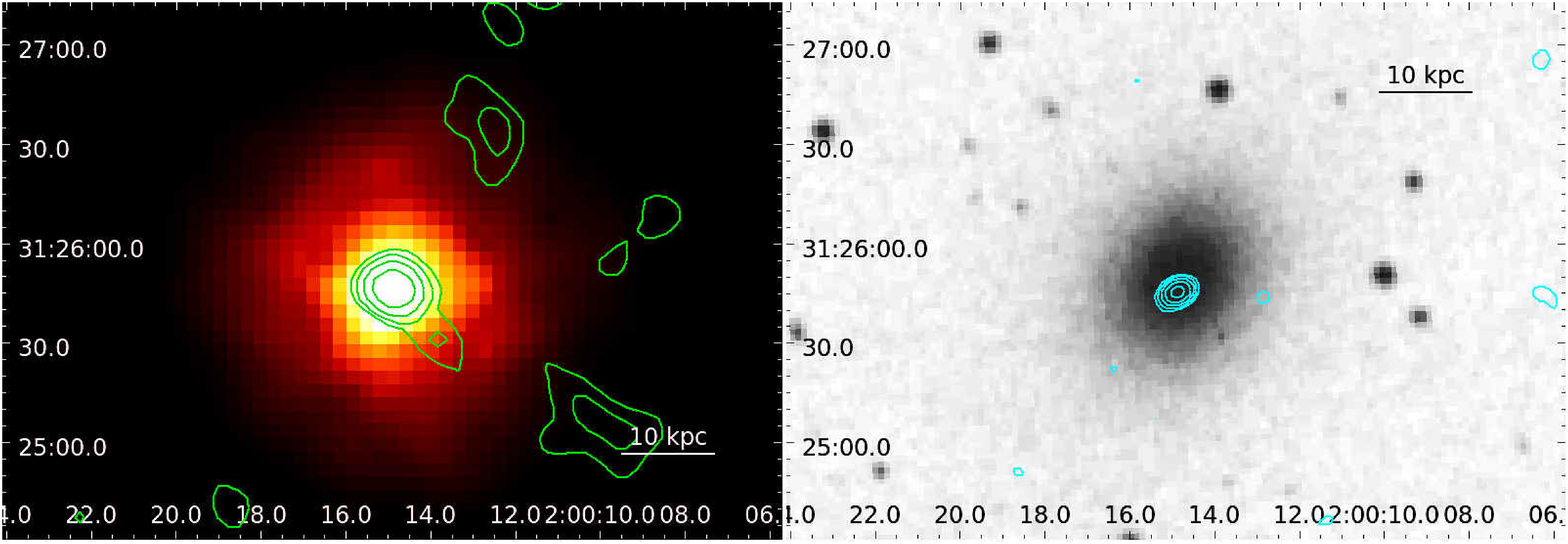}
}
\vspace{-3mm}
\caption{LGG 42 / NGC 777. \textit{Left:} GMRT 235 MHz contours in green (1$\sigma$ = 0.4 mJy beam$^{-1}$), overlaid on the adaptively smoothed 0.3$-$2.0 keV \textit{XMM-Newton} image. \textit{Right:} GMRT 610~MHz contours in cyan (1$\sigma$ = 150 $\mu$Jy beam$^{-1}$), overlaid on the \textit{DSS} optical image. In both panels the radio contours are spaced by a factor of two, starting from 3$\sigma$. For this source the scale is 0.354 kpc arcsec$^{-1}$.}
\label{fig:777}

\end{figure*}

We examined the elongated feature at 235~MHz (see~\ref{fig:777}), making images with natural weighting (equal weight to all samples, good for tracing extended emission; $robust$ parameter 5 in \textsc{aips}) and creating lower resolution images ($19.86'' \times  13.94''$, rms$\sim$0.8~mJy). This revealed that the elongated feature is related to artifacts extending from northwest to southeast surrounding the central radio source, owing to systematic noise that leads to poor calibration with the data we have in hand. The upper limit of the spectral index in this area (using a 3$\sigma$ significance at 610~MHz) is $\gtrsim$2.5 which is very steep and argues against a jet emission for NGC~777.

\subsection{NGC 940}

Previous single dish 2.38~GHz observations of NGC~940 found only a tentative detection with flux density of 5$\pm$4 mJy \citep{Dressler78}. We detect the galaxy at both GMRT frequencies, finding a point-like radio source with an inverted spectral index $\alpha_{235}^{610}$ of $\sim-0.3$. No hot gas halo is detected in the X-rays, and the galaxy is found to be point source dominated. The inverted spectral index could be attributed to absorption of the lower frequency emission, along with contribution from SF.




\subsection{NGC 924}

NGC 924 is the BGE of the X-ray faint group LGG~61 and was previously undetected in the radio regime. We detect a weak radio point source associated with NGC~924 only at 610~MHz, with an upper limit of 1.5~mJy at 5$\sigma$ level of significance for 235~MHz (Table~\ref{Sourcetable}).



\subsection{NGC 978}
		
NGC 978 is the BGE of the X-ray faint LGG~66 group. It forms a galaxy pair with MCG+05-07-017 but shows no signs of interaction or distortion in optical imaging \citep{Karachentsev72}. The galaxy has no previous radio detection but we detect a weak radio point source at 610~MHz that coincides with the optical centre of the galaxy, with a flux density of just 1.3~mJy, one of the weakest in our sample.

\subsection{NGC 1060}

\begin{figure*}
\centering{
\includegraphics[width=1.00\textwidth]{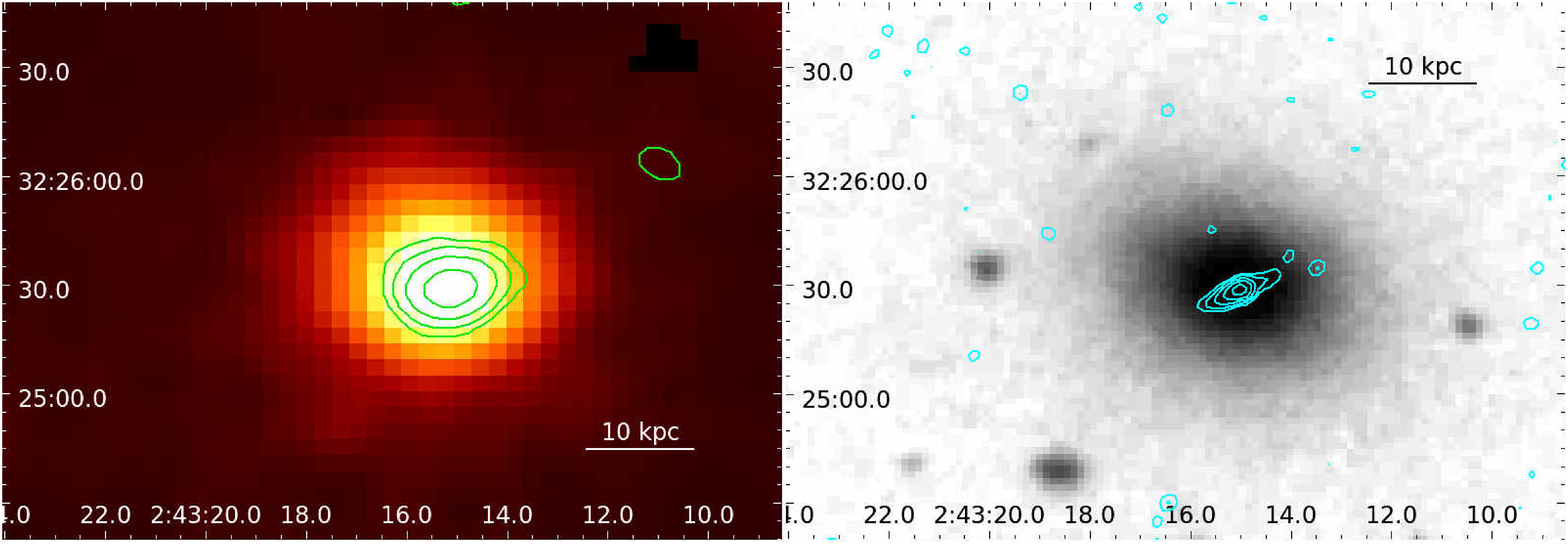}
}
\vspace{-3mm}
\caption{LGG 72 / NGC 1060. \textit{Left:} GMRT 235 MHz contours in green (1$\sigma$ = 0.5 mJy beam$^{-1}$), overlaid on the adaptively smoothed 0.3$-$2.0 keV \textit{XMM-Newton} image. \textit{Right:} GMRT 610~MHz contours in cyan (1$\sigma$ = 90 $\mu$Jy beam$^{-1}$), overlaid on the \textit{DSS} optical image. In both panels the radio contours are spaced by a factor of two, starting from 3$\sigma$. For this source the scale is 0.368 kpc arcsec$^{-1}$.}
\label{fig:1060}
\end{figure*}

NGC~1060 is the BGE in LGG~72 group, a strongly interacting, possibly merging X-ray bright system. Our 610~MHz image reveals a small scale jet (see figure~\ref{fig:1060}) coincident with the optical centre of the galaxy. At 235~MHz this structure is unresolved. The source presents a moderately steep spectral index of $\sim$0.9. The system in X-rays presents a long arc-like path connecting NGC~1060 and NGC~1066, suggestive of a group-group merger \citep[see][]{OSullivan17}, raising the possibility that tidal interactions may have triggered the nuclear activity in NGC~1060.

\subsection{NGC 1167}
		
NGC 1167 is the BGE of the X-ray faint LGG~80 group, and hosts the radio source B2~0258+35. Previous radio observations have revealed this to be a complex, multi-scale source, with multiple sets of radio lobes. The nuclear source is classed as a young, compact steep spectrum (CSS) source \citep{Sanghera95}, with VLA 8 and 22~GHz imaging showing a parsec-scale core with small plume-like lobes and no evidence of hotspots \citep{Giroletti05}. VLBI 1.6~GHz observations show that the faint compact core (7.6~mJy), is surrounded by a bright diffuse region coincident with the peak of the VLA images. The diffuse component has a spectral index of $\alpha_{0.08}^{22}\sim1.0 - 1.5$, while the core is flatter \citep[$\sim$0.6,][]{Giroletti05}. However, these small-scale structures are dwarfed by the 240~kpc double-lobed structure  detected at 1.4~GHz using the Westerbork Synthesis Radio Telescope (WSRT), which clearly indicates a much more powerful period of activity in the recent past \citep{Shulevski12}.

\begin{figure*}
\centering{
\includegraphics[width=1.00\textwidth]{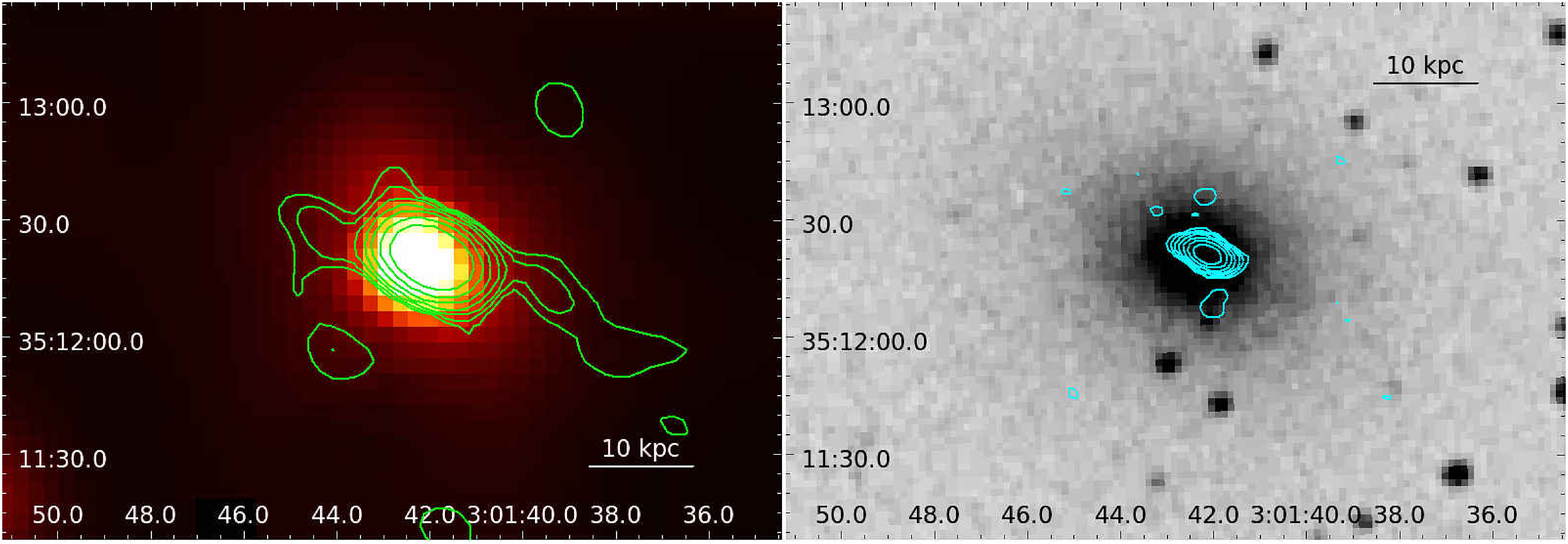}
}
\vspace{-3mm}
\caption{LGG 80 / NGC 1167. \textit{Left:} GMRT 235 MHz contours in green (1$\sigma$ = 0.6 mJy beam$^{-1}$), overlaid on the adaptively smoothed 0.3$-$2.0 keV \textit{XMM-Newton} image. \textit{Right:} GMRT 610~MHz contours in cyan (1$\sigma$ = 60 $\mu$Jy beam$^{-1}$), overlaid on the \textit{DSS} optical image. In both panels the radio contours are spaced by a factor of two, starting from 3$\sigma$. For this source the scale is 0.349 kpc arcsec$^{-1}$.}
\label{fig:1167}
\end{figure*}

Our GMRT observations detect the central source at both frequencies, with extensions to both east and west at 235~MHz (see Figure~\ref{fig:1167}), potentially representing jets associated with the large-scale remnant lobes, or some intermediate period of activity. We find a total spectral index of $\alpha_{235}^{610}=0.59$, and a value for the core similar to the one from \citet{Giroletti05} ($\alpha_{235}^{610}=0.55$).




\subsection{NGC 1453}
		
\begin{figure*}
\centering{
\includegraphics[width=1.00\textwidth]{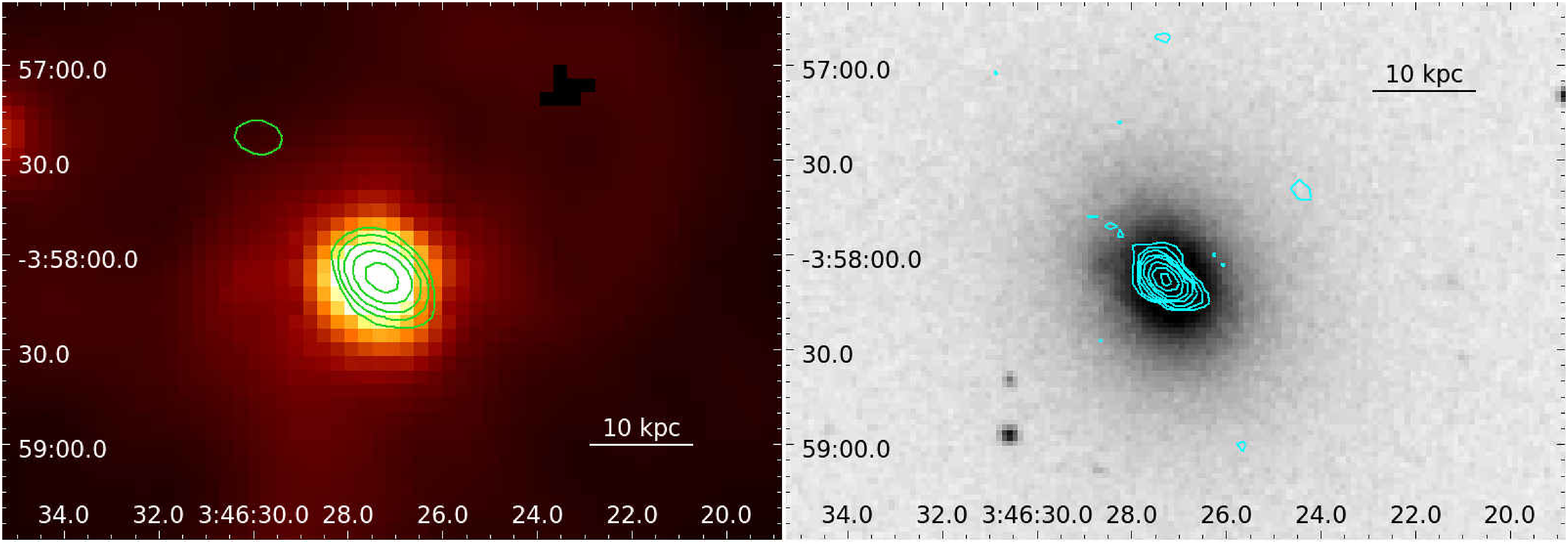}
}
\vspace{-3mm}
\caption{LGG 103 / NGC 1453.  \textit{Left:} GMRT 235 MHz contours in green (1$\sigma$ = 0.6 mJy beam$^{-1}$), overlaid on the adaptively smoothed 0.3$-$2.0 keV \textit{XMM-Newton} image. \textit{Right:} GMRT 610~MHz contours in cyan (1$\sigma$ = 60 $\mu$Jy beam$^{-1}$), overlaid on the \textit{DSS} optical image. In both panels the radio contours are spaced by a factor of two, starting from 3$\sigma$. For this source the scale is 0.305 kpc arcsec$^{-1}$.}
\label{fig:1453}
\end{figure*}

NGC~1453 is a LINER galaxy \citep{Zeilinger96}, located in the centre of the X-ray luminous LGG~103 group. Previous observations at 1.4~GHz detected a point-like radio source of 28~mJy (Table~\ref{Sourcetable}). Our images at 610~MHz and 235~MHz (see figure~\ref{fig:1453}) show a compact radio point-like source at both GMRT frequencies that coincides with the central region of the optical galaxy. The spectral index of $\alpha_{235}^{610}\sim0.2$ indicates that the radio emission possibly originates from a young AGN. 



\subsection{NGC 2563} 

NGC 2563 is the BGE of the X-ray bright LGG~158 group. We detect a weak point-like radio source at 610~MHz, coincident with the centre of the optical component, but the galaxy is undetected at 235~MHz. Previous VLA observations from \cite{Brown11} found a very weak point source at 1.4~GHz. 




\subsection{NGC 3078}

\begin{figure*}
\centering{
\includegraphics[width=1.00\textwidth]{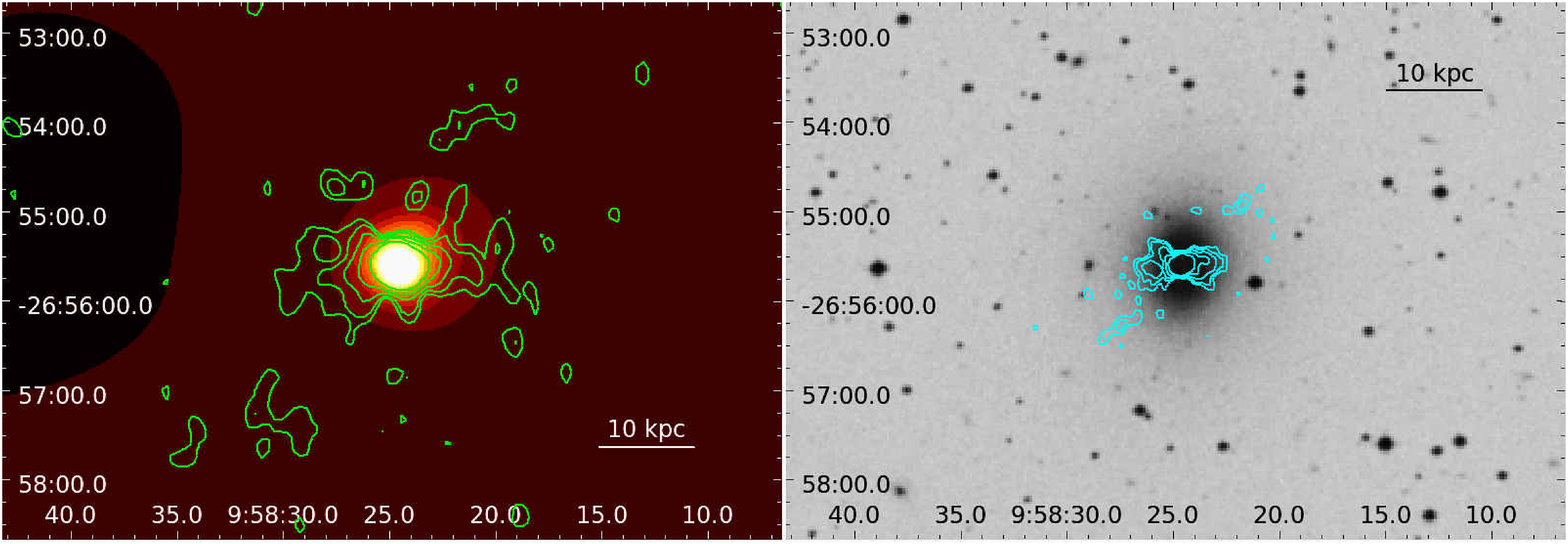}
}
\vspace{-3mm}
\caption{LGG 185 / NGC 3078. \textit{Left:} GMRT 235 MHz contours in green (1$\sigma$ = 0.5 mJy beam$^{-1}$), overlaid on the adaptively smoothed 0.3$-$2.0 keV \textit{Chandra} image. \textit{Right:} GMRT 610~MHz contours in cyan (1$\sigma$ = 200 $\mu$Jy beam$^{-1}$), overlaid on the \textit{DSS} optical image. In both panels the radio contours are spaced by a factor of two, starting from 3$\sigma$. For this source the scale is 0.165 kpc arcsec$^{-1}$.}
\label{fig:3078}
\end{figure*}

NGC 3078, in the X-ray faint LGG~185, hosts a nuclear dust disk \citep{Rest01} and a previously detected, relatively strong radio AGN (310~mJy source at 1.4~GHz). Figure~\ref{fig:3078} shows the source at both GMRT frequencies. It presents a symmetric diffuse structure along an east-west axis, with no clear indication of jets and separated lobes from the central component. The eastern structure appears somewhat broader than the west and overall the source is less extended at 610~MHz than at 235~MHz. The radio source presents a spectral index of $\alpha_{235}^{610}\sim0.5$.

\subsection{NGC 4008} 

\begin{figure*}
\centering{
\includegraphics[width=1.00\textwidth]{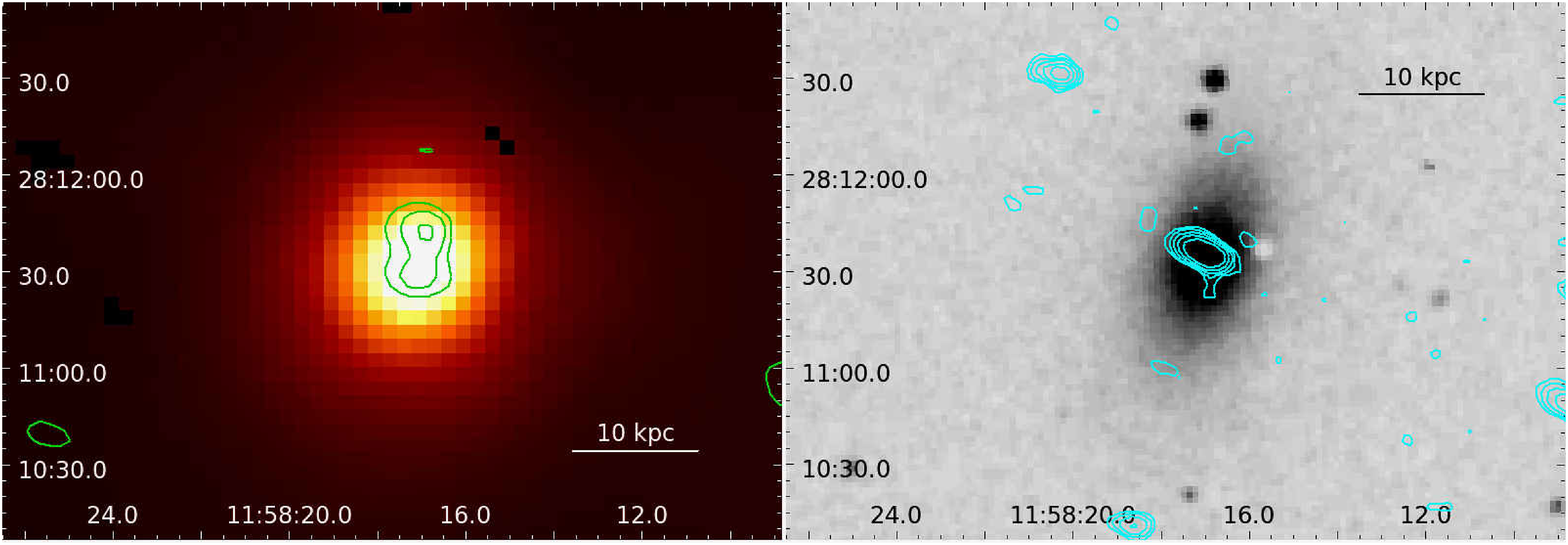}
}
\vspace{-3mm}
\caption{LGG 262 / NGC 4008. \textit{Left:} GMRT 235 MHz contours in green (1$\sigma$ = 1.3 mJy beam$^{-1}$), overlaid on the adaptively smoothed 0.3$-$2.0 keV \textit{XMM-Newton} image. \textit{Right:} GMRT 610~MHz contours in cyan (1$\sigma$ = 50 $\mu$Jy beam$^{-1}$), overlaid on the \textit{DSS} optical image. In both panels the radio contours are spaced by a factor of two, starting from 3$\sigma$. For this source the scale is 0.262 kpc arcsec$^{-1}$.}
\label{fig:4008}
\end{figure*}

NGC 4008 is the H\textsc{i}-rich \citep{Eskridge91} BGE of the X-ray faint LGG~262 group. NVSS observations detect a weak point radio source, and our GMRT images in Figure~\ref{fig:4008} show a slightly elongated point-like radio source at both frequencies. 




\subsection{NGC 4169} 

NGC~4169 is the BGE in the X-ray faint group HCG 61 (LGG~276). Previous observations from NVSS detect a very weak radio point source of $\sim$1~mJy. We detect a radio point source coincident with the center of NGC~4169 with a flux density of 3~mJy at 610 MHz, but no detection is seen at 235~MHz. The upper limit for the 235 MHz detection is placed  at $\leq6$~mJy, at 5$\sigma$ level of significance. We note here that the repetitive patterns in the radio source of NGC~4175, are most probably due to residual amplitude and/or phase calibration errors. However the angular separation of NGC~4169 is far enough for these errors to affect the measured flux density of our source at 610~MHz.

\subsection{ESO 507-25} 

ESO~507-25 is the BGE of the X-ray faint LGG~310 group. A weak (24~mJy) nuclear source was detected at 1.4~GHz \citep{Brown11}, and a weak broad CO emission line \citep{Knapp96} provides evidence of cold gas in the galaxy. At 610~MHz we observe a central point source surrounded by an asymmetric diffuse component with separated clumps of emission north and south of the central region of the galaxy. At 235~MHz no diffuse component is detected but only a compact central point source, with a hint of extension to the north-east. The source exhibits a very flat spectral index ($\alpha_{235}^{610}=0.18$).







\subsection{NGC 5084} 

NGC 5084 is the BGE in the X-ray faint LGG~345 group, and probably recently underwent a merger with a gas-rich dwarf companion \citep{Zeilinger90}, as shown by distorted inner isophotes \citep{Carignan97} and a high H\textsc{i} mass ($M_{(H\textsc{i})}/L_{(B)}\sim0.35$; \citealt{vanDriel91}). The galaxy hosts a nuclear radio point source in the NVSS, with a flux density of of 46.6 mJy.

\begin{figure*}
\centering
\includegraphics[width=1.00\textwidth]{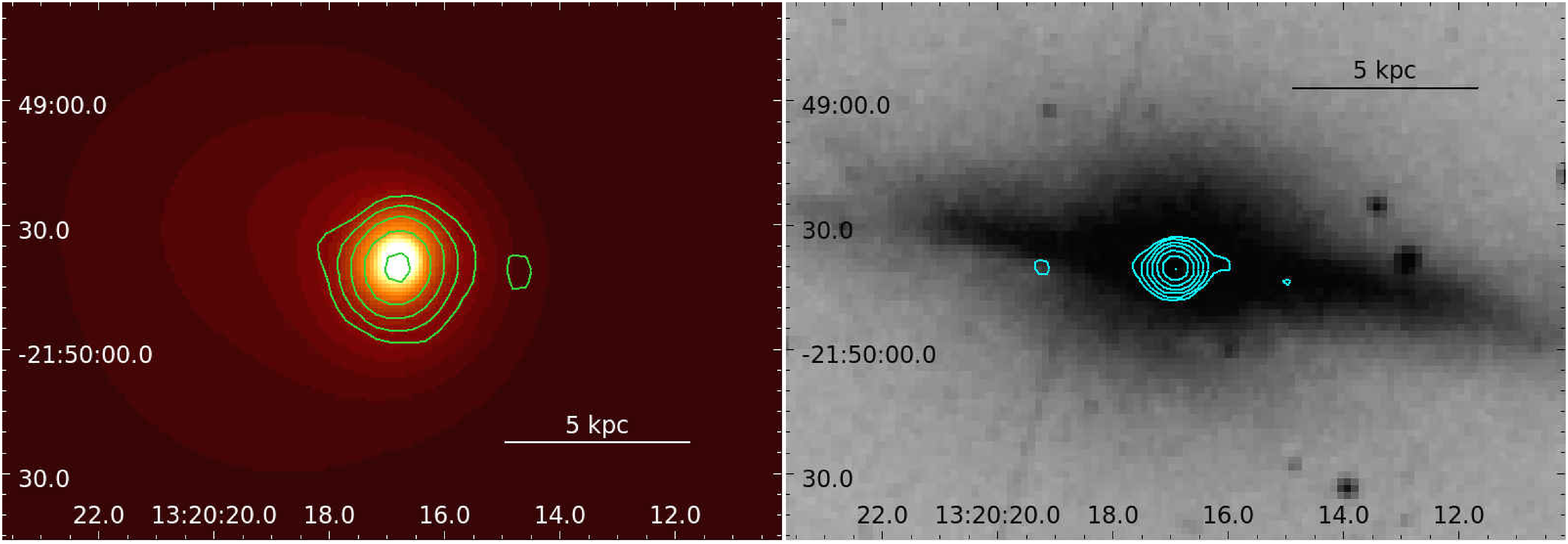}
\vspace{-3mm}
\caption{LGG 345 / NGC 5084. \textit{Left:} GMRT 235 MHz contours in green (1$\sigma$ = 0.65 mJy beam$^{-1}$), overlaid on the adaptively smoothed 0.3$-$2.0 keV \textit{Chandra} image. \textit{Right:} GMRT 610~MHz contours in cyan (1$\sigma$ = 90 $\mu$Jy beam$^{-1}$), overlaid on the \textit{DSS} optical image. In both panels the radio contours are spaced by a factor of two, starting from 3$\sigma$. For this source the scale is 0.112 kpc arcsec$^{-1}$.}
\label{fig:5084}
\end{figure*}

At both GMRT frequencies, we detect a radio point source that coincides with the central region of the galaxy, presenting some hints of extension in the east$-$west direction. However, noise features owing to imperfect calibration are responsible for the detached clumps of emission east and west of the core in Figure~\ref{fig:5084}, and we therefore consider only the central point-like structure as reliable. A flux density of 36.2 mJy at 610~MHz and 53.9 mJy at 235~MHz give a relatively flat spectral index of $\alpha\sim0.4$ for this source at this frequency range.


                                 

\subsection{NGC 5153} 

NGC 5153 is the BGE in the X-ray faint LGG~351 group and is interacting with the spiral galaxy NGC~5152 \citep{Weilbacher00}. There is no previous radio measurement for this system and no emission is also detected by our GMRT observations. 





\subsection{NGC 5353} 

\begin{figure*}
\centering
\includegraphics[width=1.00\textwidth]{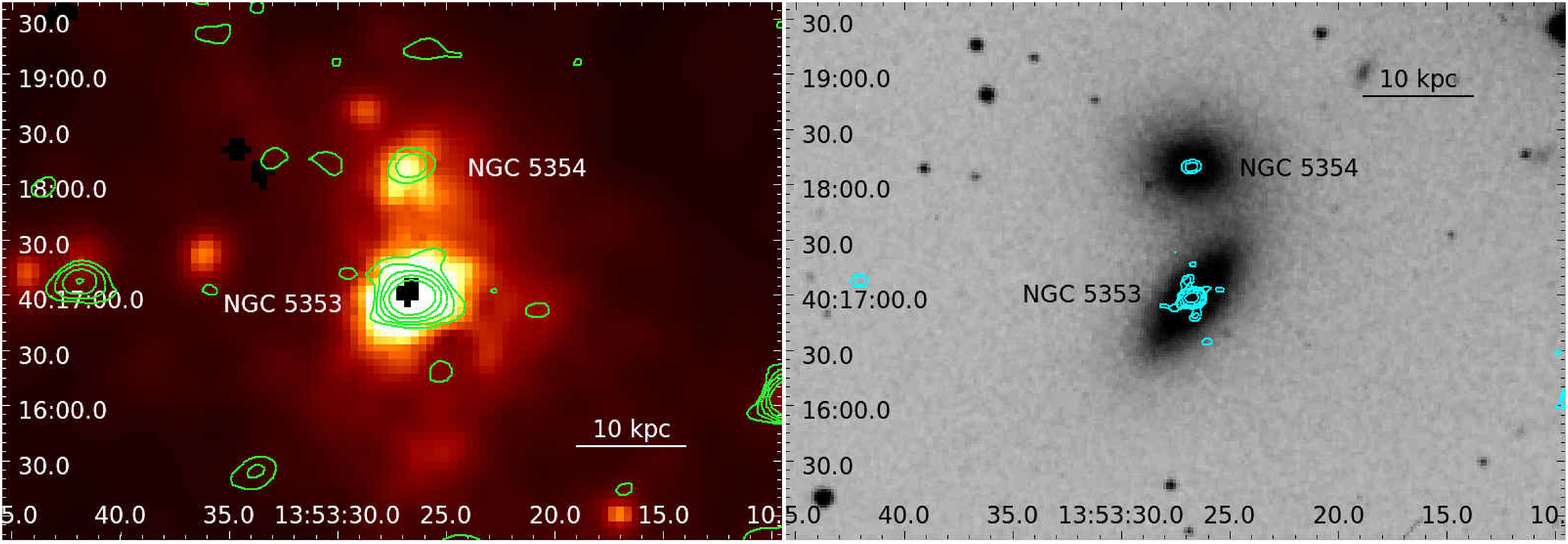}
\vspace{-3mm}
\caption{LGG 363 / NGC 5353. \textit{Left:} GMRT 235 MHz contours in green (1$\sigma$ = 0.6 mJy beam$^{-1}$), overlaid on the adaptively smoothed 0.3$-$2.0 keV \textit{XMM-Newton} image. \textit{Right:} GMRT 610~MHz contours in cyan (1$\sigma$ = 60 $\mu$Jy beam$^{-1}$), overlaid on the \textit{DSS} optical image. In both panels the radio contours are spaced by a factor of two, starting from 4$\sigma$. For this source the scale is 0.170 kpc arcsec$^{-1}$.}
\label{fig:5353}
\end{figure*}

NGC 5353 is an edge-on S0 galaxy laying in the dynamical centre of the X-ray bright LGG~363 group \citep[HCG~68,][]{Hickson82} being part of the group's dominant pair suspected to be in the process of merging \citep{Tully08}. Point radio sources with flux densities of 41~mJy in NGC~5353 and of 8.4~mJy in NGC~5354 are reported in the NVSS. Our GMRT observations detect compact point radio sources in both galaxies. The presence of a strong background source nearby produces high noise levels at 610~MHz, adding false structures symmetrically distributed in the east$-$west direction around NGC~5353. We therefore start the 610~MHz contours shown in Figure~\ref{fig:5353} at the 4$\sigma$ level of significance.

\subsection{NGC 5846}
		
NGC 5846 is the BGE in the X-ray luminous massive LGG~393 group of galaxies \citep[see][]{Mahdavi05}. Using the GMRT at 610~MHz, \citet{Simona11} detected a nuclear source with small scale jets of size of $\sim$12 kpc. In our 235~MHz observations we detect a compact point radio source that coincides with the central region of the optical galaxy, with hints of extension on an east$-$west axis, but the resolution is not sufficient to trace the jets seen at 610~MHz. X-ray observations reveal a number of structures in the group core, including cavities \citep[e.g.,][]{Allen06,Dong10} and evidence of sloshing \citep{Machacek11}.

\subsection{NGC 5982} 

\begin{figure*}
\centering{
\includegraphics[width=1.00\textwidth]{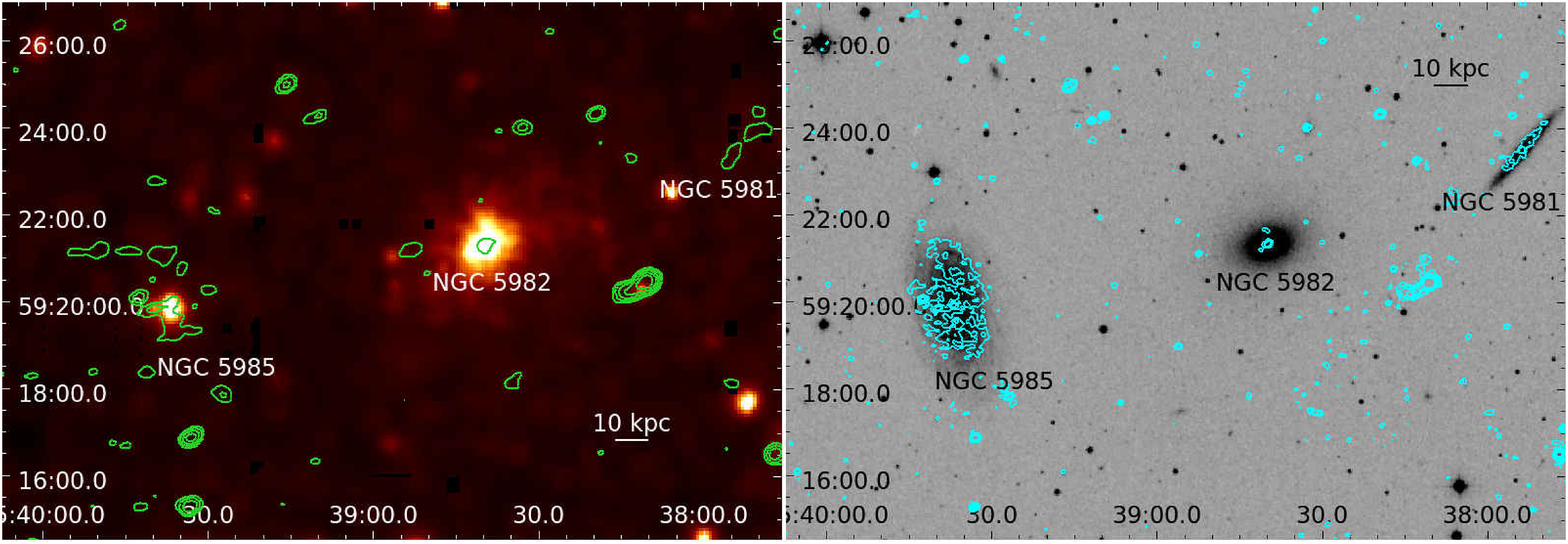}
}
\vspace{-3mm}
\caption{LGG 402 / NGC 5982  \textit{Left:} GMRT 235 MHz contours in green (1$\sigma$ = 0.4 mJy beam$^{-1}$), overlaid on the adaptively smoothed 0.3$-$2.0 keV \textit{XMM-Newton} image. \textit{Right:} GMRT 610~MHz contours in cyan (1$\sigma$ = 90 $\mu$Jy beam$^{-1}$), overlaid on the \textit{DSS} optical image. In both panels the radio contours are spaced by a factor of two, starting from 3$\sigma$. For this source the scale is 0.213 kpc arcsec$^{-1}$.}
\label{fig:5982}
\end{figure*}

LGG~402 is a relatively poor X-ray luminous group dominated by the elliptical NGC~5982 \citep[possibly post-merger;][]{DB} and the two large spiral galaxies that lie close by (the `Draco Trio'). GMRT 610 and 235~MHz observations (Figure~\ref{fig:5982}) detect a weak, point-like radio source at the center of the NGC~5982. Previous 1.4 GHz observations found only hints of this central source \citep{Brown11}.

\subsection{NGC 6658} 

\begin{figure*}
\centering{
\includegraphics[width=1.00\textwidth]{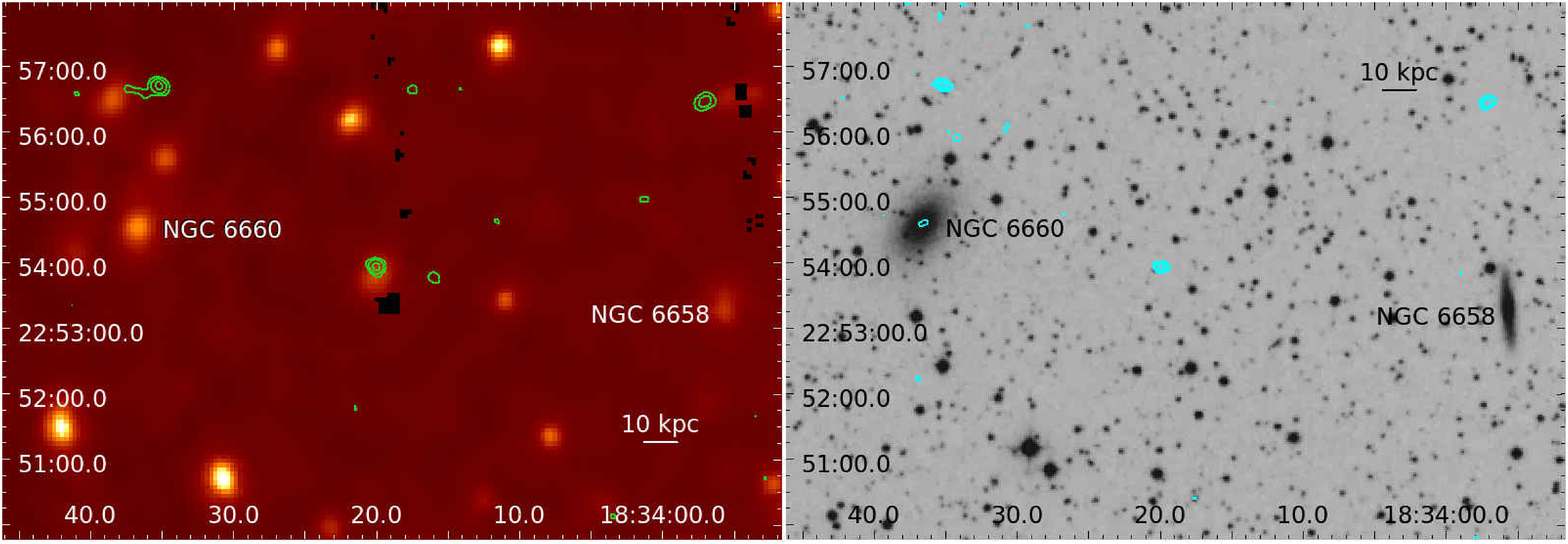}
}
\vspace{-3mm}
\caption{LGG 421 / NGC 6658.  \textit{Left:} GMRT 235 MHz contours in green (1$\sigma$ = 0.6 mJy beam$^{-1}$), overlaid on the adaptively smoothed 0.3$-$2.0 keV \textit{XMM-Newton} image. \textit{Right:} GMRT 610~MHz contours in cyan (1$\sigma$ = 50 $\mu$Jy beam$^{-1}$), overlaid on the \textit{DSS} optical image. In both panels the radio contours are spaced by a factor of two, starting from 3$\sigma$. For this source the scale is 0.305 kpc arcsec$^{-1}$.}
\label{fig:6658}

\end{figure*}

NGC 6658 is an edge-on S0 type BGE in the X-ray faint LGG~421 poor group of galaxies. The galaxy was not previously detected in the radio band and our GMRT observations find no radio detection at either frequency (fig.~\ref{fig:6658}). The flux density upper limit set for this galaxy at 5$\sigma$ level of significance is 0.3~mJy at 610~MHz and 3 mJy at 235~MHz. Figure~\ref{fig:6658} shows only a point source detected in the spiral companion NGC~6660. 
 




\begin{figure*}
\centering
\includegraphics[width=1.00\textwidth]{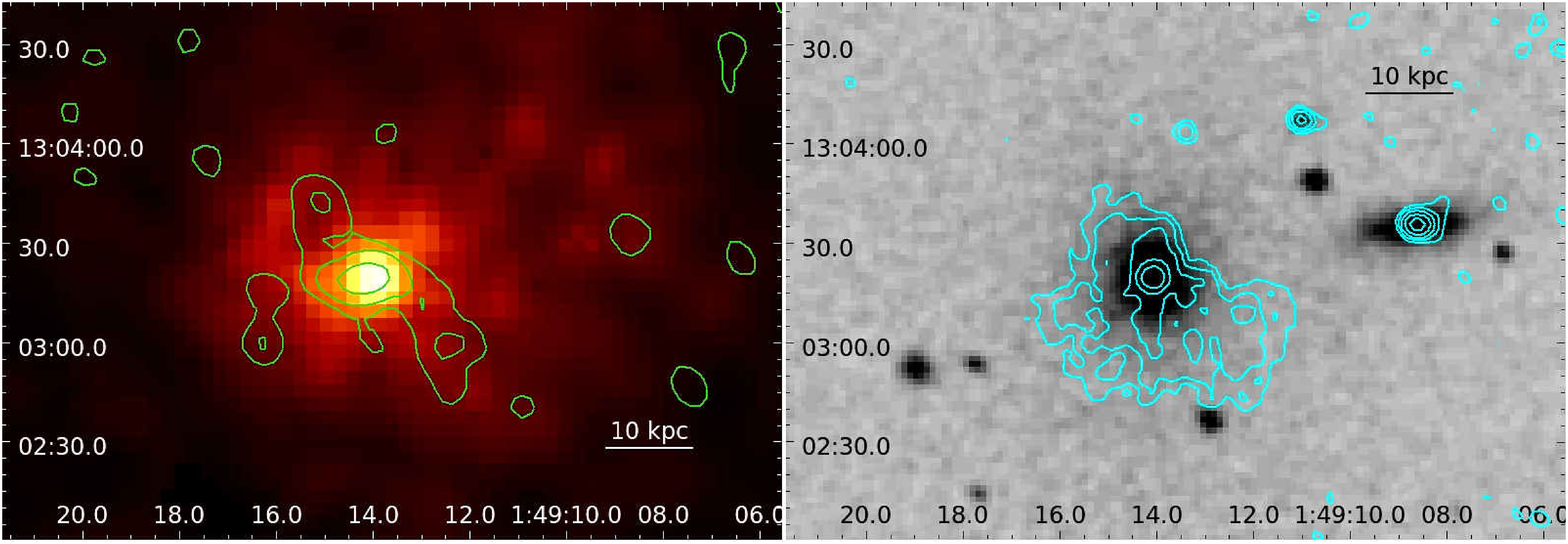}
\vspace{-3mm}
\caption{LGG 31 / NGC 677. \textit{Left:} GMRT 235 MHz contours in green (1$\sigma$ = 1.2 mJy beam$^{-1}$), overlaid on the adaptively smoothed 0.3$-$2.0 keV \textit{XMM-Newton} image. \textit{Right:} GMRT 610~MHz contours in cyan (1$\sigma$ = 40 $\mu$Jy beam$^{-1}$), overlaid on the \textit{DSS} optical image. In both panels the radio contours are spaced by a factor of two, starting from 3$\sigma$. For this source the scale is 0.378 kpc arcsec$^{-1}$.}
\label{fig:677}
\end{figure*}

\begin{figure*}
\centering{
\includegraphics[width=1.00\textwidth]{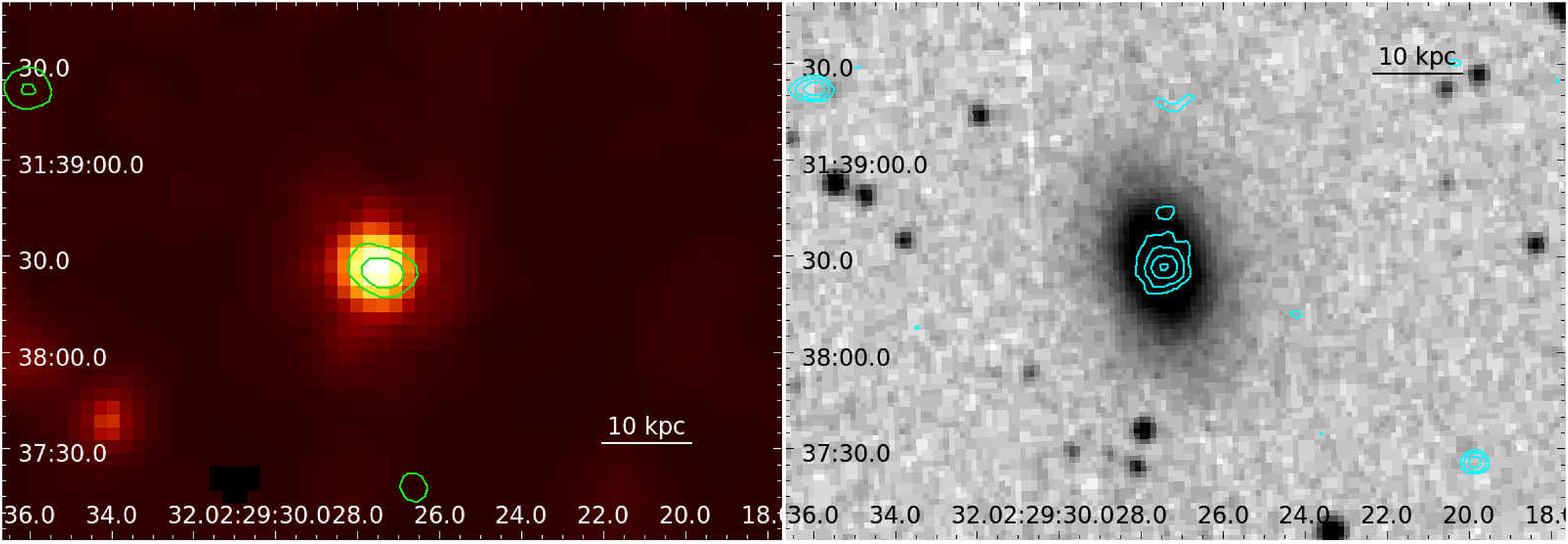}
}
\vspace{-3mm}
\caption{LGG 58 / NGC 940. \textit{Left:} GMRT 235 MHz contours in green (1$\sigma$ = 0.3 mJy beam$^{-1}$), overlaid on the adaptively smoothed 0.3$-$2.0 keV \textit{XMM-Newton} image. \textit{Right:} GMRT 610~MHz contours in cyan (1$\sigma$ = 60 $\mu$Jy beam$^{-1}$), overlaid on the \textit{DSS} optical image. In both panels the radio contours are spaced by a factor of two, starting from 3$\sigma$. For this source the scale is 0.354 kpc arcsec$^{-1}$.}
\label{fig:940}
\end{figure*}

\begin{figure*}
\centering
\includegraphics[width=1.00\textwidth]{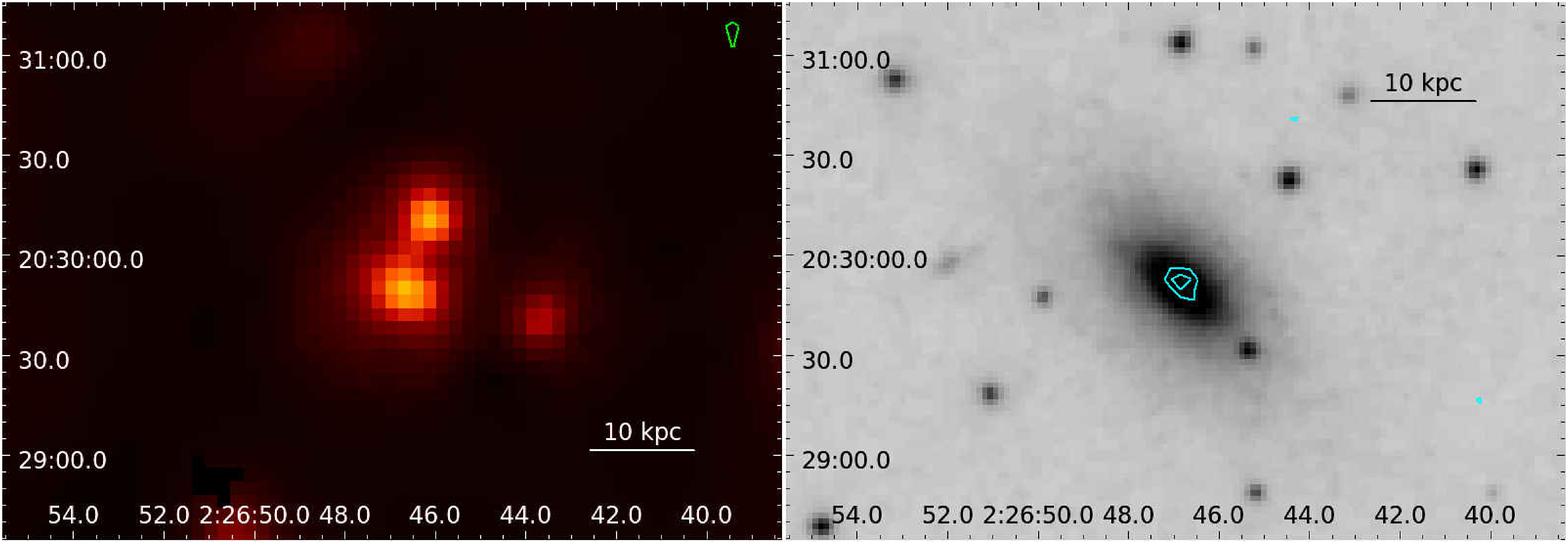}
\vspace{-3mm}
\caption{LGG 61 / NGC 924. \textit{Left:} GMRT 235 MHz contours in green (1$\sigma$ = 0.3 mJy beam$^{-1}$), overlaid on the adaptively smoothed 0.3$-$2.0 keV \textit{XMM-Newton} image. \textit{Right:} GMRT 610~MHz contours in cyan (1$\sigma$ = 50 $\mu$Jy beam$^{-1}$), overlaid on the \textit{DSS} optical image. In both panels the radio contours are spaced by a factor of two, starting from 3$\sigma$. For this source the scale is 0.310 kpc arcsec$^{-1}$.}
\label{fig:924}
\end{figure*}

\begin{figure*}
\centering{
\includegraphics[width=1.00\textwidth]{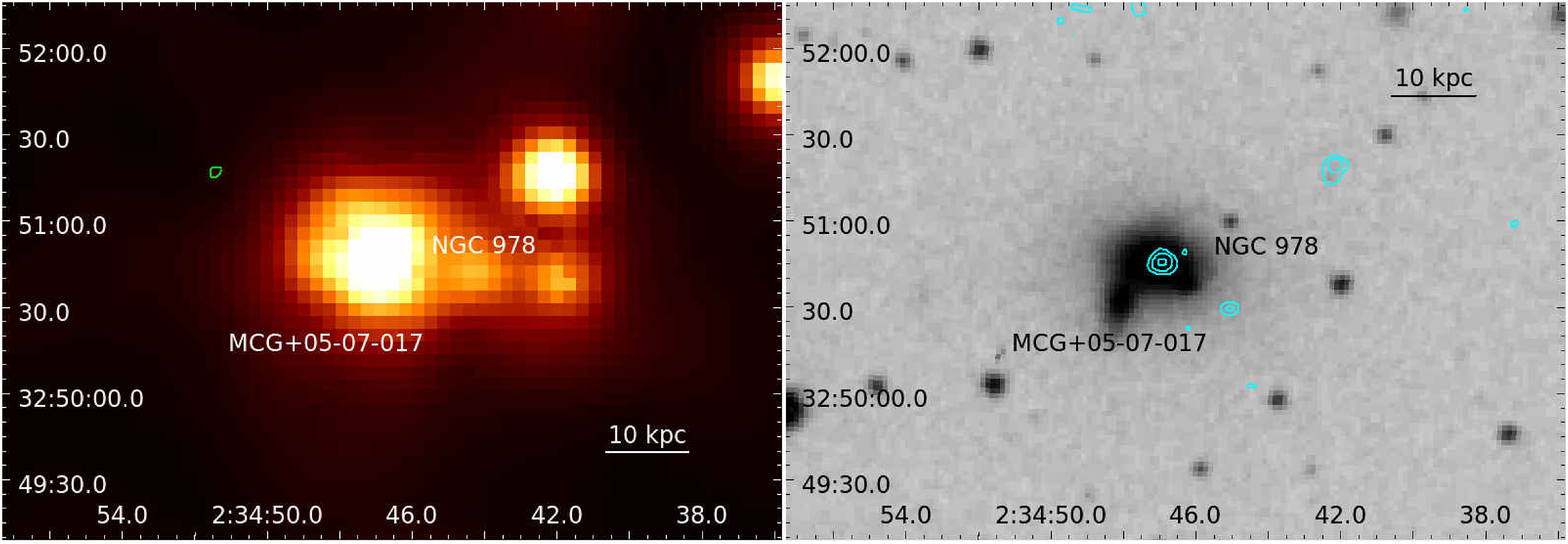}
}
\vspace{-3mm}
\caption{LGG 66 / NGC 978. \textit{Left:} GMRT 235 MHz contours in green (1$\sigma$ = 0.4 mJy beam$^{-1}$), overlaid on the adaptively smoothed 0.3$-$2.0 keV \textit{XMM-Newton} image. \textit{Right:} GMRT 610~MHz contours in cyan (1$\sigma$ = 60 $\mu$Jy beam$^{-1}$), overlaid on the \textit{DSS} optical image. In both panels the radio contours are spaced by a factor of two, starting from 3$\sigma$. For this source the scale is 0.334 kpc arcsec$^{-1}$.}
\label{fig:978}
\end{figure*}

\begin{figure*}
\centering{
\includegraphics[width=1.00\textwidth]{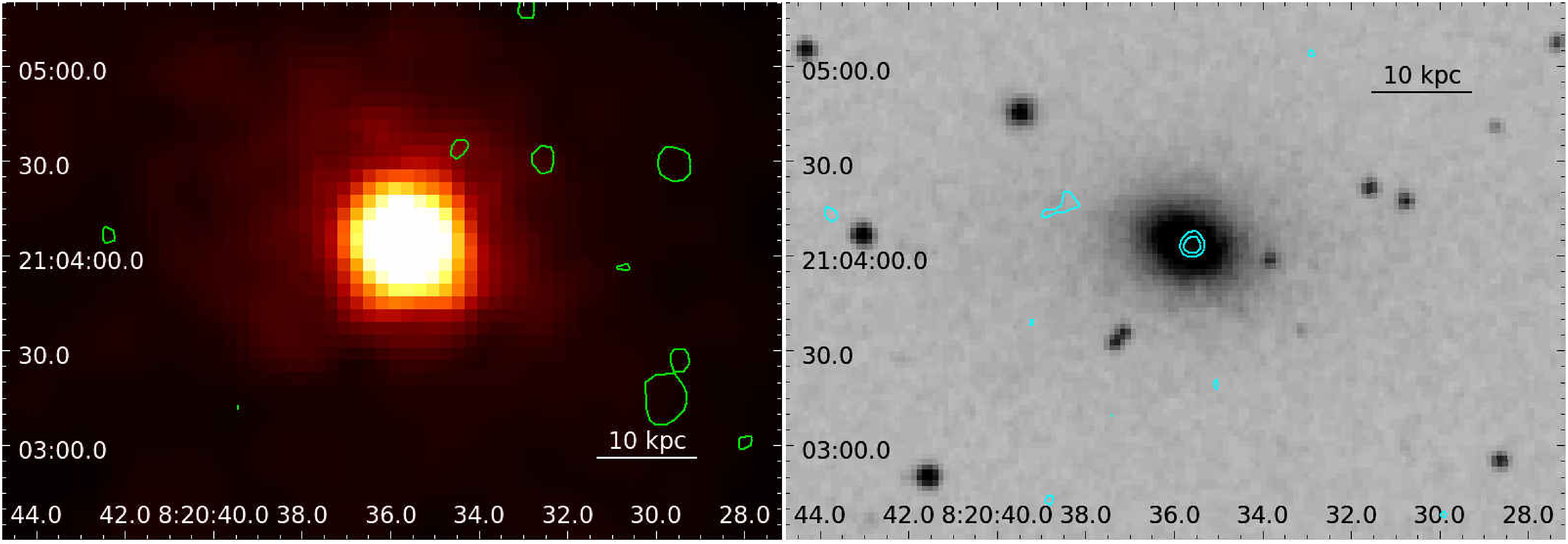}
}
\vspace{-3mm}
\caption{LGG 158 / NGC 2563. \textit{Left:} GMRT 235 MHz contours in green (1$\sigma$ = 0.3 mJy beam$^{-1}$), overlaid on the adaptively smoothed 0.3$-$2.0 keV \textit{XMM-Newton} image. \textit{Right:} GMRT 610~MHz contours in cyan (1$\sigma$ = 70 $\mu$Jy beam$^{-1}$), overlaid on the \textit{DSS} optical image. In both panels the radio contours are spaced by a factor of two, starting from 3$\sigma$. For this source the scale is 0.315 kpc arcsec$^{-1}$.}
\label{fig:2563}
\end{figure*}

\begin{figure*}
\centering{
\includegraphics[width=1.00\textwidth]{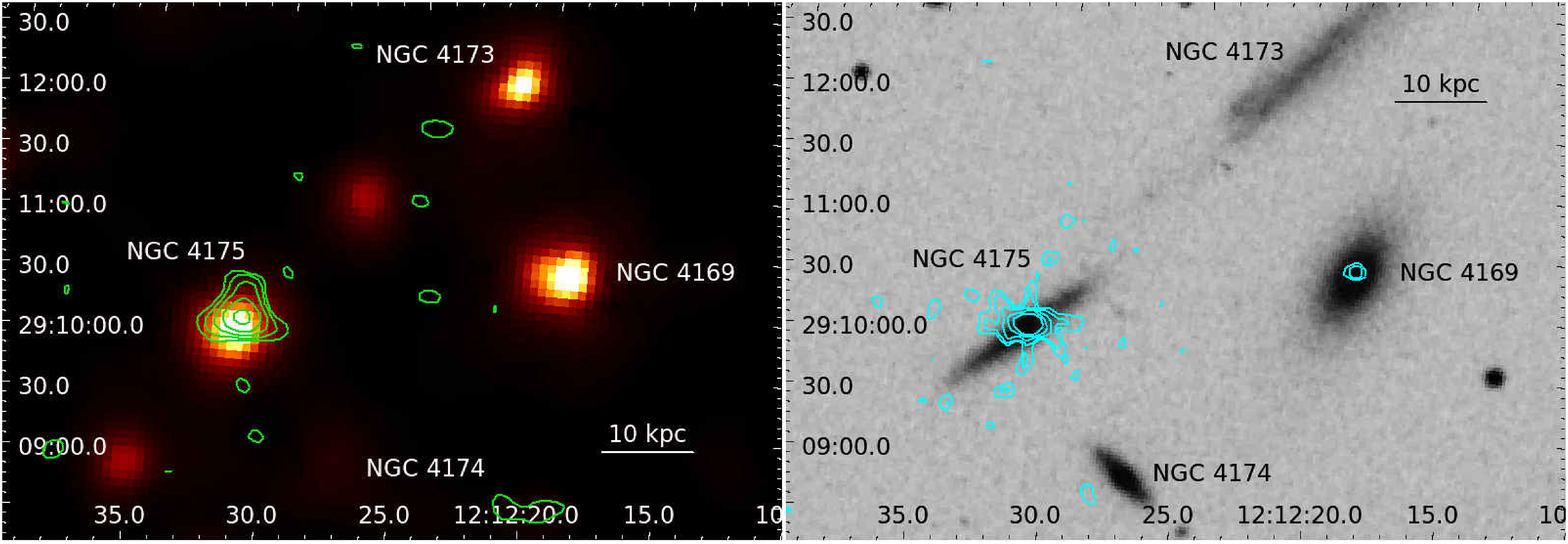}
}
\vspace{-3mm}
\caption{LGG 276 / NGC 4169. \textit{Left:} GMRT 235 MHz contours in green (1$\sigma$ = 1.2 mJy beam$^{-1}$), overlaid on the adaptively smoothed 0.3$-$2.0 keV \textit{XMM-Newton} image. \textit{Right:} GMRT 610~MHz contours in cyan (1$\sigma$ = 80 $\mu$Jy beam$^{-1}$), overlaid on the \textit{DSS} optical image. In both panels the radio contours are spaced by a factor of two, starting from 3$\sigma$. For this source the scale is 0.218 kpc arcsec$^{-1}$.}
\label{4169}
\end{figure*}

\begin{figure*}
\centering{
\includegraphics[width=1.00\textwidth]{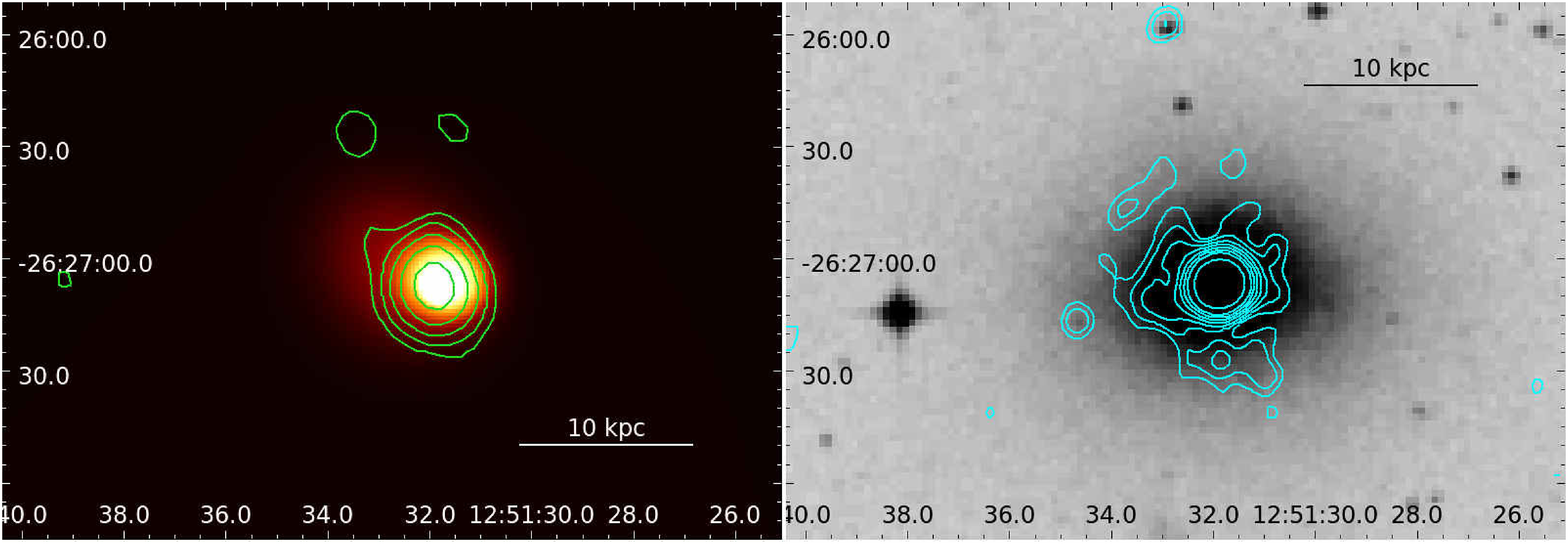}
}
\vspace{-3mm}
\caption{LGG 310 / ESO 507-25. \textit{Left:} GMRT 235 MHz contours in green (1$\sigma$ = 0.5 mJy beam$^{-1}$), overlaid on the adaptively smoothed 0.3$-$2.0 keV \textit{Chandra} image. \textit{Right:} GMRT 610~MHz contours in cyan (1$\sigma$ = 100 $\mu$Jy beam$^{-1}$), overlaid on the \textit{DSS} optical image. In both panels the radio contours are spaced by a factor of two, starting from 3$\sigma$. For this source the scale is 0.218 kpc arcsec$^{-1}$.}
\label{fig:ESO}
\end{figure*}

\begin{figure*}
\centering
\includegraphics[width=1.00\textwidth]{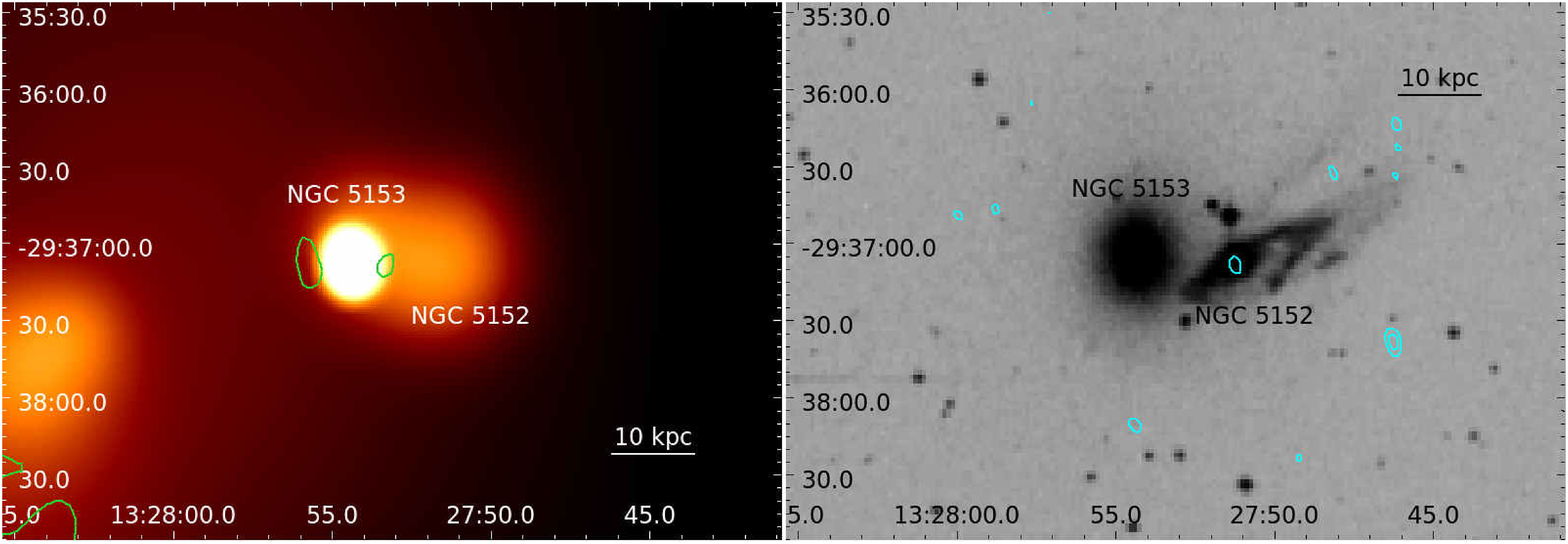}
\vspace{-3mm}
\caption{LGG 351 / NGC 5153. \textit{Left:} GMRT 235 MHz contours in green (1$\sigma$ = 0.3 mJy beam$^{-1}$), overlaid on the adaptively smoothed 0.3$-$2.0 keV \textit{Chandra} image. \textit{Right:} GMRT 610~MHz contours in cyan (1$\sigma$ = 60 $\mu$Jy beam$^{-1}$), overlaid on the \textit{DSS} optical image. In both panels the radio contours are spaced by a factor of two, starting from 3$\sigma$. For this source the scale is 0.291 kpc arcsec$^{-1}$.}
\label{fig:5153}
\end{figure*}

\begin{figure*}
\centering{
\includegraphics[width=1.00\textwidth]{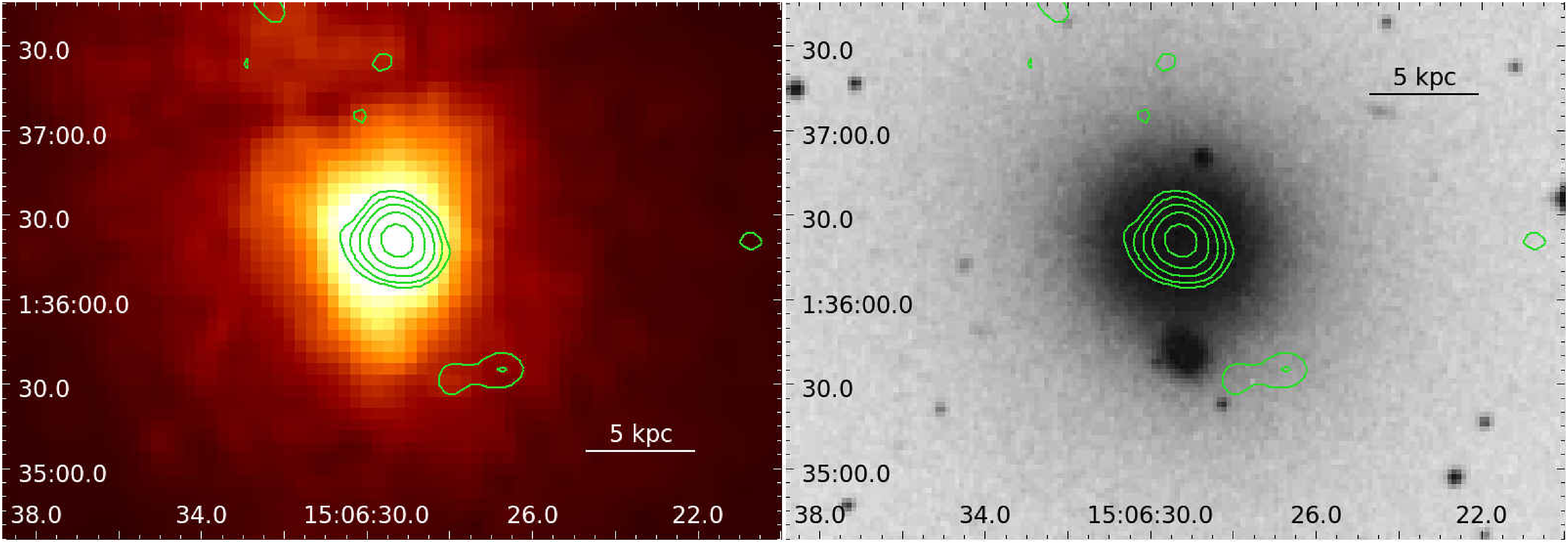}
}
\vspace{-3mm}
\caption{LGG 393 / NGC~5846.  GMRT 235 MHz contours in green (1$\sigma$ = 0.5 mJy beam$^{-1}$), overlaid on the adaptively smoothed 0.3$-$2.0 keV \textit{XMM-Newton} image (left) and on the \textit{DSS} optical image (right). In both panels the radio contours are spaced by a factor of two, starting from 3$\sigma$. For this source the scale is 0.126 kpc arcsec$^{-1}$. X-ray cavities have been identified by \citet{Allen06} and confirmed by \citet{Dong10} at a distance $<$1 kpc from the group centre. The low resolution of the 235~MHz image does not profit for a direct comparison as the X-ray cavities are placed in the central area covered by the radio point source.}
\label{fig:5846}
\end{figure*}

\onecolumn

\section{Star formation rates and expected radio powers}

Table~\ref{SFR} lists the FUV SFR for those galaxies where star formation could potentially affect our radio flux density measurements, and its estimated contribution to 610~MHz radio power.

\begin{table}\mbox{} \mbox{} \vspace{50mm}

 \caption{Star formation rates from FUV and the expected radio power at 610~MHz due to star formation for diffuse and point radio sources in our sample with FUV data. For each group we note the SFR$_{FUV}$, the expected radio power at 610~MHz P$_{610~expected}$ from the calculated SFR$_{FUV}$, the radio power at 610~MHz P$_{610~observed}$, and the radio morphology of the source. 
 }
 \label{SFR}
\begin{center}
\begin{tabular}{cccccc}
\hline 
Group &  BGE &  SFR$_{FUV}$  &  P$_{610~expected}$ & P$_{610~observed}$ & Morphology  \\
LGG   &      & ($10^{-2}$ M$_{\odot}$ yr$^{-1}$) &  (10$^{21}$ W Hz$^{-1}$)&   (10$^{21}$ W Hz$^{-1}$) &  \\ \\
\hline \hline\\
 18 & NGC 410   & 7.5  & 0.26  & 9.6    & point \\
 27 & NGC 584   & 1.3 &  0.05  &  0.08$^{a}$ &  point  \\
 31 & NGC 677   & 3.4& 0.12   & 33.1    & diffuse  \\
 42 & NGC 777   & 9.3 & 0.33   &  6.5    & point    \\
 58 & NGC 940   & 9.8  & 0.34   &  2.8   & point    \\
 61 & NGC 924   & 4.8  & 0.17   &   0.8  & point    \\
 66 & NGC 978   & 2.2 & 0.08   &   0.7  & point    \\
103 & NGC 1453  & 2.7  & 0.09   &   19.0  & point    \\
117 & NGC 1587  & 2.2 & 0.08   &   69.2  & diffuse  \\
158 & NGC 2563   & 4.1 & 0.15   &   0.7  & point    \\
185 & NGC 3078   & 1.4 & 0.05   &   52.9  & diffuse  \\
262 & NGC 4008   & 1.7  & 0.06   &   5.7  & point    \\
310 & ESO 507-25 & 11.1 & 0.39   &  11.2   & diffuse  \\
345 & NGC 5084   & 1.8  & 0.07   &   2.3  & point    \\
363 & NGC 5353   & 3.9  & 0.14   &   6.7 & point    \\
402 & NGC 5982   & 4.5 & 0.16   &   0.4  & point    \\
473 & NGC 7619   & 3.1 & 0.11   &   11.2  & point    \\ \\

\hline

\multicolumn{6}{c}{$^a$ \scriptsize{P$_{610~observed}$ was calculated here by extrapolating the 610~MHz flux density from the available 1.4~GHz emission, using a spectral index of 0.8}}
\end{tabular}
\end{center}

\vspace{-150mm}
\end{table}

 \vspace{140mm}
\section{Radio spectra of CLoGS BGEs}

\begin{figure*}
\centering{
\caption{Flux density distribution of CLoGS central galaxies over 235, 610 and 1400 MHz}}
\label{power}
\begin{tabular}{ccc}
\includegraphics[width=0.22\textwidth]{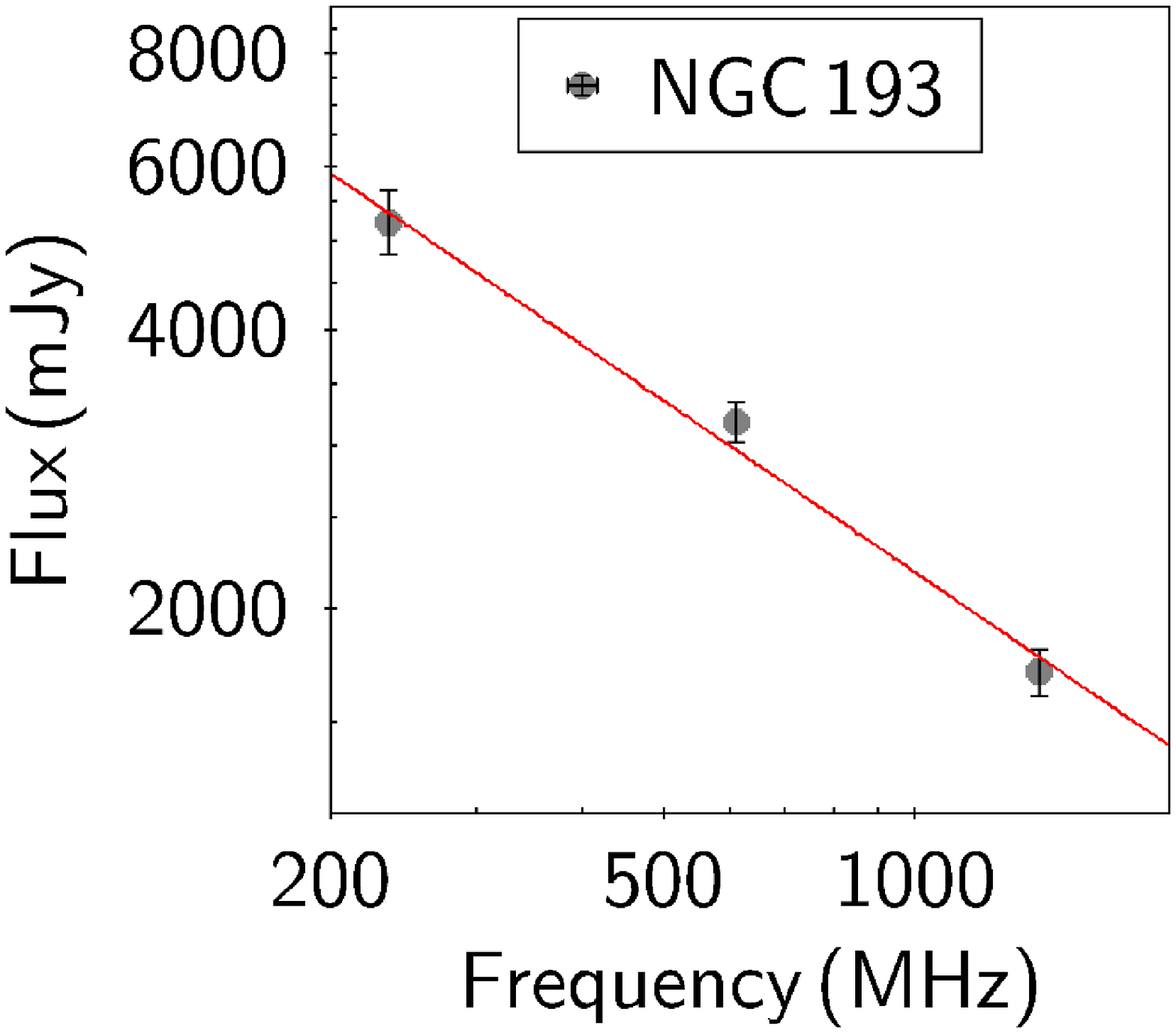}& 
\includegraphics[width=0.21\textwidth]{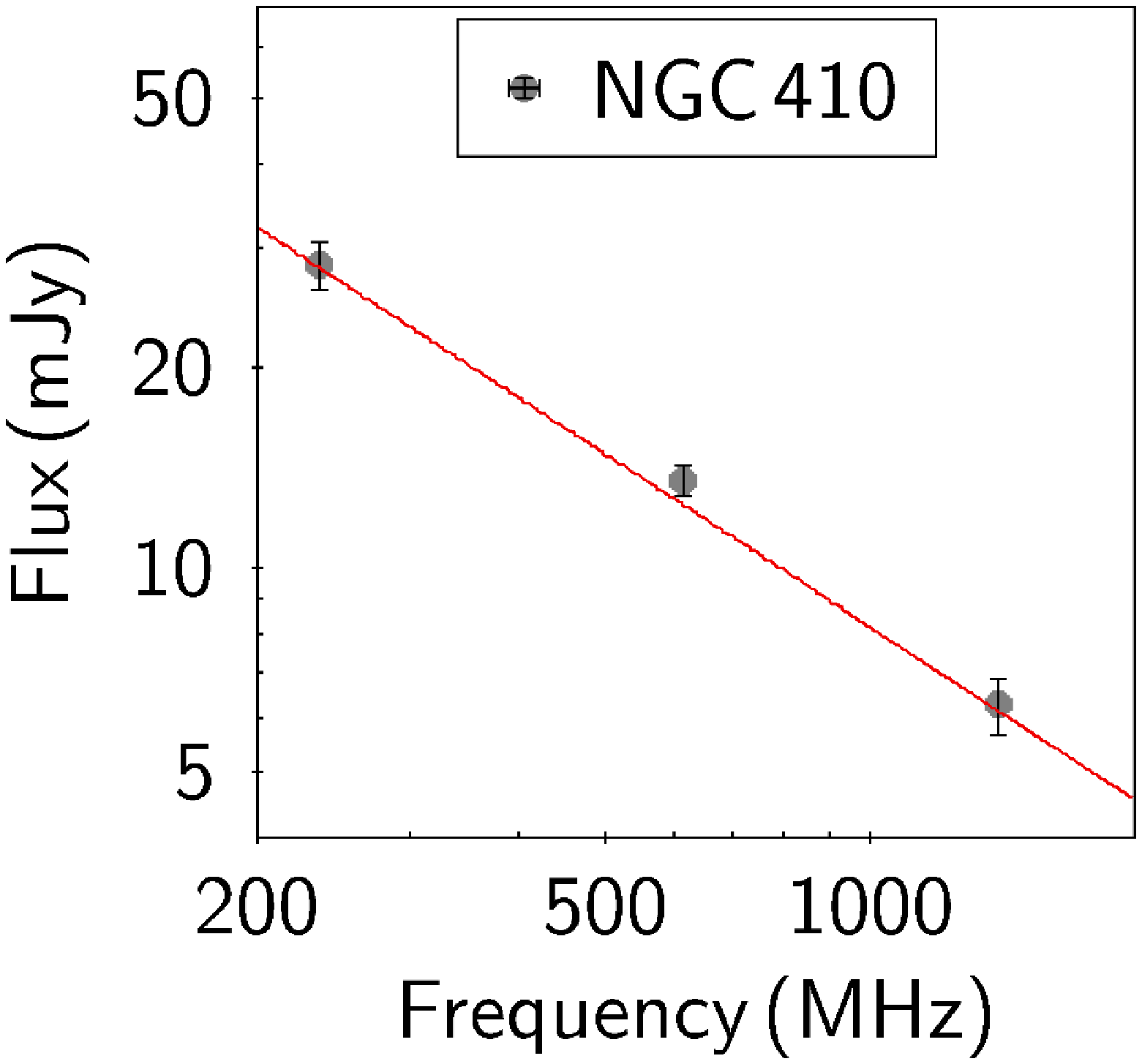}&  
\includegraphics[width=0.21\textwidth]{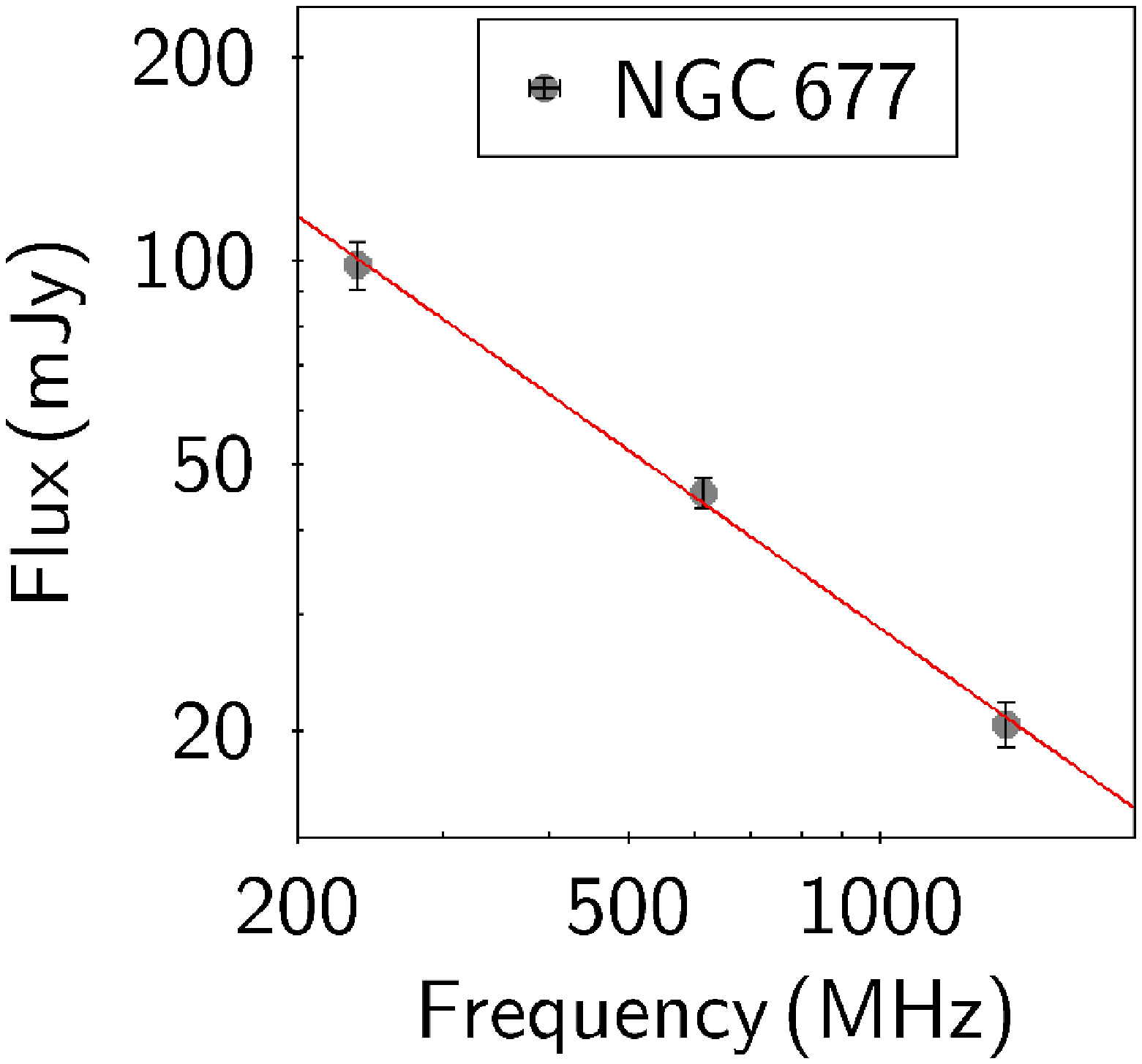}\\
\includegraphics[width=0.21\textwidth]{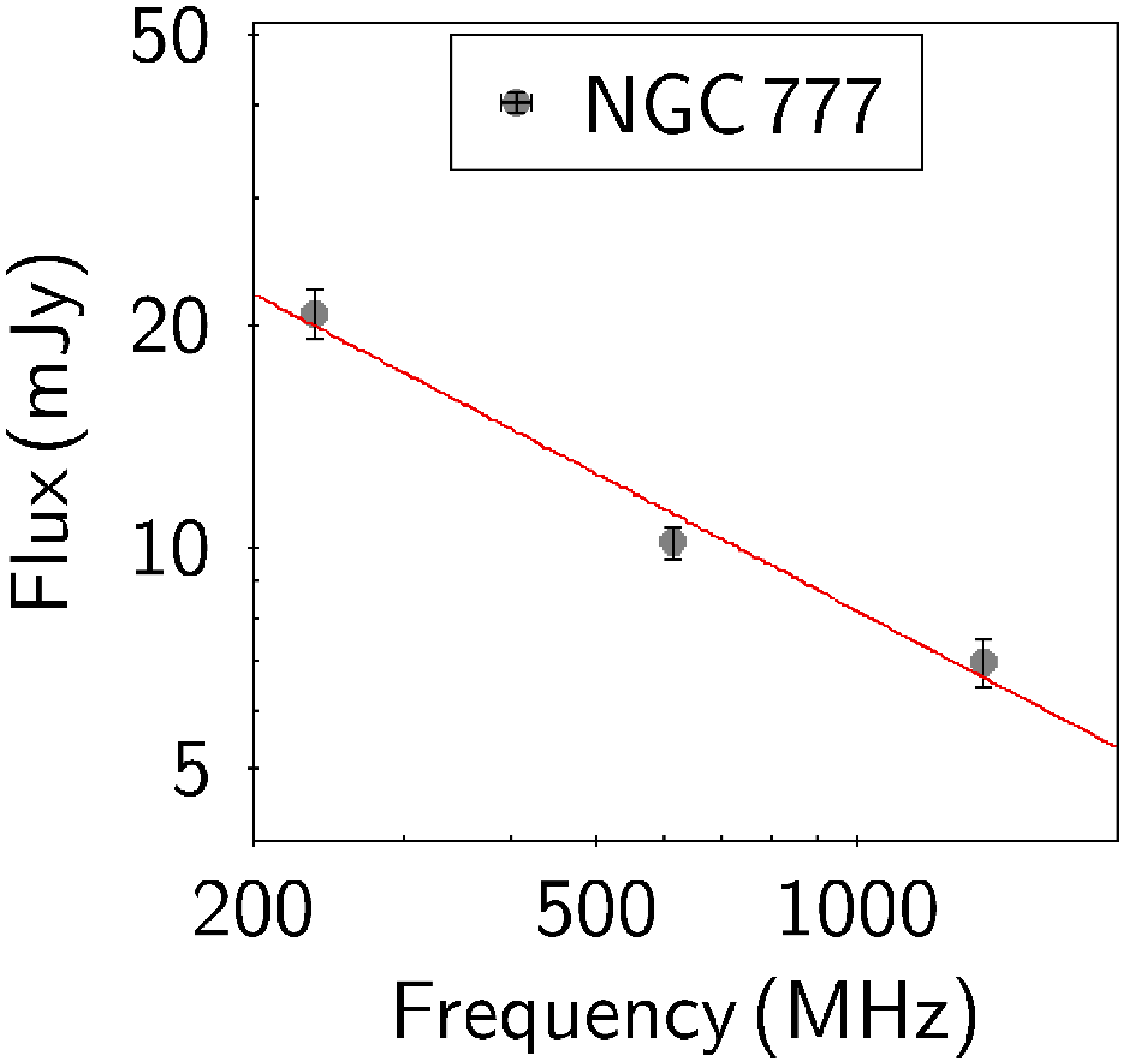}&      
\includegraphics[width=0.21\textwidth]{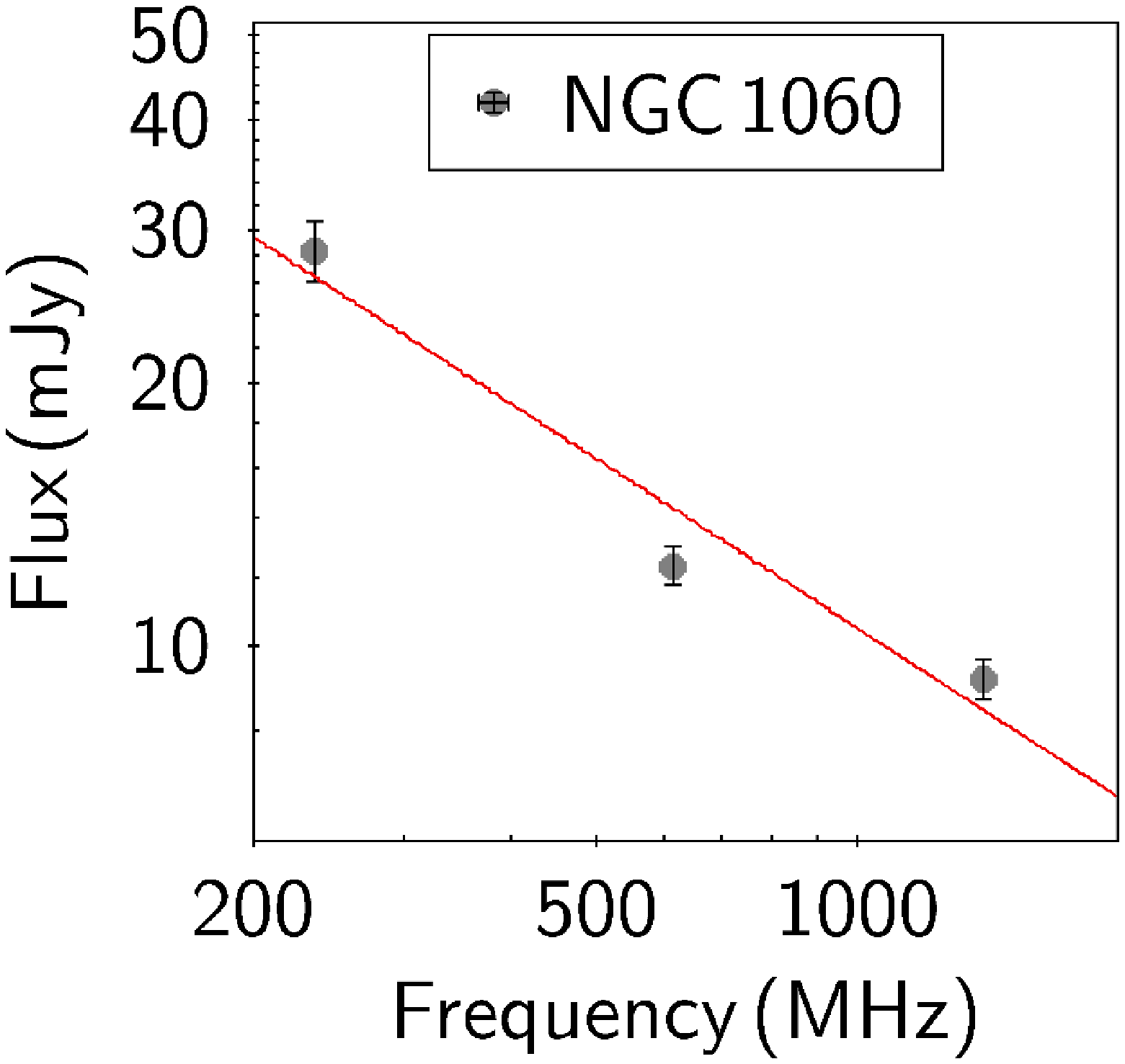}&   
\includegraphics[width=0.22\textwidth]{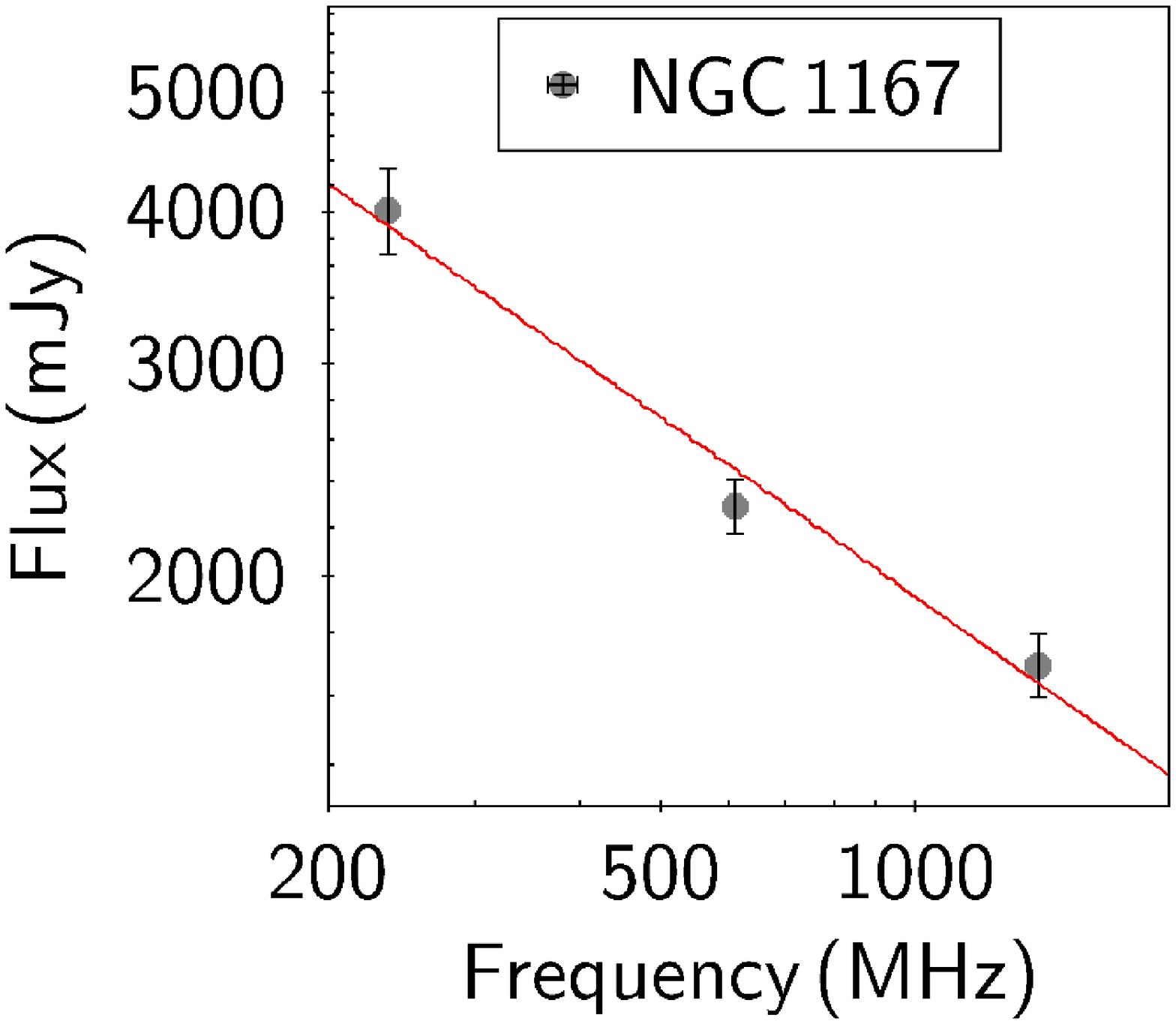}\\
\includegraphics[width=0.21\textwidth]{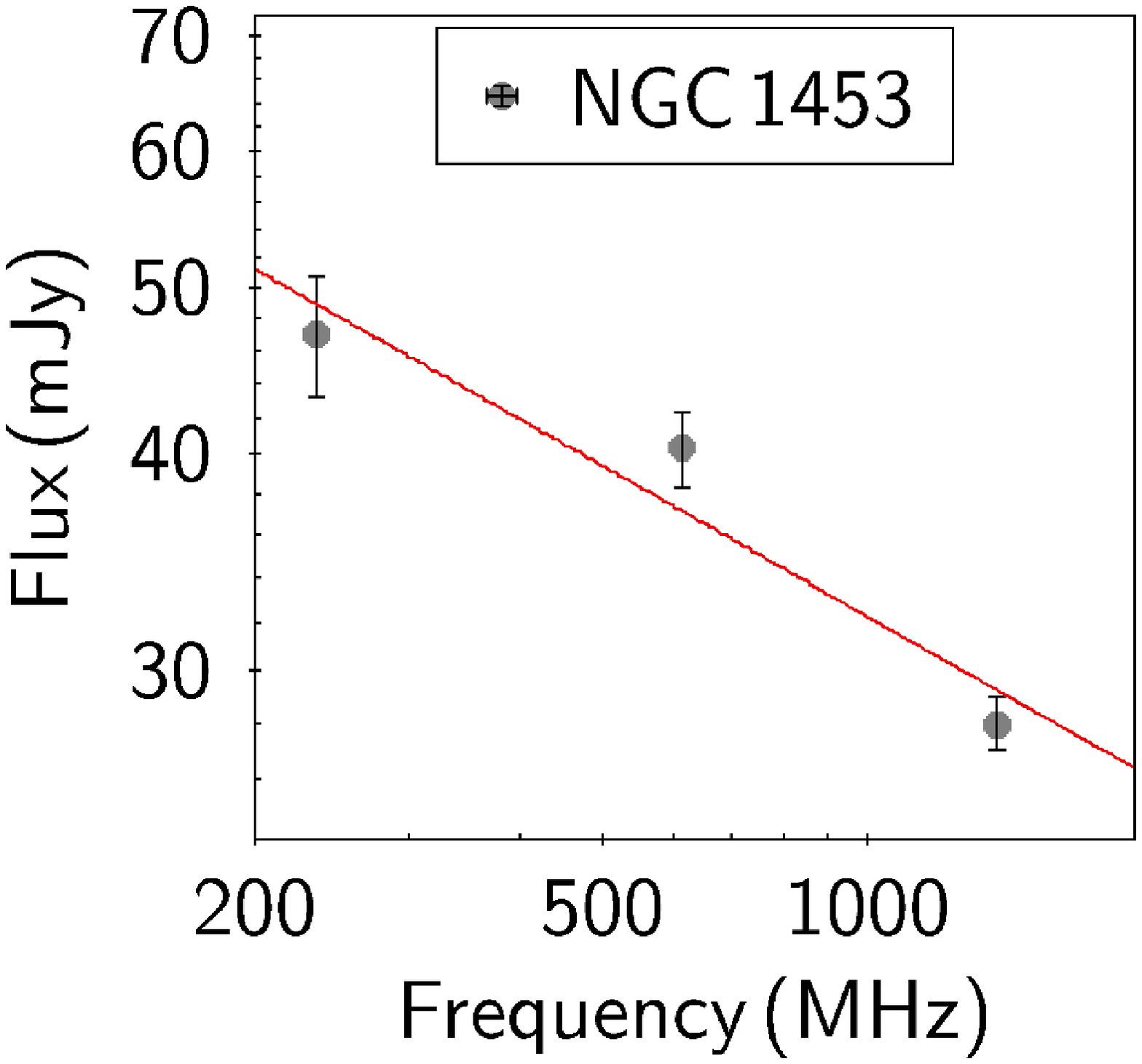}&
\includegraphics[width=0.215\textwidth]{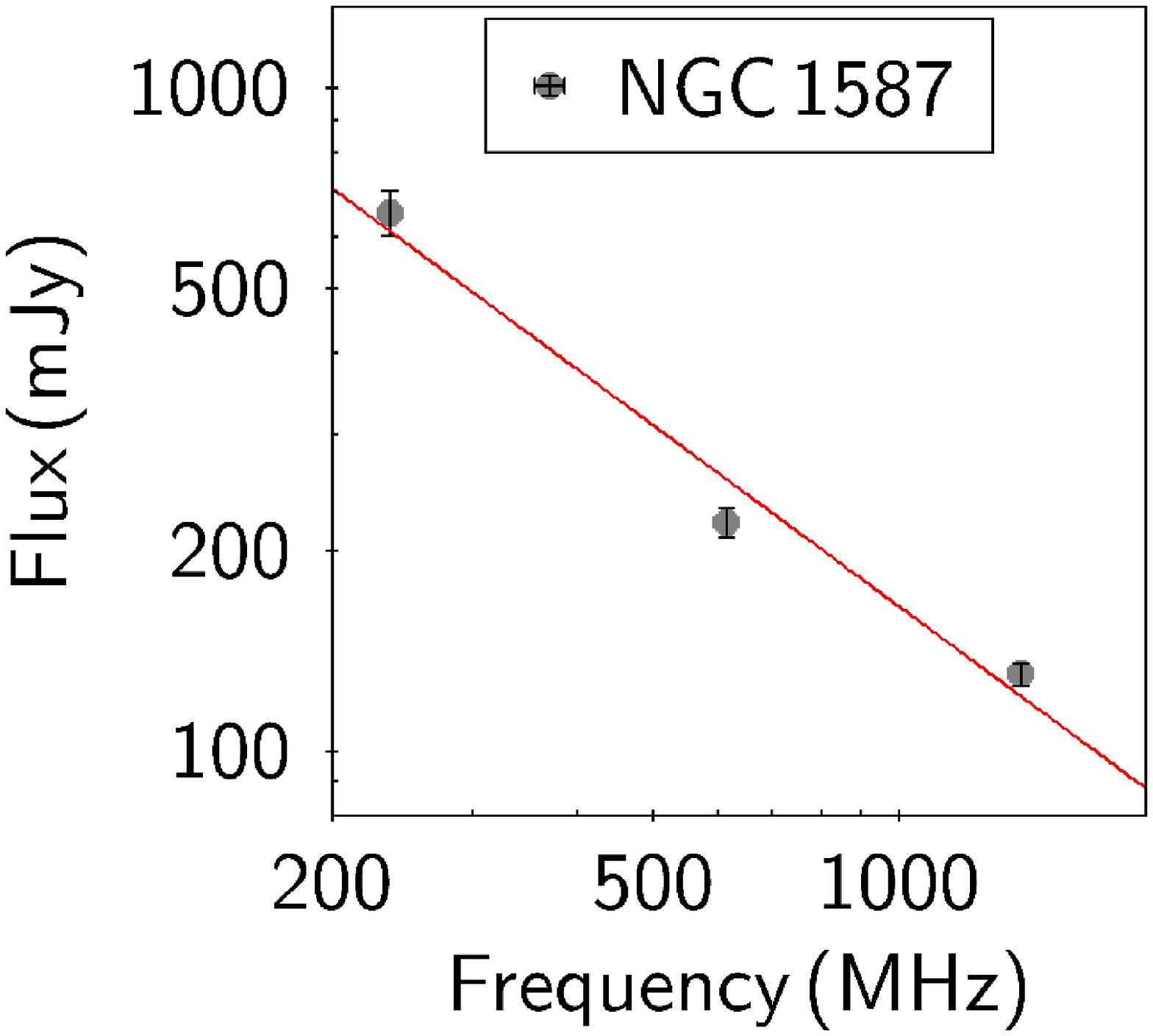}& 
\includegraphics[width=0.215\textwidth]{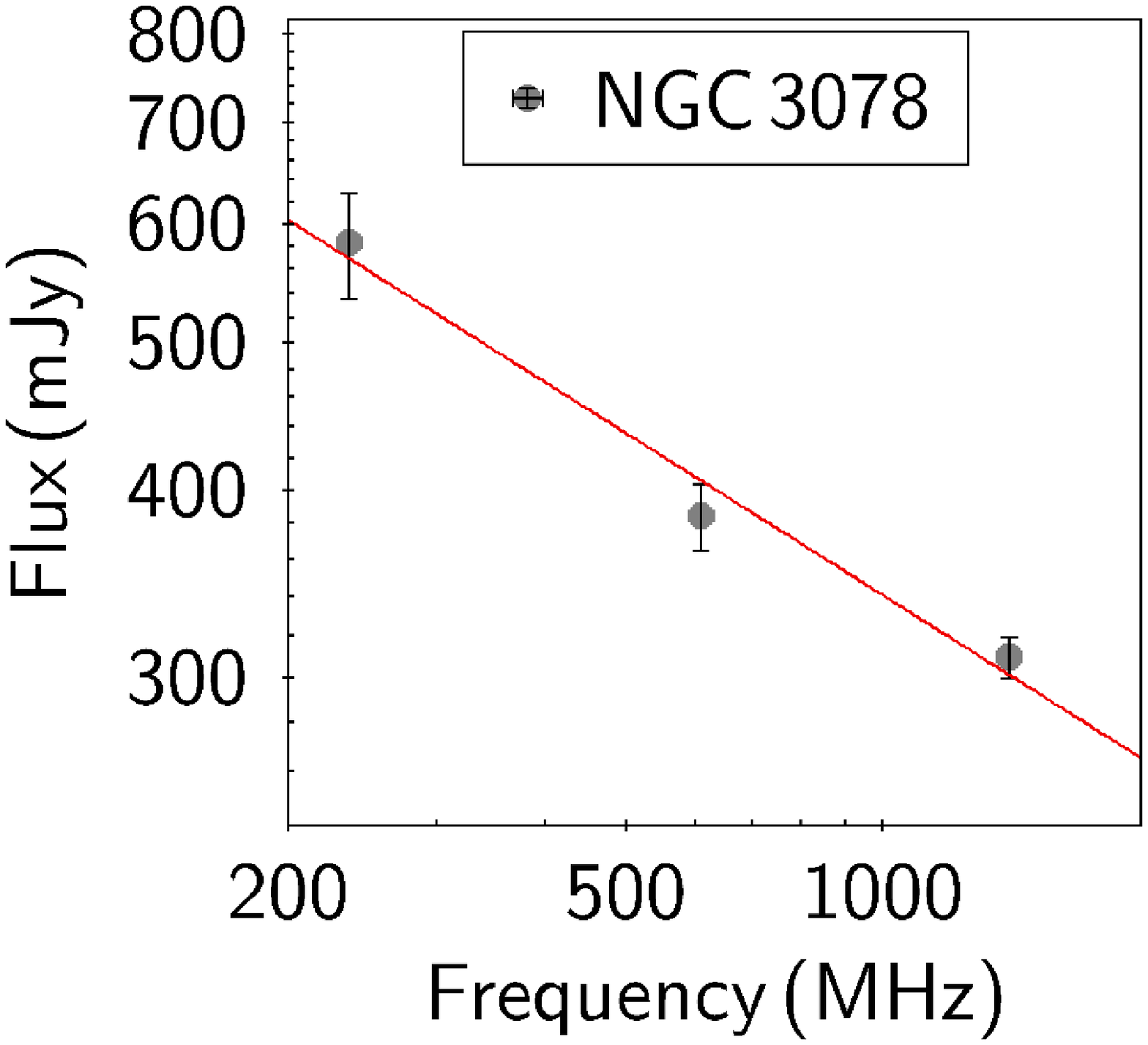}\\
\includegraphics[width=0.21\textwidth]{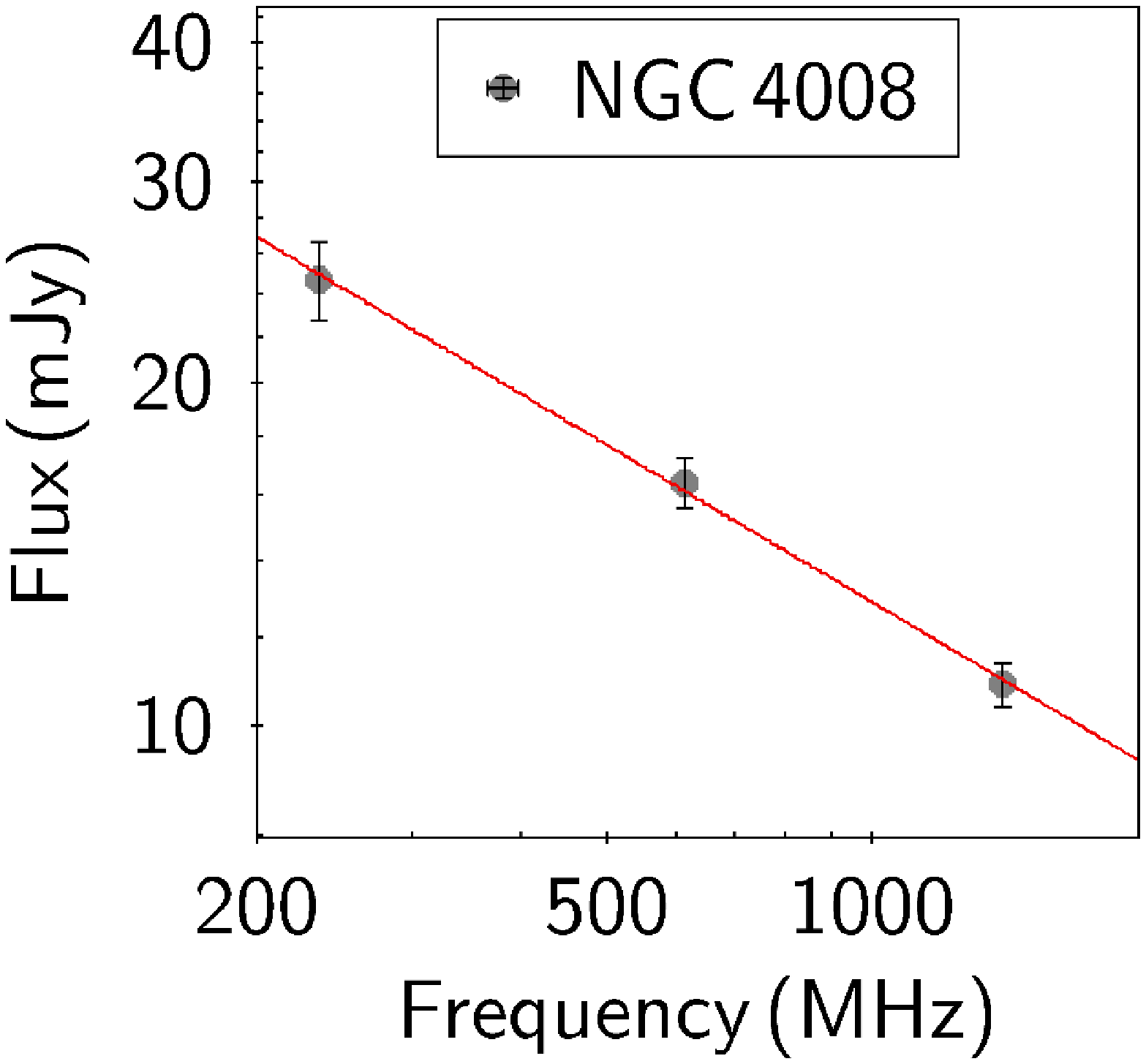}& 
\includegraphics[width=0.215\textwidth]{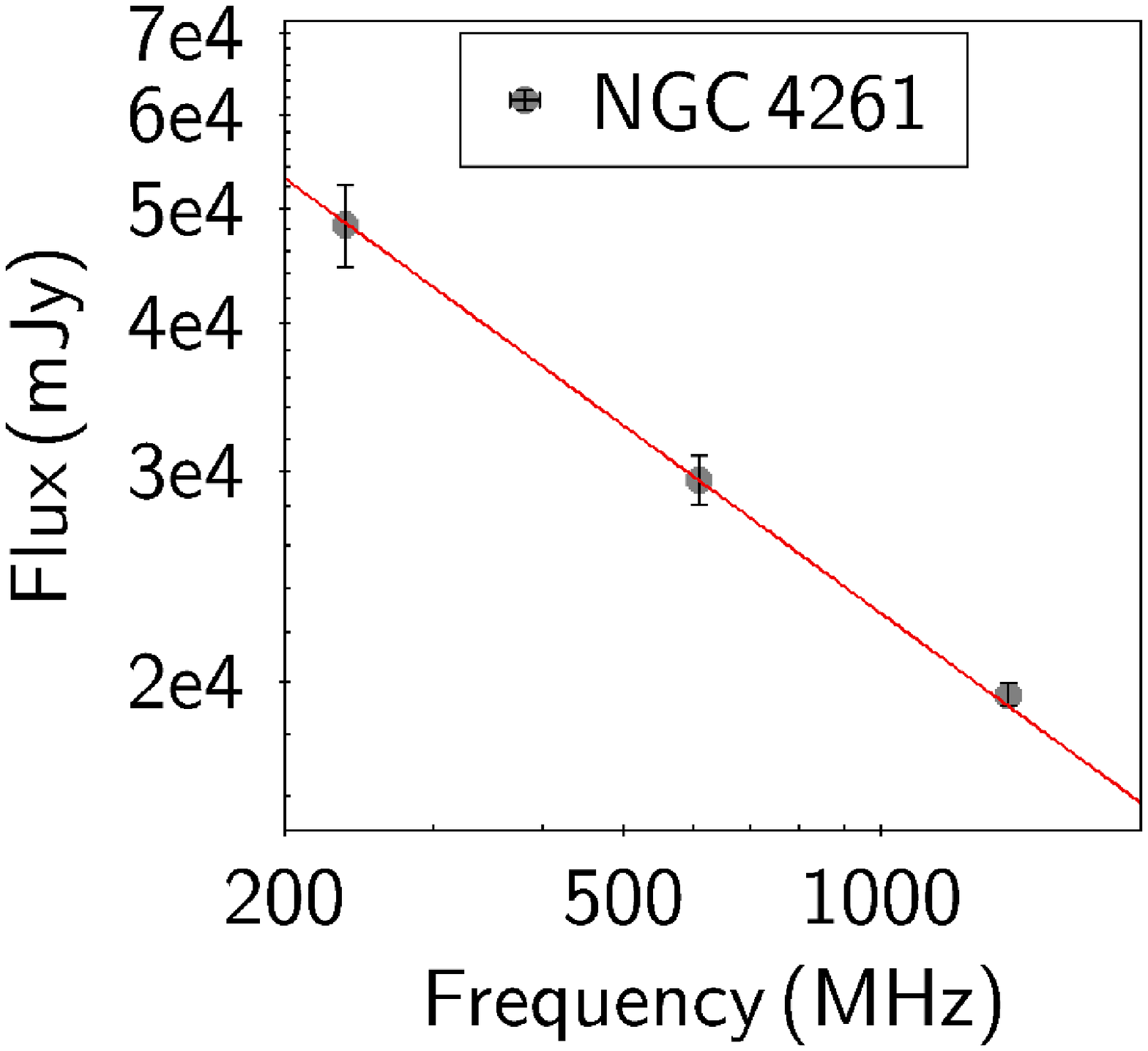}&
\includegraphics[width=0.21\textwidth]{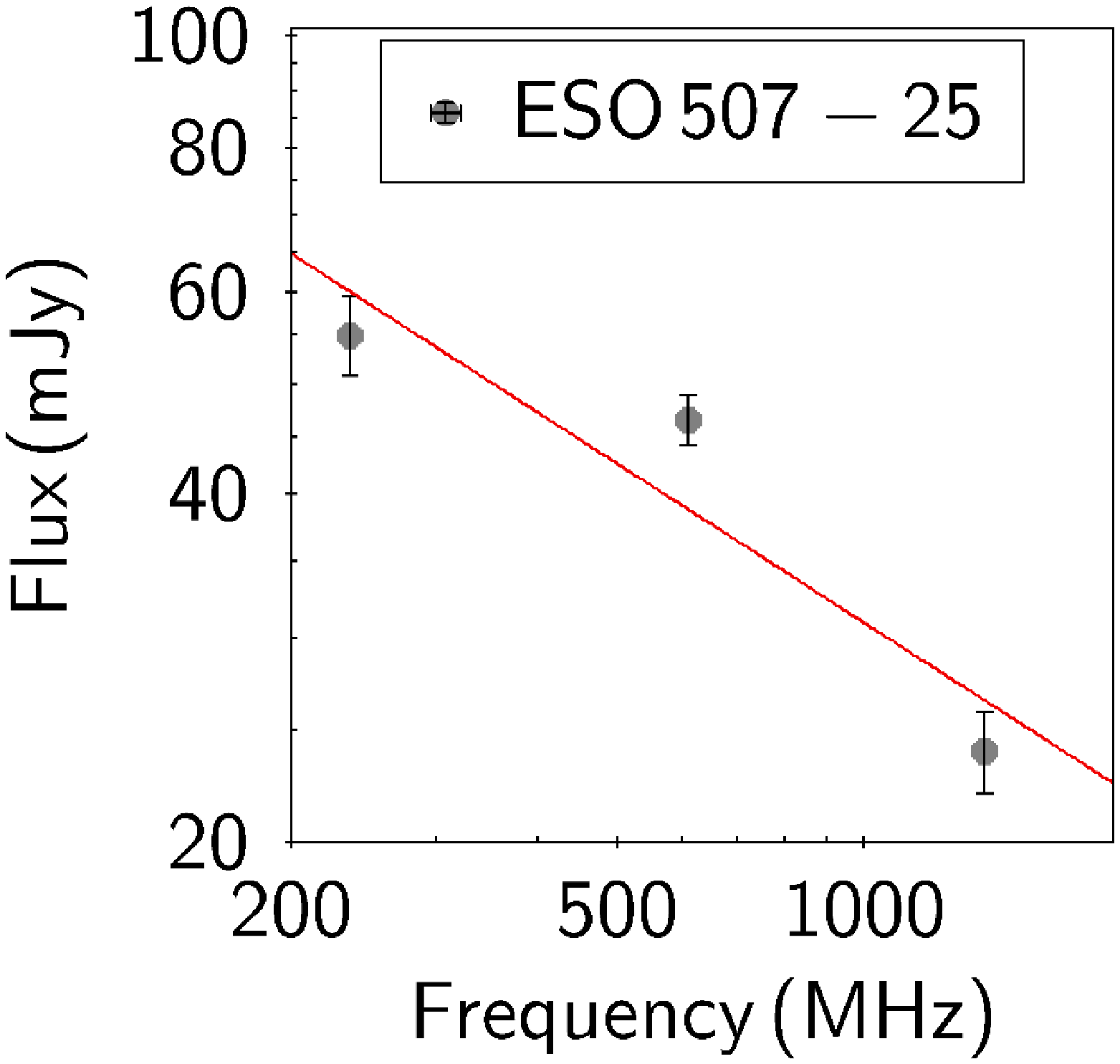}\\  
\includegraphics[width=0.215\textwidth]{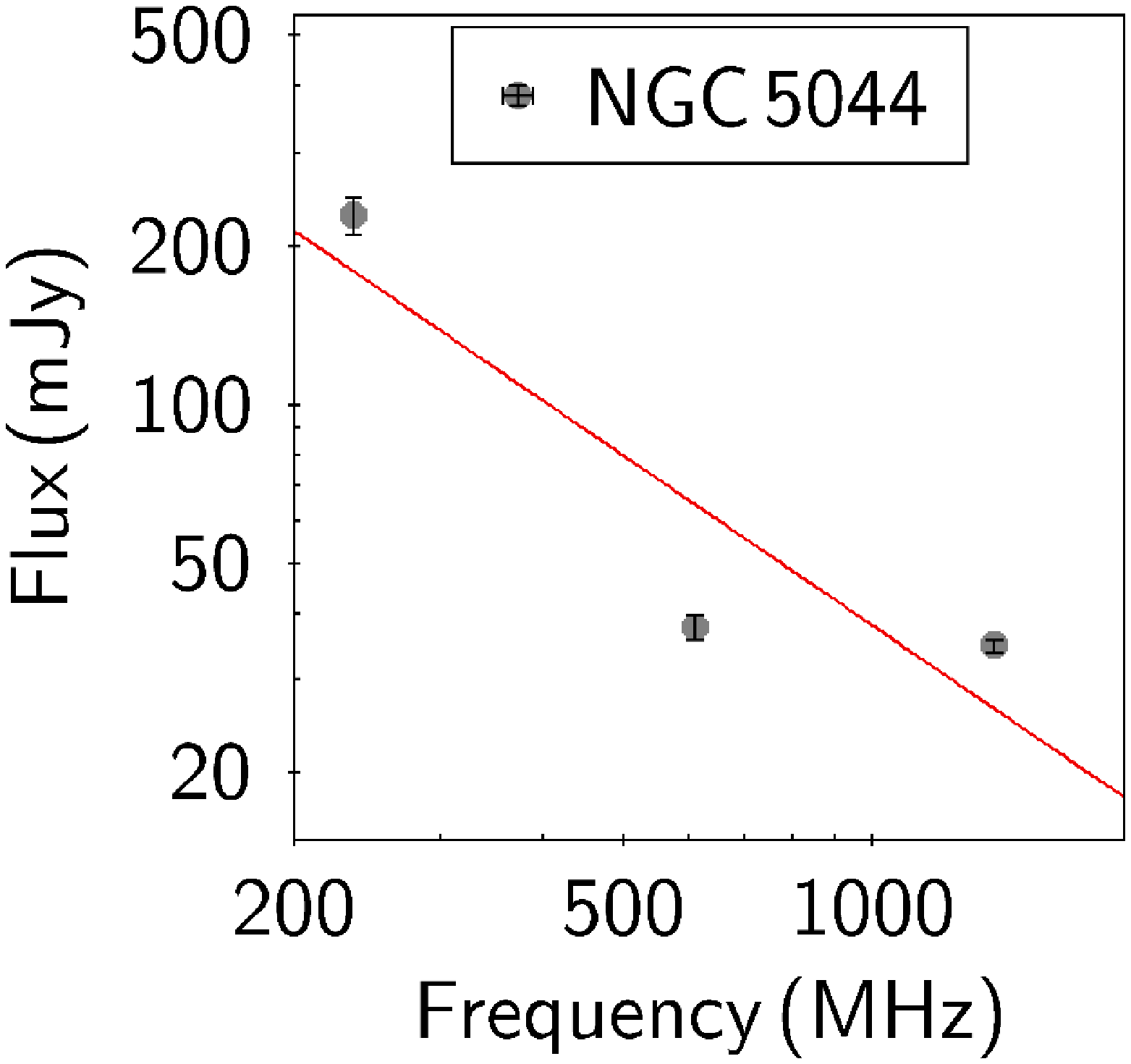}&
\includegraphics[width=0.23\textwidth]{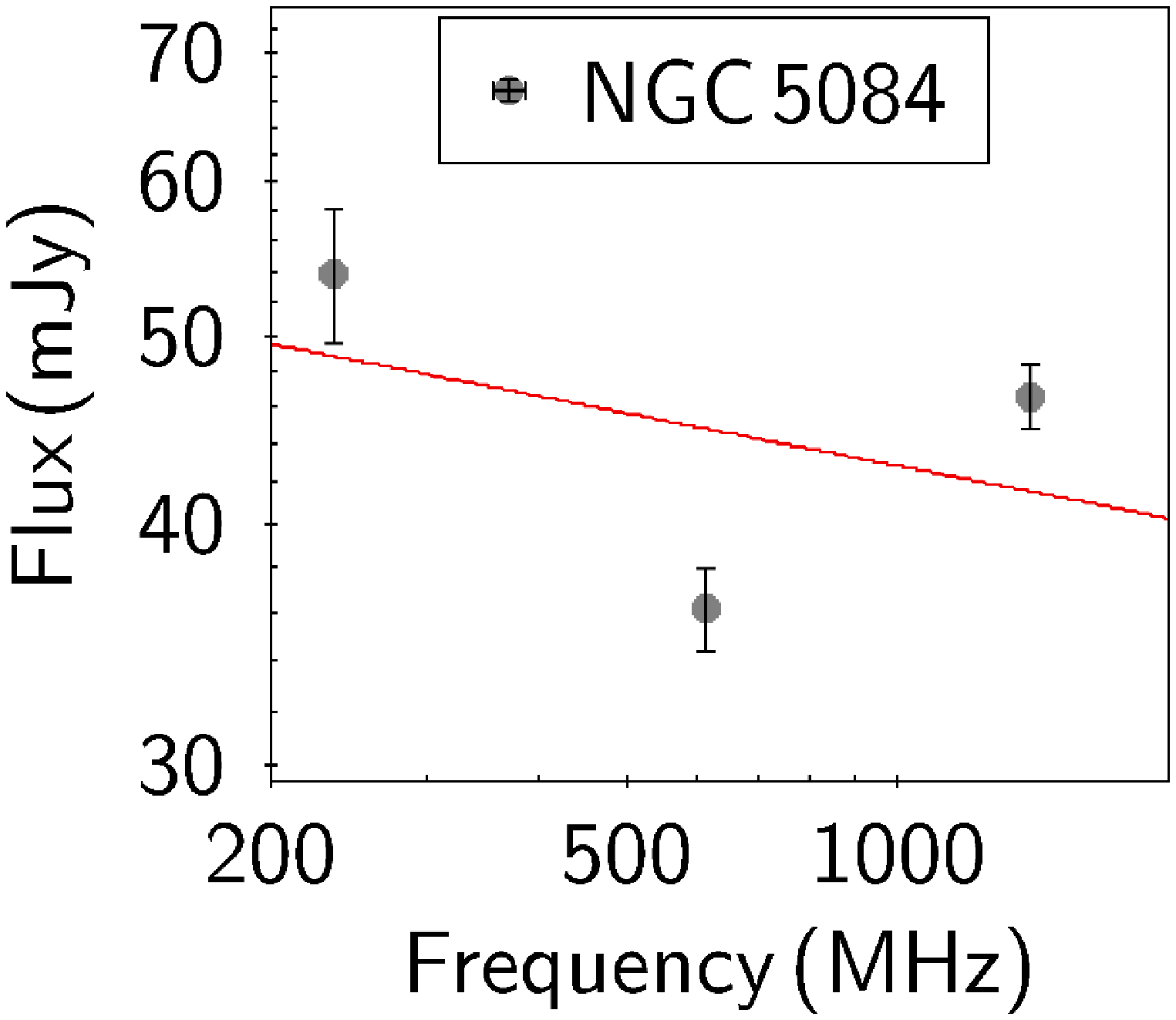}& 
\includegraphics[width=0.215\textwidth]{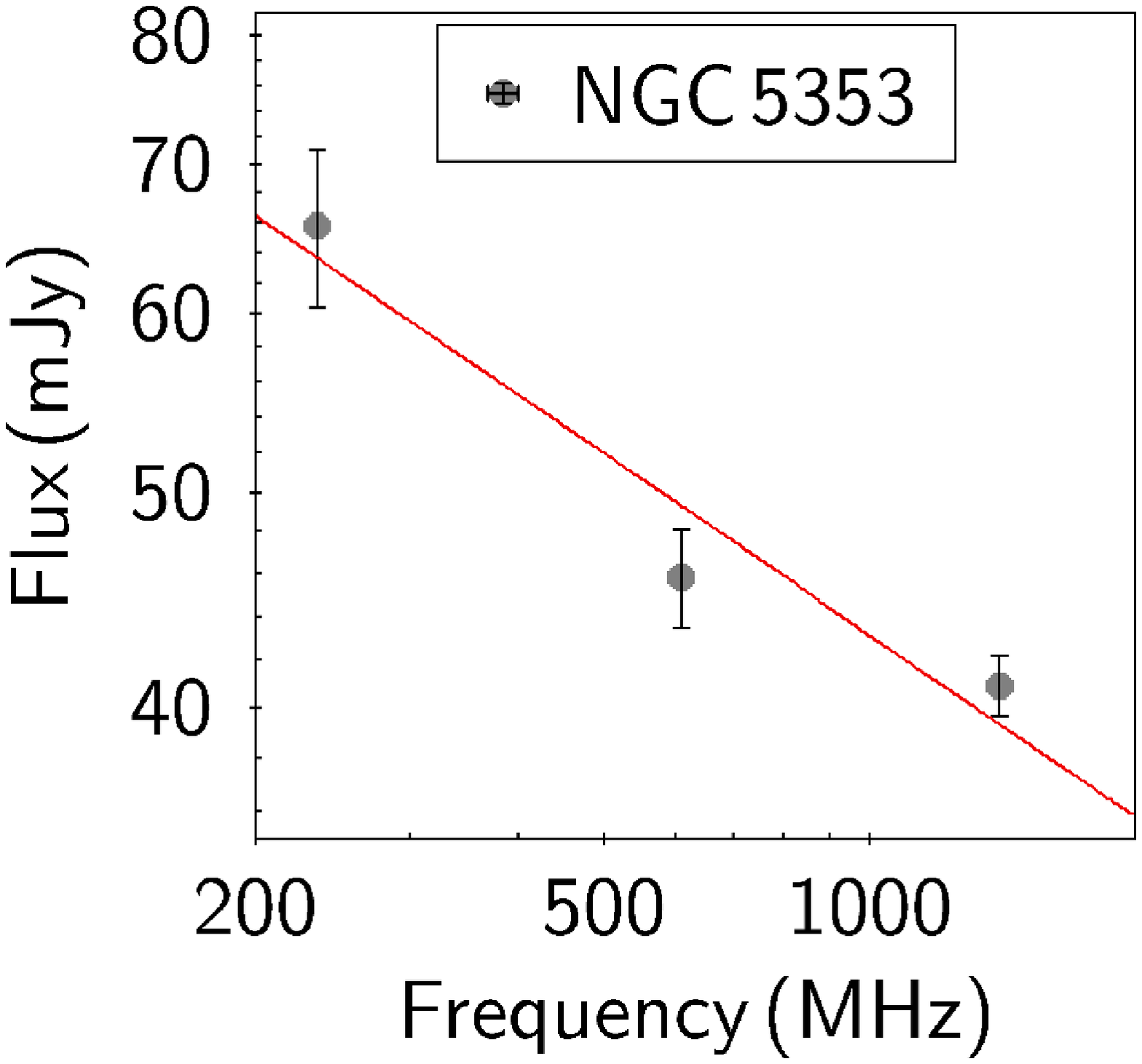}\\
\includegraphics[width=0.21\textwidth]{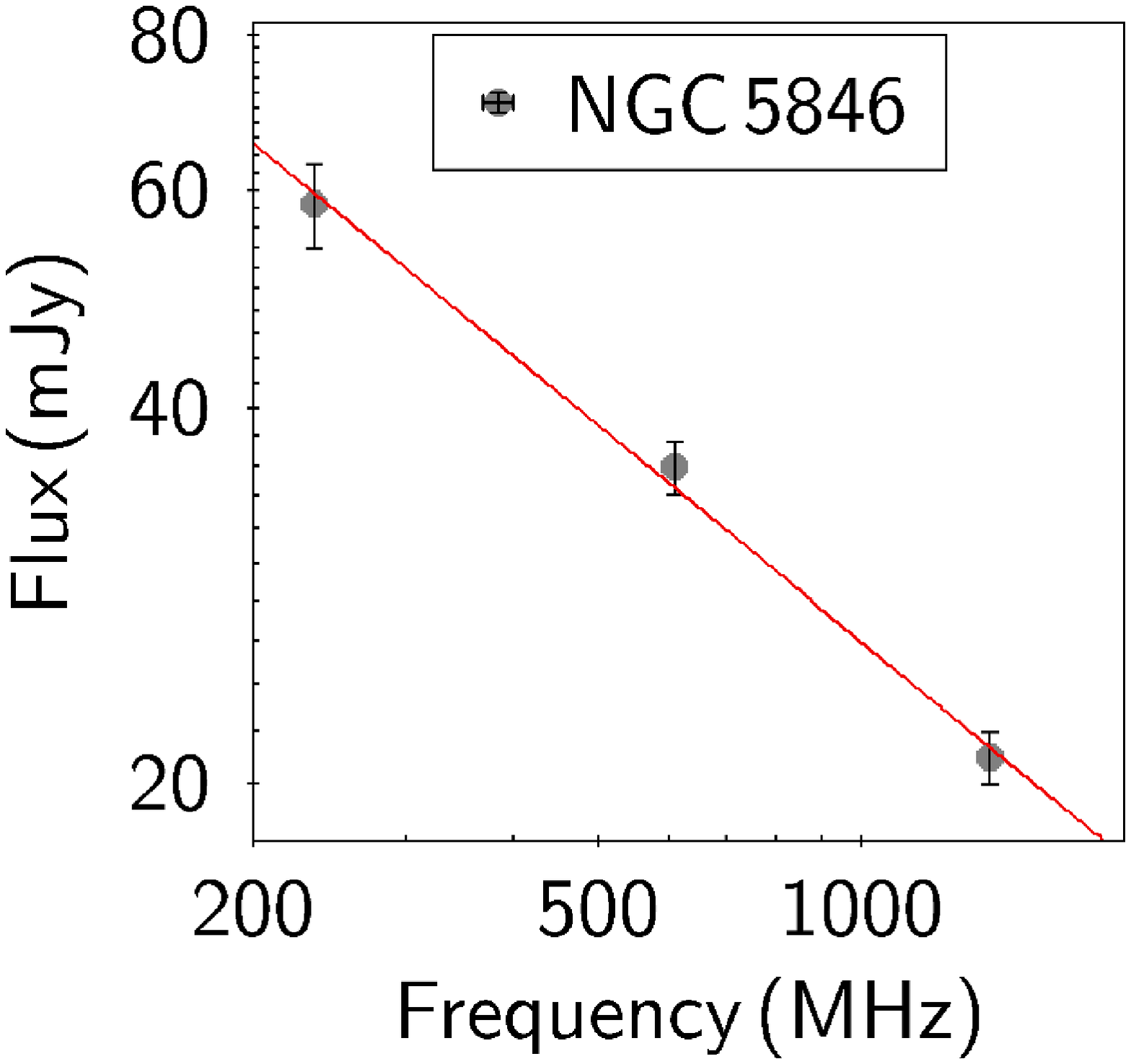}& 
\includegraphics[width=0.21\textwidth]{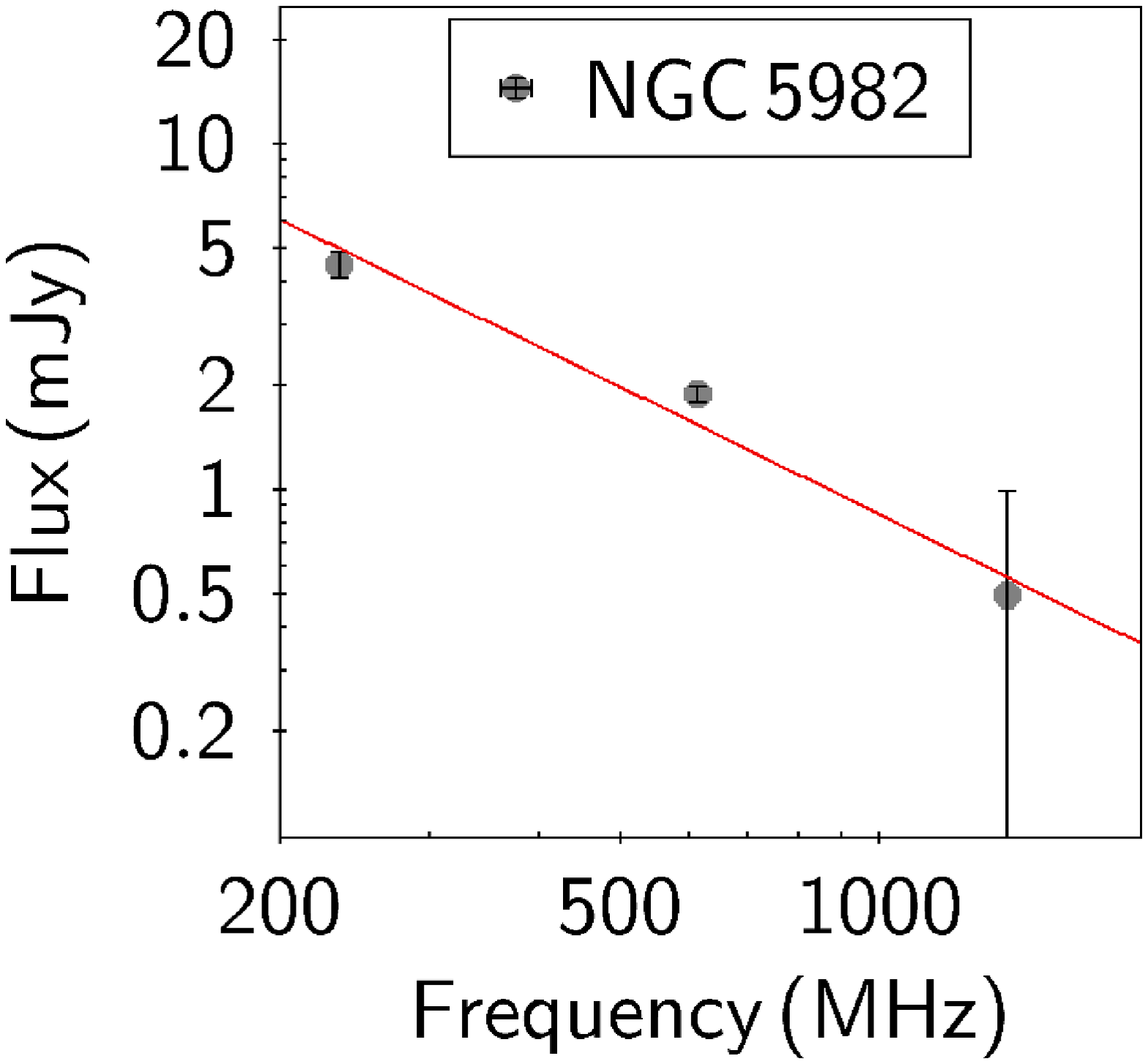}&
\includegraphics[width=0.21\textwidth]{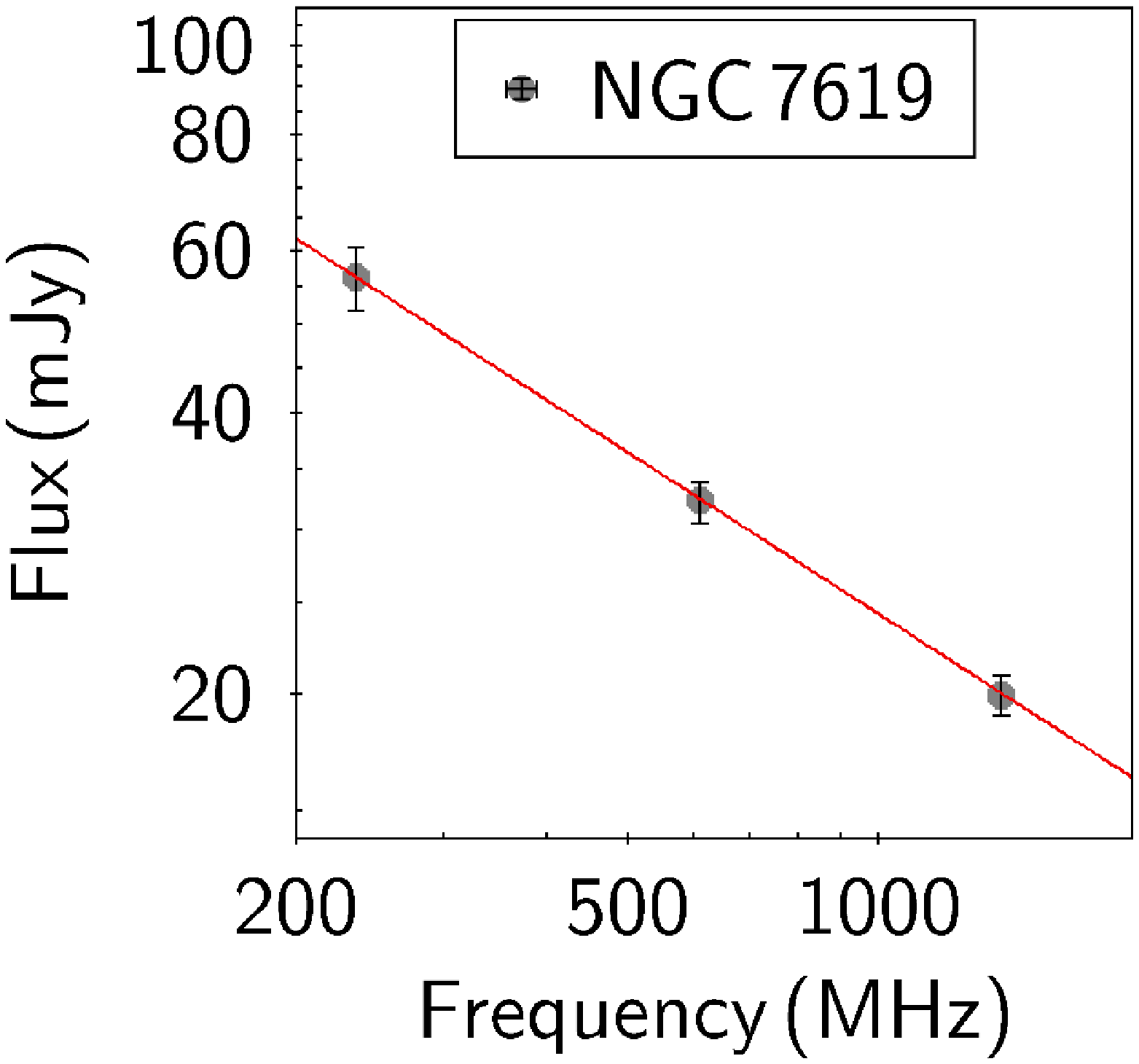}
\end{tabular}
\end{figure*}

Figure~\ref{power} shows the radio spectra of the BGEs detected at 235, 610 and 1400~MHz.

\bsp	
\label{lastpage}
\end{document}